\documentclass[12pt]{article}

  \usepackage{graphicx}
  \usepackage{epsfig}
  \usepackage{amssymb}
\usepackage{cite}
\usepackage{url}
\evensidemargin - .5truecm
\begin{document}
\newcommand{\pdf}{PDF~}
\newcommand{\pdfs}{PDFs~}

\newcommand{\la}{\left\langle}
\newcommand{\ra}{\right\rangle}
\newcommand{\lc}{\left[}
\newcommand{\rc}{\right]}
\newcommand{\lp}{\left(}
\newcommand{\rp}{\right)}

\newcommand{\rep}{\mathrm{rep}}

\newcommand{\be}{\begin{equation}}
\newcommand{\ee}{\end{equation}}
\newcommand{\nn}{\nonumber}
\newcommand{\bea}{\begin{eqnarray}}
\newcommand{\eea}{\end{eqnarray}}
\newcommand{\bfig}{\begin{figure}}
\newcommand{\efig}{\end{figure}}
\newcommand{\bc}{\begin{center}}
\newcommand{\ec}{\end{center}}
\def\ad{\dot{\alpha}}
\def\ov{\overline}
\def\hlf{\frac{1}{2}}
\def\qrt{\frac{1}{4}}
\def\as{\alpha_s}
\def\at{\alpha_t}
\def\ab{\alpha_b}
\def\sq2{\sqrt{2}}
\newcommand{\smallz}{{\scriptscriptstyle Z}} %
\newcommand{\mz}{M_\smallz}
\newcommand{\smallw}{{\scriptscriptstyle W}}
\newcommand{\mw}{m_\smallw} 
\newcommand{\smallh}{{\scriptscriptstyle H}}
\newcommand{\mh}{m_\smallh}
\newcommand{\mt}{m_t}
\newcommand{\mb}{m_b}
\newcommand{\wh}{w_\smallh}
\def\th{t_\smallh}
\def\zh{z_\smallh}
\newcommand{\text}[1]{#1}
\newcommand{\Mvariable}[1]{#1}
\newcommand{\Mfunction}[1]{#1}
\newcommand{\Muserfunction}[1]{#1}
\newcommand{\smartpap}{p\hskip-7pt\hbox{$^{^{(\!-\!)}}$}}

\newenvironment{appendletterA}
 {
  \typeout{ Starting Appendix \thesection }
  \setcounter{section}{0}
  \setcounter{equation}{0}
  \renewcommand{\theequation}{A\arabic{equation}}
 }{
  \typeout{Appendix done}
 }
\newenvironment{appendletterB}
 {
  \typeout{ Starting Appendix \thesection }
  \setcounter{equation}{0}
  \renewcommand{\theequation}{B\arabic{equation}}
 }{
  \typeout{Appendix done}
 }

%
%
%
%
%

\begin{titlepage}
\nopagebreak
\begin{flushright}
IFUM-956-FT
\end{flushright}  
\renewcommand{\thefootnote}{\fnsymbol{footnote}}
\vskip 2.5cm
\begin{center}
\boldmath
{\Large\bf The impact of PDF and $\alpha_s$ uncertainties on  \\[7pt]
Higgs Production in gluon--fusion at hadron colliders}\unboldmath
\vskip 1.5cm
{\large Federico Demartin$^a$, 
Stefano Forte$^{a,b}$,
Elisa Mariani$^a$,\\
Juan Rojo$^b$ and
Alessandro Vicini$^{a,b}$
\vskip .2cm
{\it 
$^a$Dipartimento di Fisica, Universit\`a di Milano, and \\
$^b$INFN, Sezione di Milano, \\[2mm]
Via Celoria 16, I--20133 Milano, Italy
} 
}
\end{center}
\vskip 0.7cm

\begin{abstract}
We present a systematic study of 
uncertainties due to parton distributions and the strong
coupling on the gluon--fusion  production cross section of the 
Standard Model Higgs at the Tevatron and LHC colliders. 
We compare procedures and results when three recent sets of
PDFs are used, CTEQ6.6, MSTW08 and NNPDF1.2, and we discuss
specifically the way PDF and strong coupling uncertainties are combined.
We find that  results obtained from different
PDF sets are in reasonable agreement if a common value of the strong coupling
is adopted. We show that the addition in quadrature of PDF and
$\alpha_s$  uncertainties
provides  an adequate approximation 
to the full result with exact error propagation. We discuss a simple
recipe to determine a conservative PDF+$\alpha_s$ uncertainty from
available global parton sets, and we use it to estimate this
uncertainty on the given process
to be about 10\% at the Tevatron and 5\% at the LHC for a
light Higgs.

\vskip .4cm

\end{abstract}
\vfill
\end{titlepage}    
\setcounter{footnote}{0}

\tableofcontents
\clearpage

\section{Introduction}
\label{sec:intro}

At the dawn of experimentation of LHC it is important to assess
carefully the expected accuracy of standard candle signal and
background measurements.  Standard model Higgs production is clearly
one such process. The main production mechanism for a scalar Higgs 
boson at the LHC is the gluon--fusion process 
($pp \to H+X$)~\cite{H2gQCD0}.  This process is also an especially
interesting test case to study QCD uncertainties: on the one hand, it
starts at $O(\alpha_s^2)$ and it undergoes large $O(\alpha_s^3)$
corrections which almost double the
cross  section~\cite{H2gQCD1,QCDg2}. 
On the other hand,
it is driven by the gluon distribution, which is only determined
starting at  $O(\alpha_s)$ (unlike quark distributions which can be
determined from parton--model processes). Therefore, this process is
quite sensitive both to parton distributions (PDFs)
and $\alpha_s$ uncertainties, and also on
their interplay.

Our knowledge of PDFs (see
Refs.~\cite{Dittmar:2005ed,Dittmar:2009ii}) and of
$\alpha_s$ (see
Refs.~\cite{Amsler:2008zzb,Bethke:2009jm}) have considerably improved
in the last several years. However, they remain the main source of
phenomenological  uncertainty related to the treatment of the strong
interaction: they limit the accuracy in a way  which cannot be improved
upon by increasing the theoretical accuracy.  
It is the purpose of this work to explore this uncertainty using Higgs
production in gluon--fusion as a test case.
Our goal is fivefold:
\begin{itemize}
\item We would like to compare the procedure recommended by various
  groups to combine PDF uncertainties and $\alpha_s$ uncertainties
  (and specifically Hessian--based approaches with Monte Carlo
  approaches) both in terms of procedure and in terms of results. 
\item We would like to assess the impact of the correlation between
  the value of $\alpha_s$ and the PDFs both when determining central
  values and uncertainty bands, and specifically understand how much
  results change when this correlation is taken into account in
  comparison to  the case in which $\alpha_s$ and PDF variations are
  done independently with results added in quadrature.
\item We would like to assess how much of the difference in results
  found  when using different PDF sets  (both for central values and
  uncertainty bands) is due to the PDFs, and how much is due to the
  choice of value of $\alpha_s$. 
\item We would like to compare how much of the total uncertainty is
  due to $\alpha_s$ and how much is due to PDFs.
\item We would like to arrive at an assessment of the value and
  PDF+$\alpha_s$ 
uncertainty for this cross section and more in general at a procedure
  to estimate them.
\end{itemize}
For each of these issues Higgs production through gluon--fusion is an
interesting test case in that it is likely to provide 
a worst--case scenario: differences between
results obtained using different PDF sets or following different
procedures for the combination of uncertainties 
are likely to be smaller for many other relevant processes. For
instance, in processes involving quark PDFs the correlation between
PDFs and the value of $\alpha_s$ is likely to be weaker, and thus the
results found adding uncertainties in quadrature are likely to differ
less from those obtained when the correlation between PDFs and
$\alpha_s$ is fully accounted for.

The studies performed here will be done using PDF sets from the 
CTEQ, MSTW and
NNPDF Collaborations, specifically the PDF sets
CTEQ6.6~\cite{Nadolsky:2008zw}, MSTW08~\cite{Martin:2009iq} and 
NNPDF1.2~\cite{Ball:2009by,Ball:2009mk}.  In order to account for the 
$\alpha_s$
dependence, we will use PDF sets with varying $\alpha_s$ which have
been published by CTEQ~\cite{Nadolsky:2008zw} and
MSTW~\cite{Martin:2009bu}, as well as NNPDF1.2 sets with varying
$\alpha_s$~\cite{Mariani}
which will be
presented here for the first time. Comparison between CTEQ and MSTW on
one side and NNPDF on the other side will enable us to contrast
results obtained in the Hessian approach of the former with those
found in the Monte Carlo approach of the latter. 
Computations will be performed at
next-to-leading order (NLO) in the strong coupling $\alpha_s$,
because, even though  NNLO results for the process we study are
available~\cite{H2gQCD2}, global parton fits with a full  treatment
of both DIS and hadronic data only exist at NLO (for instance, the
MSTW08 set~\cite{Martin:2009iq} only treats DIS fully at NNLO, while 
Drell-Yan is described using  $K$--factors, and jets using NLO theory).
There are of course several other sources of uncertainty on standard
candles at colliders, such as electroweak uncertainties and further
QCD uncertainties unrelated to PDFs, but their study goes beyond the
scope of this work: here we concentrate on PDF uncertainties, and on
the $\alpha_s$ uncertainty which is tangled with them.

The outline of the paper is as follows: in Sect.~\ref{sec:setup} we
summarize the computation of the Higgs production cross section and
the choice of  value of $\alpha_s$.
In Sect.~\ref{sec:nnpdf} we discuss and compare PDF sets 
with varying $\alpha_s$, and specifically present the NNPDF1.2 sets
with varying $\alpha_s$, which allow for a direct computation of the
correlation between $\alpha_s$ and the gluon. We then turn to a
comparison of predictions obtained using different PDF sets: first, in 
Sect.~\ref{sec:pdf-unc} we study PDF uncertainties and compare
predictions for the cross section and the PDF uncertainty on it
obtained using different sets; then in Sect.~\ref{sec:as} we discuss
$\alpha_s$ uncertainties and their combination with PDF uncertainties;
finally in Sect.~\ref{sec:finres} we compare final results and discuss
a procedure to construct a combined prediction from the available sets.
Conclusions are drawn in  Sect.~\ref{sec:conclusions}.

\clearpage
\section{The Higgs boson production cross section}

\label{sec:setup}
\subsection{The hadron--level cross section}

The hadronic total cross section for the production of a Standard Model
 Higgs of mass $m_H$ via gluon--fusion at 
center-of-mass energy $\sqrt{s}$ is
\bea
\sigma(h_1 + h_2 \to H+X) & = & 
          \sum_{a,b}\int_0^1 dx_1 dx_2 \,\,f_{a,h_1}(x_1,\mu_F^2)\,
         f_{b,h_2}(x_2,\mu_F^2) \nonumber\\
& & \times
\int_0^1 dz~ \delta \left(z-\frac{\tau_H}{x_1 x_2} \right)
\hat\sigma_{ab}(z) \, ,
\label{sigmafull}
\eea
where $\tau_H= \mh^2/s$, $\mu_F$ is the factorization scale,
$f_{a,h_i}(x,\mu_F^2)$ are the PDFs 
for parton $a, \,(a = g,q,\bar{q})$ of hadron $h_i$, and
$\hat\sigma_{ab}$ is 
the 
cross section for the partonic subprocess $ ab \to H +X$ at the center-of-mass 
energy  $\hat{s}=x_1 x_2 s=\mh^2/z$. The latter can be written as
\be
\hat\sigma_{ab}(z,\mu_R^2)=
\sigma^{(0)}\lp \mu_R^2 \rp\,z \, G_{ab}(z,\mu_R^2) \, ,
\label{Geq}
\ee
where the Born cross section is
\be
\sigma^{(0)}\lp \mu_R^2 \rp  =  
\frac{G_\mu \alpha_s^2 (\mu_R^2)}{512\, \sqrt{2} \, \pi}
\left|
   \sum_{q} 
\, 
{\mathcal G}^{(1l)}_{q}
       \right|^2 
\label{ggh}
\ee
and the sum runs over all quark flavors that appear in the amplitude
${\cal G}^{(1l)}$,
with 
\be
{\mathcal G}^{(1l)}_{q}  =  - 4 y_{q}
 \left[ 2 - \left( 1 -4 y_{q} \right)  \, H(0,0,x_{q}) \right] \, .
\label{eq:G1l} 
\ee
In Eq.~({\ref{eq:G1l}) we have defined
\be
y_q \equiv \frac{m_q^2}{\mh^2}, ~~~~~~~~
x_q \equiv \frac{\sqrt{1- 4 y_q} - 1}{\sqrt{1- 4 y_q} + 1} ,~~~~~\,
H (0,0,z ) = \frac{1}{2} \log^2 (z)~,
\label{defx}
\ee
with the standard notation for Harmonic Polylogarithms (HPLs).

Up to NLO,
\be
G_{a b}(z)  =  G_{a b}^{(0)}(z) 
       + \frac{\alpha_s (\mu^2_R)}{\pi} \, G_{a b}^{(1)}(z) \, 
\ee
where $a,b$ stand for any allowed parton.
Exact analytic results,
with the full dependence on the masses of the quarks running in the loop,
have been obtained for the NLO coefficient function $G_{ab}^{(1)}$
in~\cite{QCDg2} and more recently in terms of HPLs in~\cite{ABDV}. 

All numerical results presented in this paper have been obtained using a
code based on the expressions of Ref.~\cite{ABDV}. 
We will consider only the gluon--gluon channel, and evaluate
the total cross  section at NLO-QCD, with
the running of the strong coupling constant $\alpha_s(\mu_R^2)$
implemented as discussed in Sect.~\ref{sec:alphaspdfs} below.
The default choice for the renormalization scale is $\mu_R=m_H$.
The cross sections have been computed including the contributions due to
the top and the bottom quark running in the fermion loop, with masses 
$\mt=172~$GeV and $\mb=4.6~$GeV; the value of the Fermi constant is
$G_{\mu}=1.16637~10^{-5}$ GeV$^{-2}$.
The top mass has been renormalized in the on--shell scheme~\cite{QCDg2,ABDV}.


\subsection{The strong coupling $\as$}
\label{sec:alphaspdfs}

Even though the  strong coupling $\alpha_s$ can be determined by a
parton fit~\cite{Martin:2009bu}, 
its most accurate determination is arrived at by
combining results from many
high--energy processes, most of which do not depend on PDFs at all.
A recent combined determination~\cite{Bethke:2009jm} is
\be
\alpha_s(M_Z)=0.1184\pm0.0007 ,
\label{asbethke}
\ee
while the currently published\footnote{The 2009 web PDG update~\cite{pdgnew} no
  longer provides a combined determination of $\alpha_s$, and refers
  to Ref.~\cite{Bethke:2009jm}.}
 PDG average~\cite{Amsler:2008zzb}
has a rather more conservative
assessment of the uncertainty:
\be
\alpha_s(M_Z)=0.1176\pm0.002.
\label{aspdg}
\ee
Both these uncertainties should be understood as
one--$\sigma$, {\it i.e.}
68\% confidence levels.

Because it starts at order $\alpha_s^2$ and it undergoes sizable
$O(\alpha_s^3)$ corrections,  
the Higgs cross section is very sensitive to the central value of the
strong coupling and it is thus important to compare results obtained
using the same value of
$\alpha_s$. However, different values of $\alpha_s$ are adopted by
various parton fitting groups. Specifically, 
for the PDF sets we are interested in, the reference values are
\bea
\as(\mz)&=&0.118 \quad{\rm for~ CTEQ6.6},  \nonumber  \\
\as(\mz)&=&0.119 \quad{\rm for~ NNPDF1.2}, \label{asrefcoll}\\
\as(\mz)&=&0.12018 \quad{\rm for~ MSTW08}.\nonumber
\eea
Therefore, in order to obtain a meaningful comparison, 
we must study the dependence of results obtained with different sets
when the value of $\alpha_s$ is varied about the central values of
Eq.~(\ref{asrefcoll}).

For comparison of relative uncertainties, which are less sensitive to
central value of $\alpha_s$, but of course very sensitive to the range
in which $\alpha_s$ is varied, we will assume  that the
one--$\sigma$ and 90\% confidence level variations of $\alpha_s$ are
respectively
given by
\bea
\label{eq:alphasref}
\Delta^{(68)}\alpha_s&=&0.0012=0.002/c_{90}  \label{sas68} \\
\Delta^{(90)}\as&=&0.002,  \label{sas90} 
\eea
where $c_{90}=1.64485...$ is the number of standard deviations for a
gaussian distribution that correspond to a 90\%~C.L. interval. Our choice
Eqs.~(\ref{sas68}-\ref{sas90}) is thus  intermediate between the
choices in
Eqs.~(\ref{aspdg}--\ref{asbethke}).
A reassessment of the value $\alpha_s$ and
and its uncertainty go beyond the scope of this work: these
values are chosen as a reasonable reference, and ensure that
our results will also be valid  for other reasonable choices in the
same ballpark.

\begin{figure}
\begin{center}
\includegraphics[width=.5\linewidth]{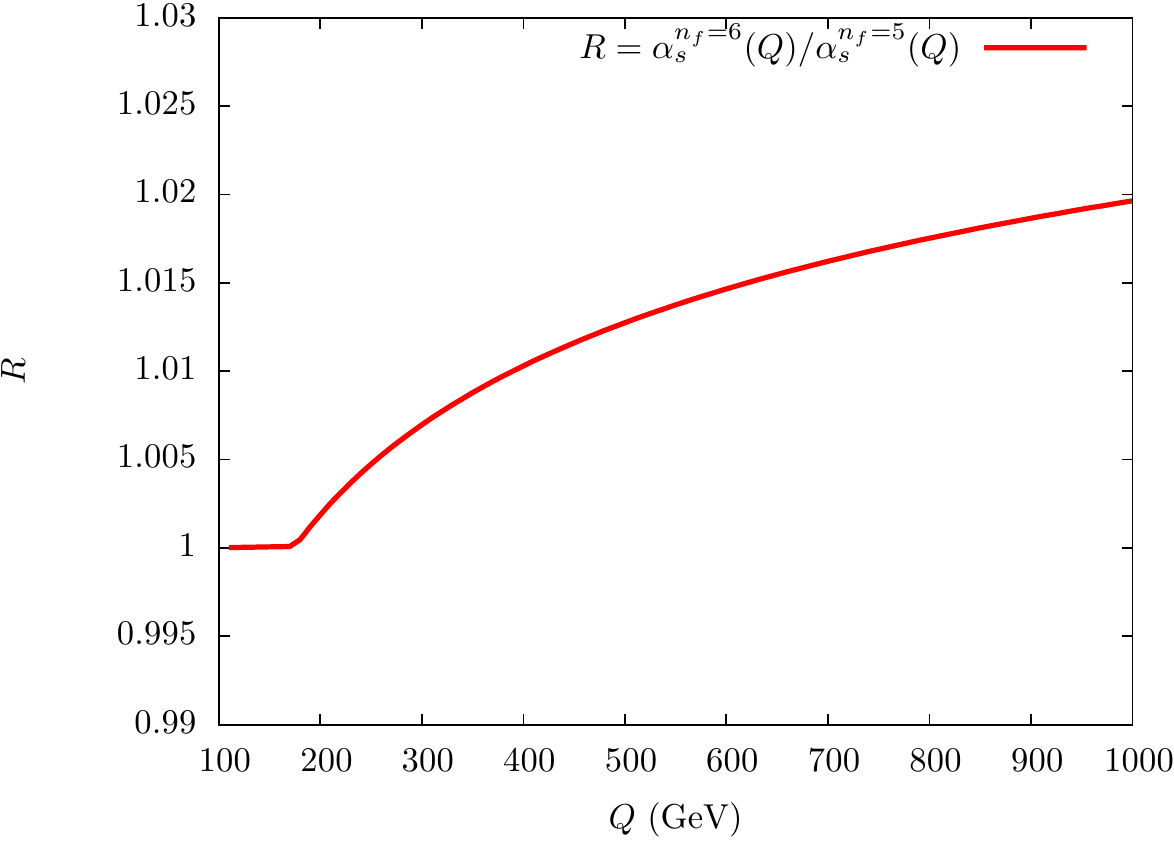}
\caption{\label{fig:as_nf56} \small
Ratio of the strong coupling determined with a number of flavours that varies
from $n_f=5$ to  $n_f=6$ at $Q>\mt$ to that with $n_f=5$ at all scales.}
\end{center}
\end{figure}

An important subtlety is the number of active flavours in the
running of the strong coupling, as implemented
by various PDF analyses. Indeed, QCD calculations are usually
performed in a decoupling scheme~\cite{Collins:1978wz}, in which
heavy flavours decouple at
scales much lower than their mass. When studying a process like Higgs
production for a wide range of the Higgs mass one must then specify
what to do above top threshold, and specifically fix the scale at
which heavy quarks decouple, which amounts to a choice of
renormalization scheme. This choice should be used consistently in the
running of $\alpha_s$, the evolution equations for PDFs, and the
computation of hard matrix elements. In Refs.~\cite{QCDg2,ABDV}, a
scheme in which the number of active flavours becomes $n_f=6$ at
$Q^2=m_t^2$ is adopted; this scheme is also used by NNPDF. However,
CTEQ and MSTW instead use a scheme in which $n_f=5$ even when
$Q^2>m_t^2$ both in the running of the strong coupling and in PDF
evolution. 

The effect of this on the running of $\alpha_s$ is small but non--negligible:
in Fig.~\ref{fig:as_nf56} we plot the ratio of the variable--flavor
$\alpha_s$ to the fixed--flavor one, and show that for $Q=2m_t$ the
discrepancy is already of order of 1\%, and thus the effect on a
quantity which depends on higher powers of $\alpha_s$ accordingly
larger. Of course,  this scheme dependence cancels to a large extent
once PDFs and hard cross sections are consistently combined, and a
fully consistent comparison would require the use of the same scheme
everywhere. This is difficult in practice because of the different
choices adopted by NNPDF on the one hand, and CTEQ and MSTW on the
other hand. For the sake of comparisons below, we will use the NLO
Higgs cross section from Refs.~\cite{QCDg2,ABDV} and, consistently
$\alpha_s$ which runs with $n_f$ that varies at $Q=m_t$: this is then
also consistent with the evolution equations used to construct the
NNPDF set, but not with those used to
construct the CTEQ and MSTW set. It should be born in mind that this
incomplete cancellation of the scheme dependence
above the top threshold
may lead to a spurious difference between central values at the
percent level between NNPDF and other groups.

\clearpage

\section{Parton sets with variable strong coupling}
\label{sec:nnpdf}

\subsection{PDFs and $\alpha_s$ in a Hessian approach}
\label{sec:mstwcteqseries}

PDF sets
with varying $\alpha_s$ have been presented by the
CTEQ~\cite{Nadolsky:2008zw} and 
MSTW~\cite{Martin:2009bu} collaborations. Specifically, CTEQ has
released the CTEQ6.6alphas sets, which add to
the central  CTEQ6.6 fit~\cite{Nadolsky:2008zw}, which has
$\alpha_s=0.118$, four more sets with
$\alpha_s=0.116,\>0.117,\>0.119,\>0.120$. In these fits, $\alpha_s$ is
taken as an external parameter which is fixed each time to the given
value along with all other physical parameters in the fit (such as,
say, the fine--structure constant or the $W$ mass). However,
eigenvector PDF sets for the computation of PDF uncertainties are only
provided for the central CTEQ6.6 set. Therefore, it is possible to
study the correlation between the value of  $\alpha_s$ and PDFs, but
not the correlation with their uncertainties.

 The MSTW collaboration instead
has performed a simultaneous determination of PDFs and $\alpha_s$,
which is thus not treated as a fixed external parameter, but rather as
a fit parameter, which leads  to the central value quoted in
Eq.~(\ref{asrefcoll}}). Furthermore, 
MSTW has also released sets of PDFs, analogous to the
CTEQ sets discussed above, in which  $\alpha_s$ is
taken as an external parameter, and varied in steps of $0.001$ for
$0.110\le \alpha_s\le 0.130$. The sets in which PDFs and $\alpha_s$ are
determined simultaneously may be used
for a determination of the
  correlation between the value of $\alpha_s$ and both PDF central
  values and PDF uncertainties, though with the limitation that
the value and uncertainty on $\alpha_s$ found in the fit must
be used.

\begin{figure}
\begin{center}
\includegraphics[width=0.48\textwidth]{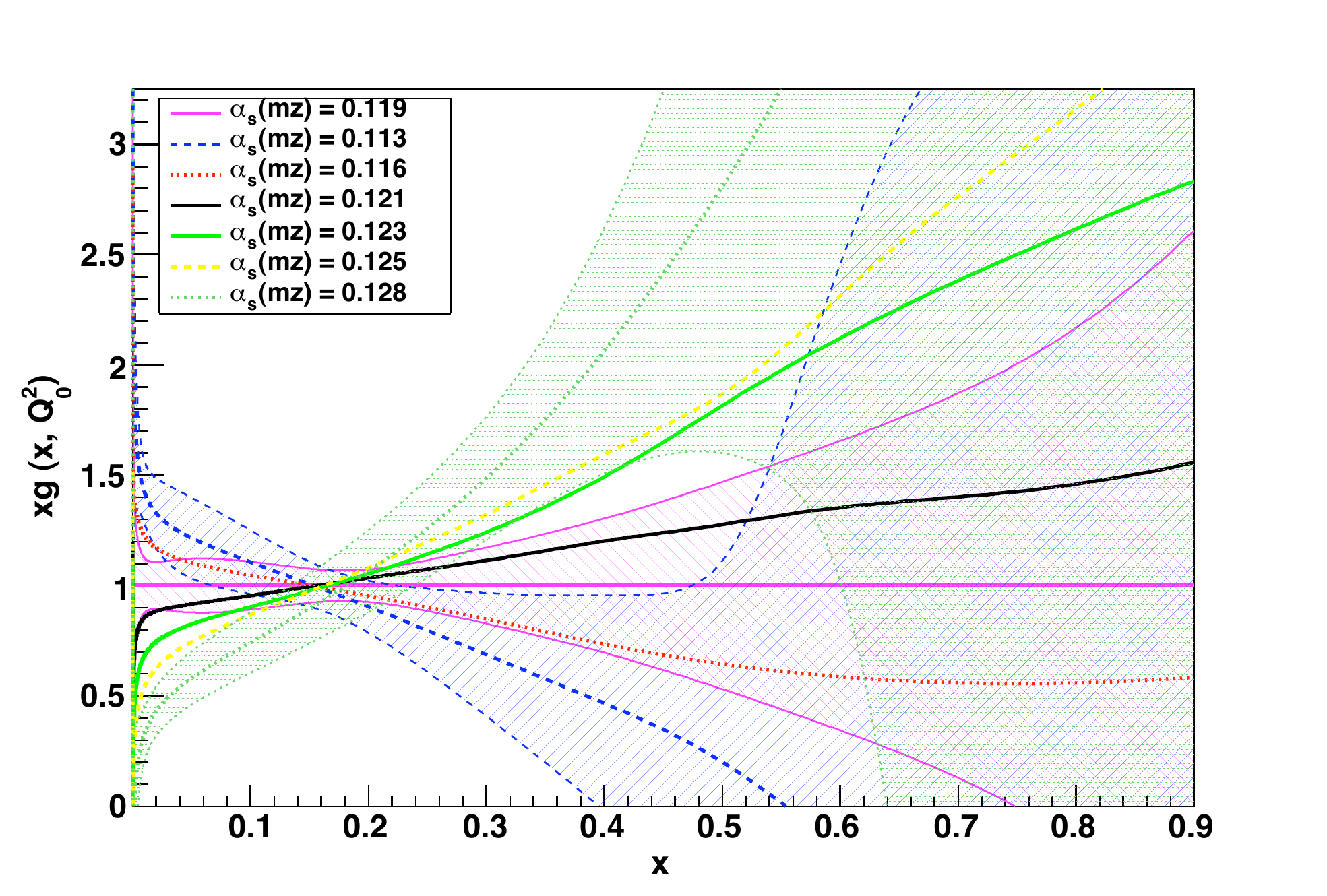}
\includegraphics[width=0.48\textwidth]{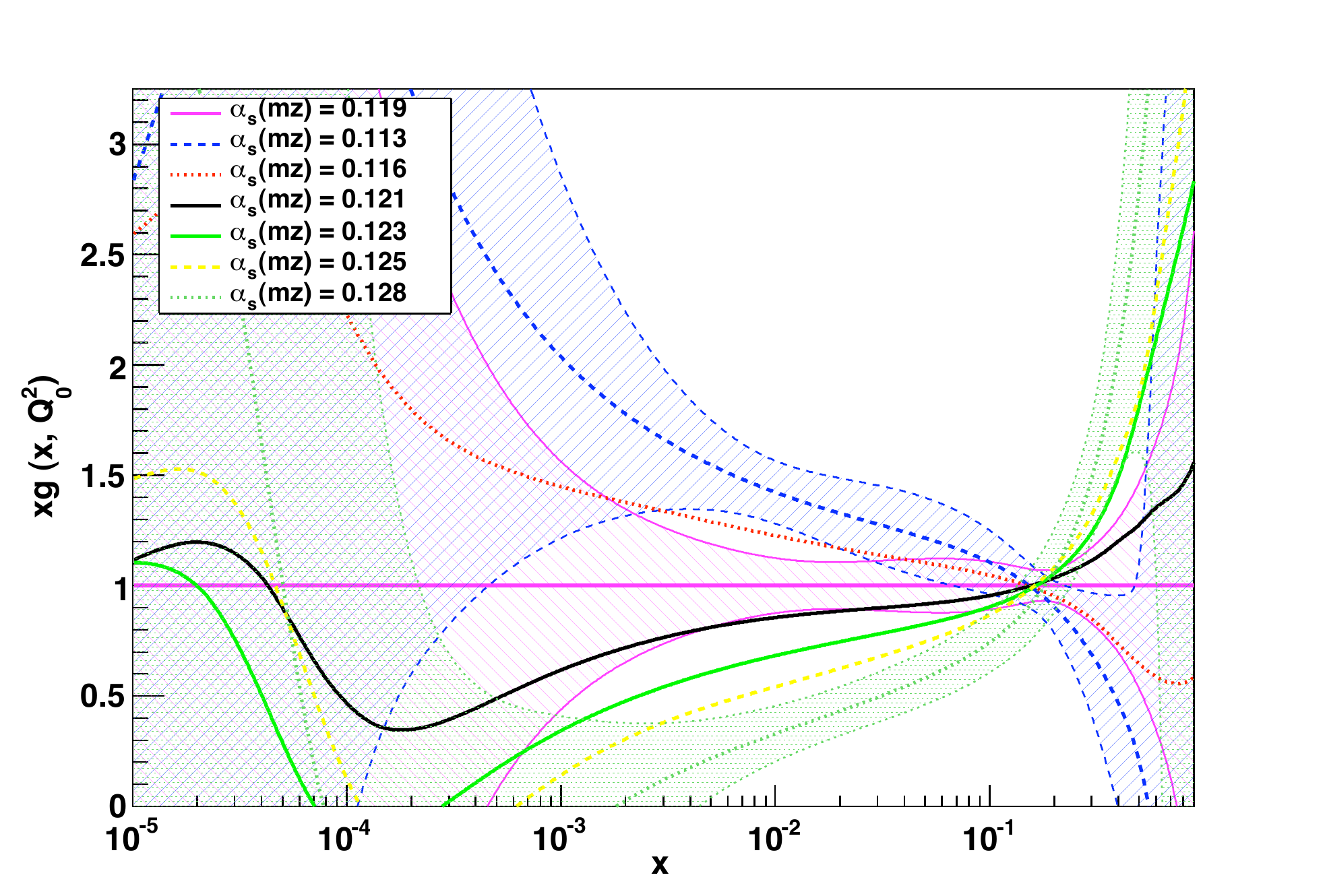}
 \caption{\small The ratio of the central gluons obtained in NNPDF1.2
fits when $\alpha_s$ is varied, divided by the reference
NNPDF1.2 gluon at the initial evolution scale
$Q^2_0=2$ GeV$^2$. The comparison
is shown both in a linear (left) and logarithmic (right)
scales. One--$\sigma$ uncertainty  bands are shown for the central,
highest and lowest values of $\alpha_s$.
}
\label{fig:xg_comp}
\end{center}
\end{figure}
\subsection{PDFs and $\alpha_s$ in a Monte Carlo approach}
\label{sec:nnpdfseries}

In the NNPDF parton determination, $\alpha_s$ is  taken as a fixed
parameter as in the CTEQ fits; 
the correlation between NNPDF1.2 PDFs and the value of
$\alpha_s$ has been discussed recently in Ref.~\cite{leshouches}.
For the present work, we have constructed
a family of NNPDF1.2 PDF sets using different
values of $\alpha_s$.
The very recent NNPDF2.0 PDF set~\cite{Ball:2010de} also includes PDFs
determined with different fixed valued of $\alpha_s$.\footnote{
The NNPDF1.2 sets with variable $\alpha_s$ are available upon
request. The NNPDF2.0 sets with variable $\alpha_s$ are available
from the webpage of the NNPDF Collaboration,
\url{htpp://sophia.ecm.ub.es/nnpdf},
and will also be available through the LHAPDF interface.}

We have repeated the NNPDF1.2 PDF determination  with $\alpha_s$
varied in the range $0.113\le\alpha_s\le0.128$ and all other aspects
of the parton determination unchanged: for each value of $\alpha_s$ we
have produced a set of $100$ PDF replicas.
In Fig.~\ref{fig:xg_comp} we show
 the ratios of the central gluons obtained in these
fits compared to the reference
NNPDF1.2 gluon with $\alpha_s\lp M_Z\rp=0.119$, 
together with the PDF uncertainty band which corresponds to 
the reference value. 

The qualitative behaviour of the gluon
in Fig.~\ref{fig:xg_comp} can be understood as follows. 
In NNPDF1.2, the gluon is determined by scaling violations of
deep--inelastic structure functions, {\it i.e.} mostly from medium and small
$x$ HERA data, with the large $x$ gluon constrained by the momentum
sum rule. With a given amount of  scale dependence seen in the data, 
smaller values of $\alpha_s$ require a larger small $x$ gluon, and
thus because of the sum rule a smaller large $x$ gluon.
In global 
fits~\cite{Martin:2009bu,Nadolsky:2008zw,Ball:2010de} the 
behaviour is essentially
 the same, up to the fact that some extra constraint on the large $x$
 gluon is provided by Tevatron jet data, as quantified 
in~\cite{Ball:2010de}. 

The size of this correlation of the gluon with the value of $\alpha_s$ shown
in Fig.~\ref{fig:xg_comp} is clearly statistically significant;
however, when $\alpha_s$ is varied within its uncertainty range,
Eq.~(\ref{eq:alphasref}), the
change in gluon distribution is generally 
smaller than the uncertainty on the gluon itself.
It is interesting to note that the size of the uncertainty for values
of $\alpha_s$ which are away from the best fit is often larger than
the uncertainty when $\alpha_s$ is at or close to its best fit value:
this is to be contrasted to what happens in a 
Hessian approach, where linear error
propagation inevitably implies that the PDF uncertainty shrinks when
$\alpha_s$ moves away from its best--fit value~\cite{Martin:2009bu}.
\begin{figure}
\begin{center}
\includegraphics[width=.49\linewidth]{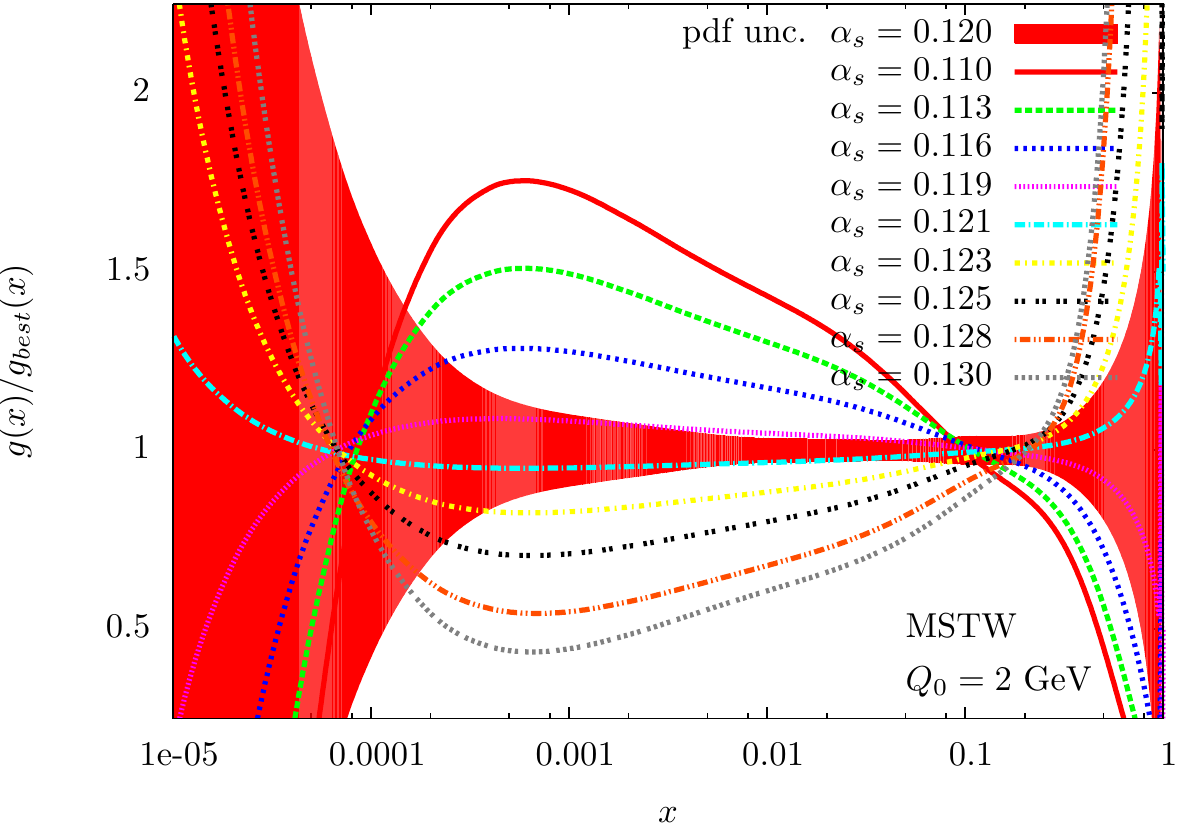}
\includegraphics[width=.49\linewidth]{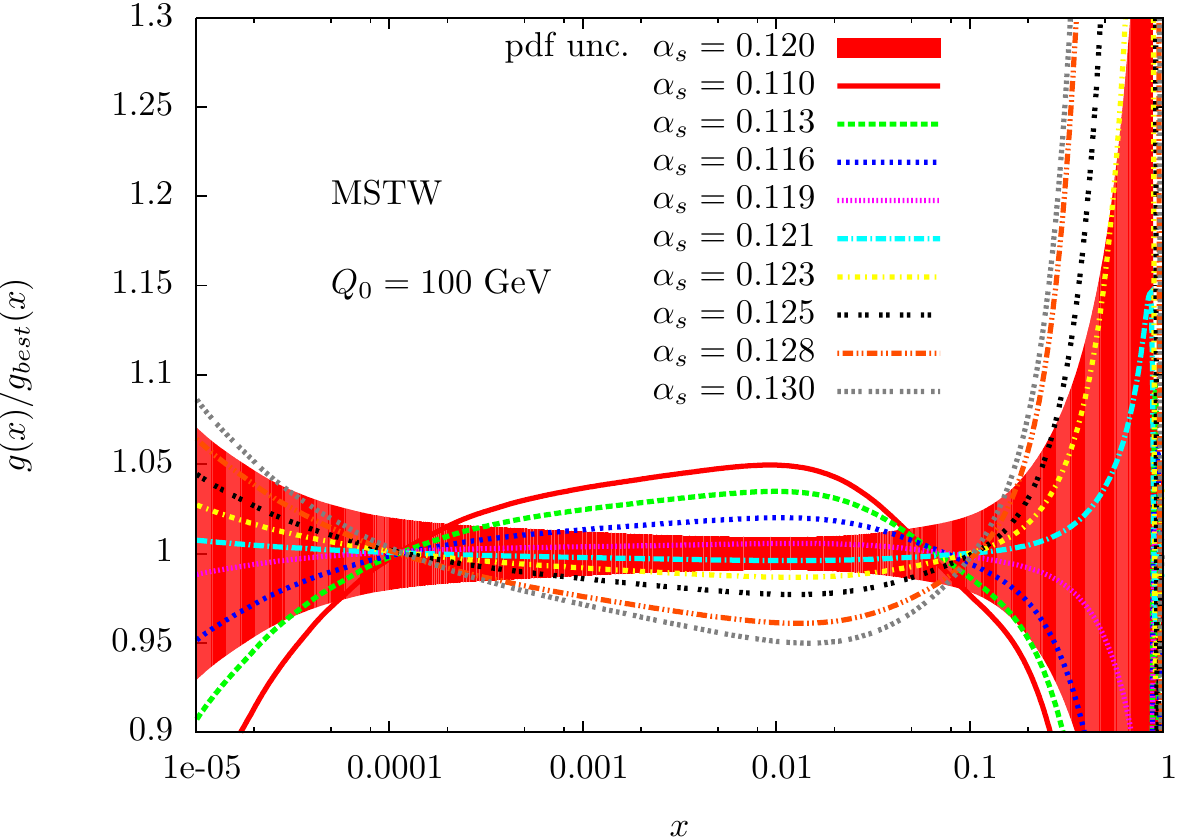}\\
\includegraphics[width=.49\linewidth]{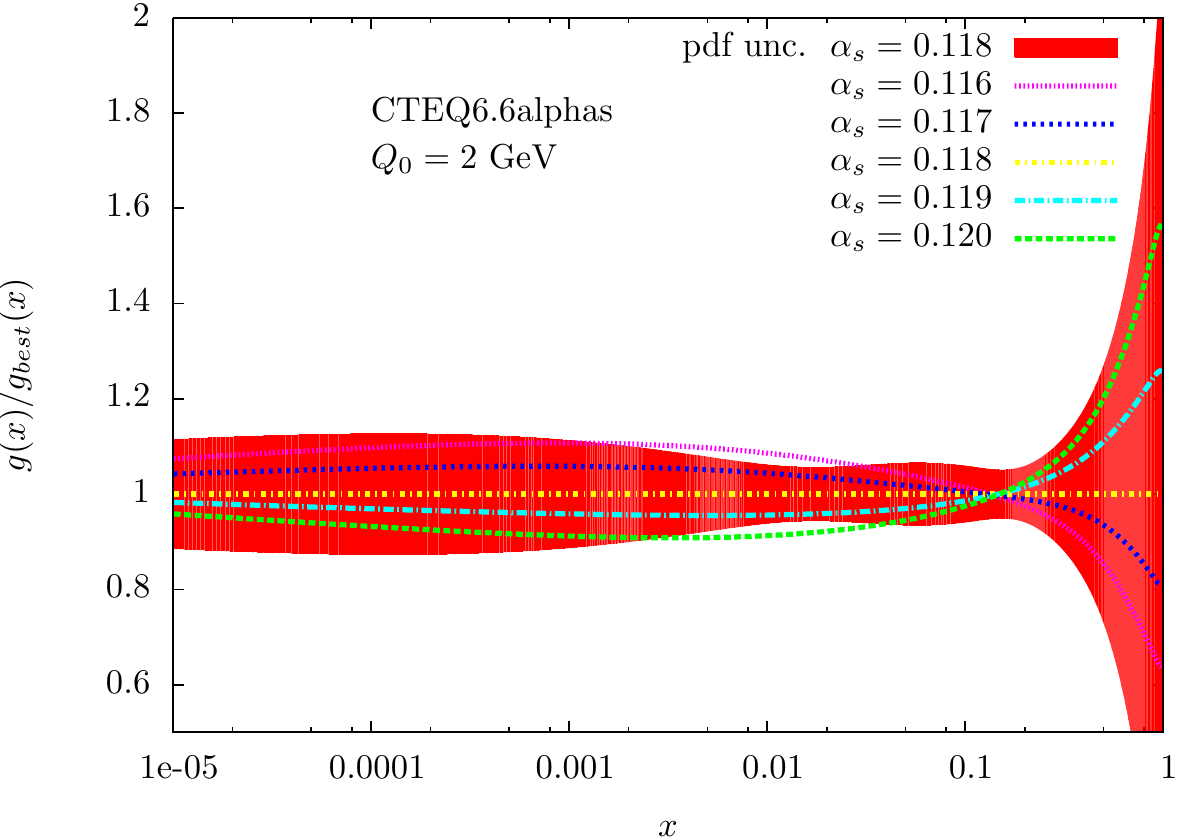}
\includegraphics[width=.49\linewidth]{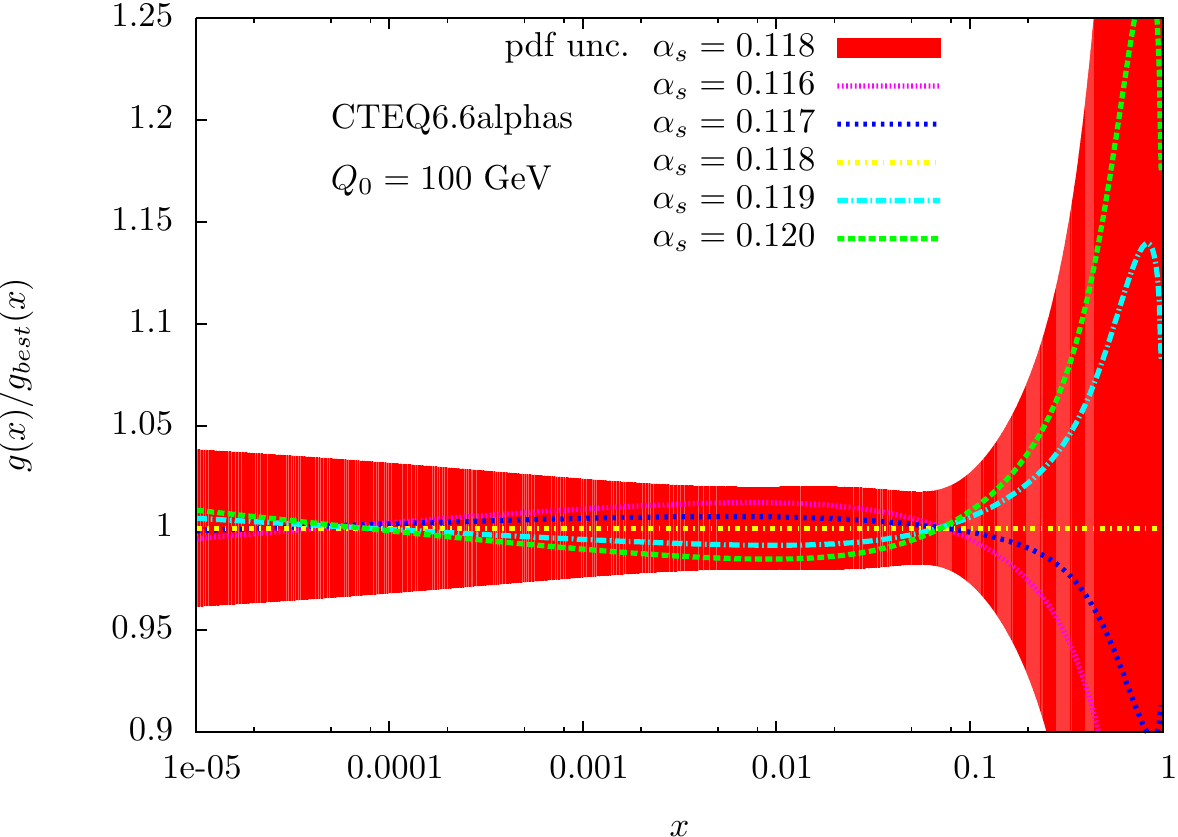}\\
\includegraphics[width=.49\linewidth]{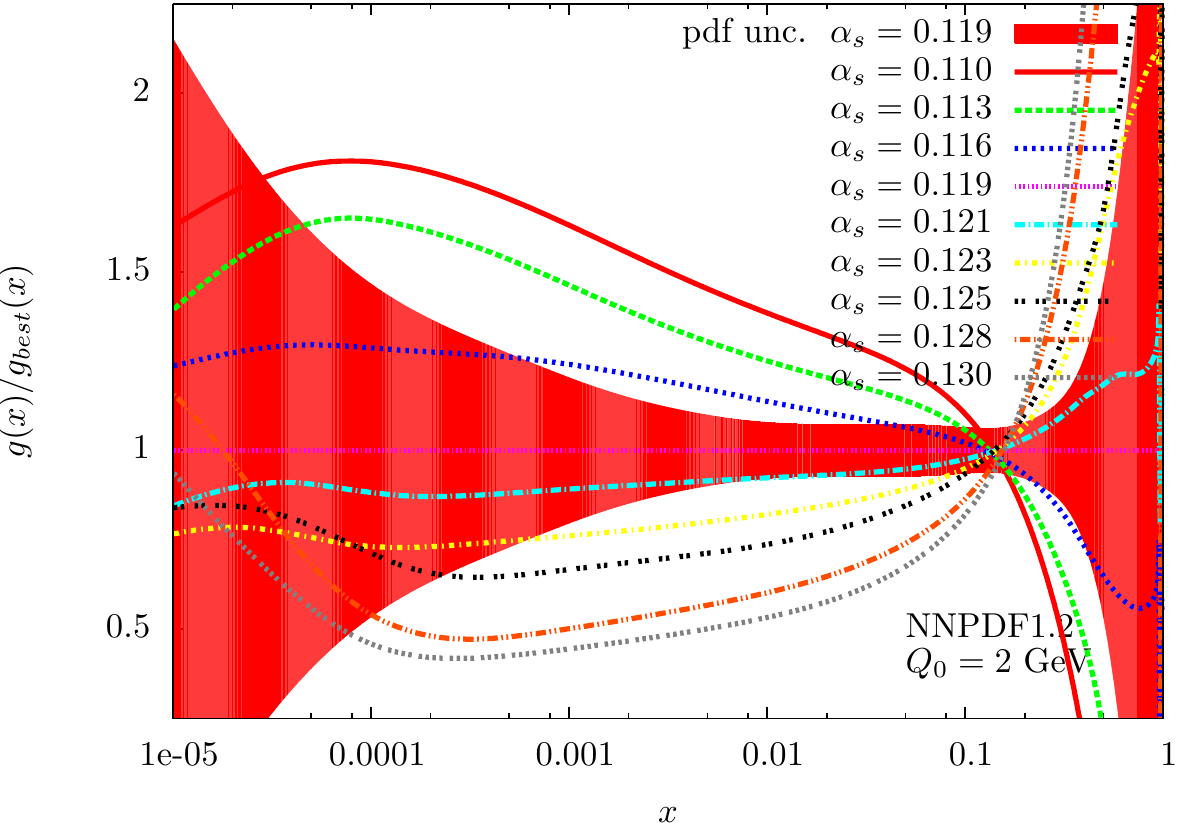}
\includegraphics[width=.49\linewidth]{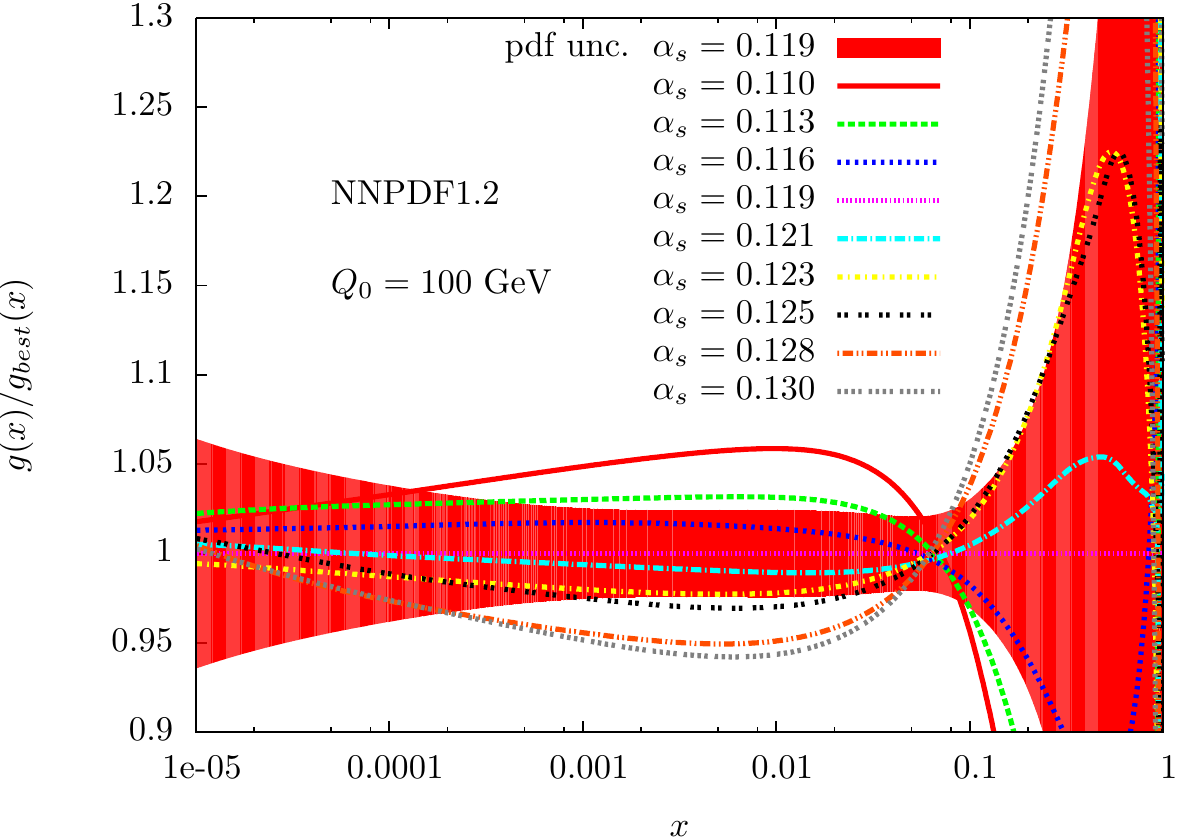}
\end{center}
\caption{\label{fig:pdf2-100} \small
Comparison of the gluon PDFs from
MSTW08 (top), CTEQ6.6 (center) 
and NNPDF1.2 (bottom) at the scales $Q^2=4$~GeV$^2$ and
$Q^2=10^4$~GeV$^2$ as $\alpha_s$ is varied, normalized to the
corresponding central sets, determined with the value of $\as$ listed in
Eq.~(\ref{asrefcoll}). The one--$\sigma$ uncertainty band for the central set is
also shown in each case.} 
\end{figure}

\begin{figure}[h]
\begin{center}
\includegraphics[width=.49\linewidth]{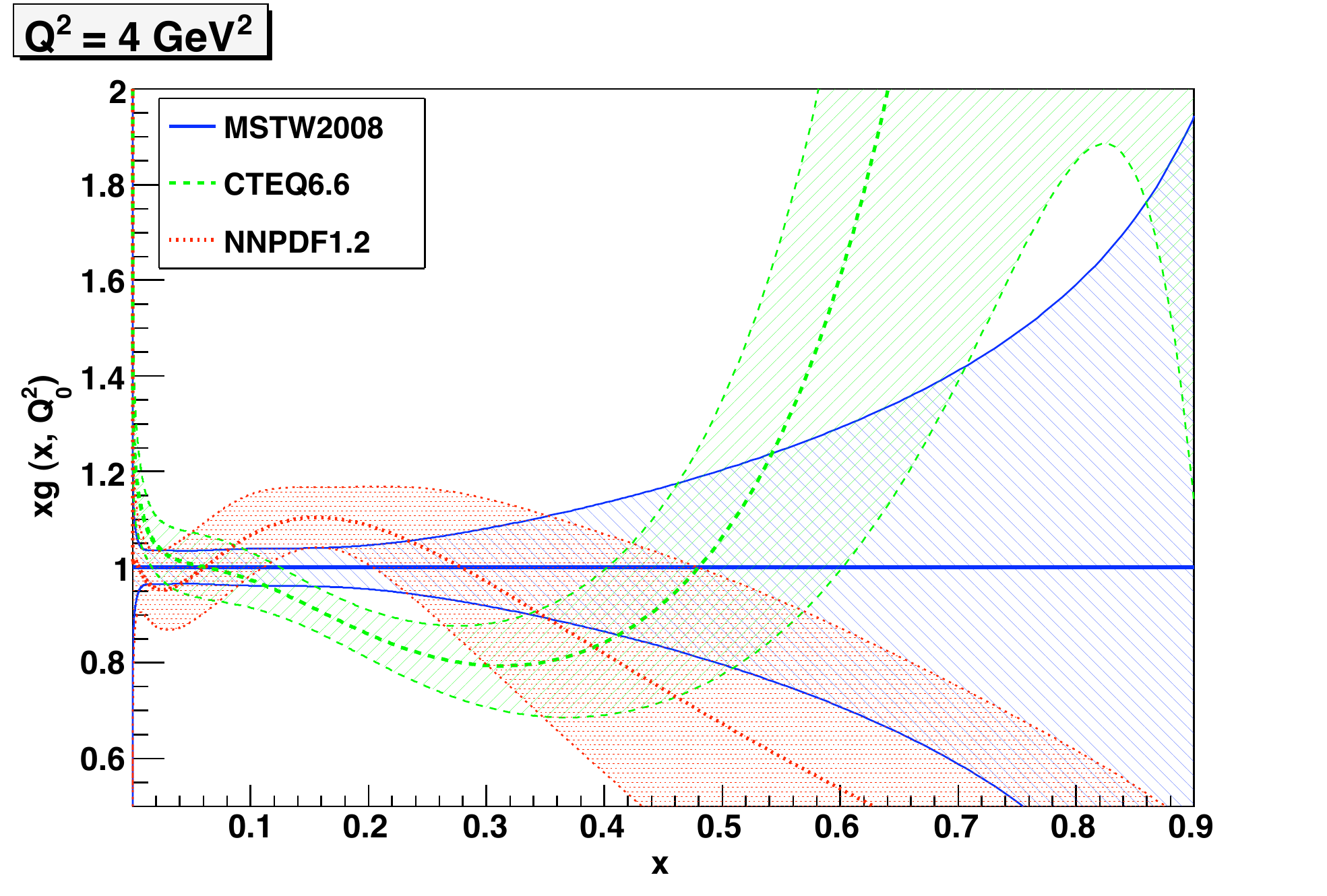}
\includegraphics[width=.49\linewidth]{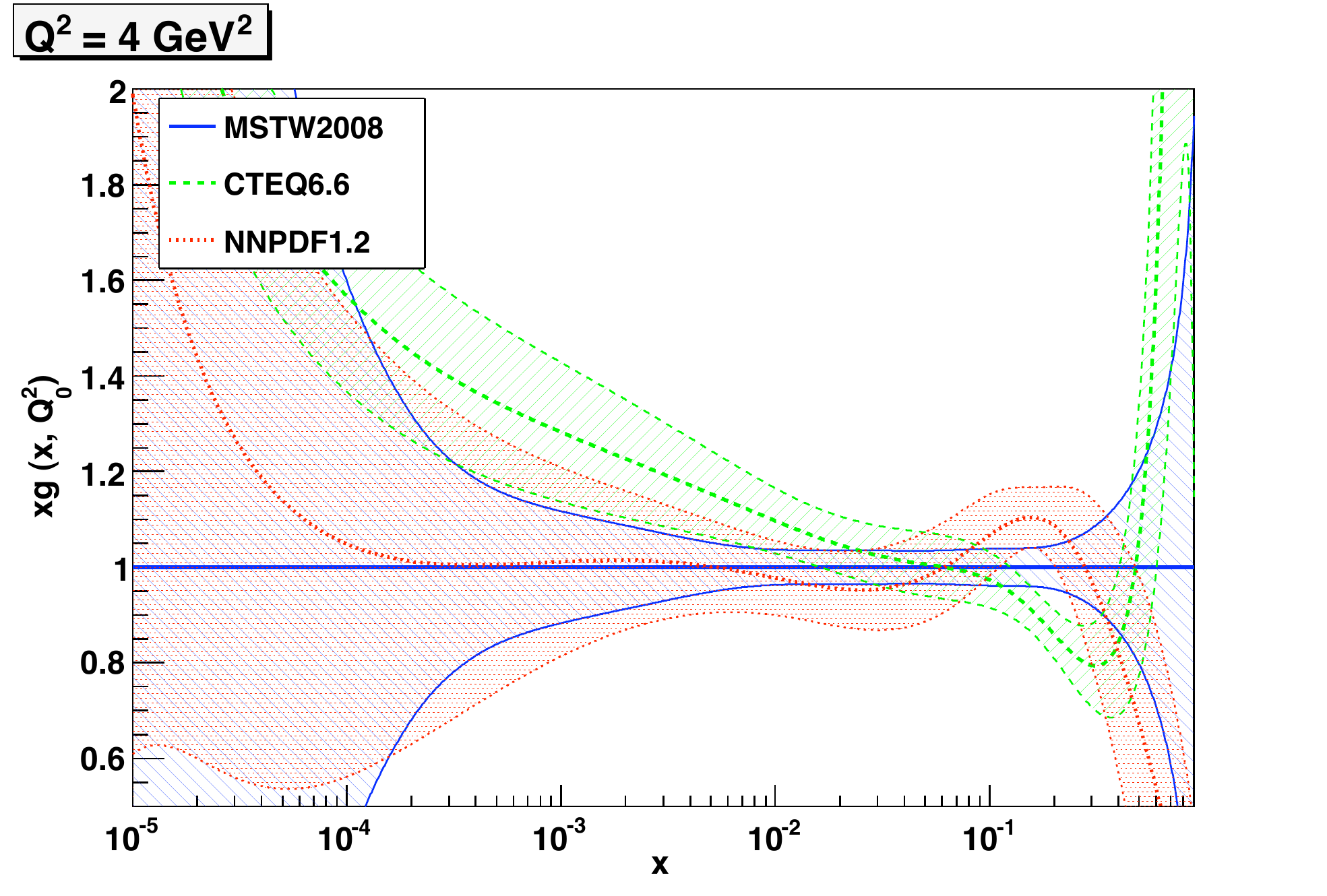}\\
\includegraphics[width=.49\linewidth]{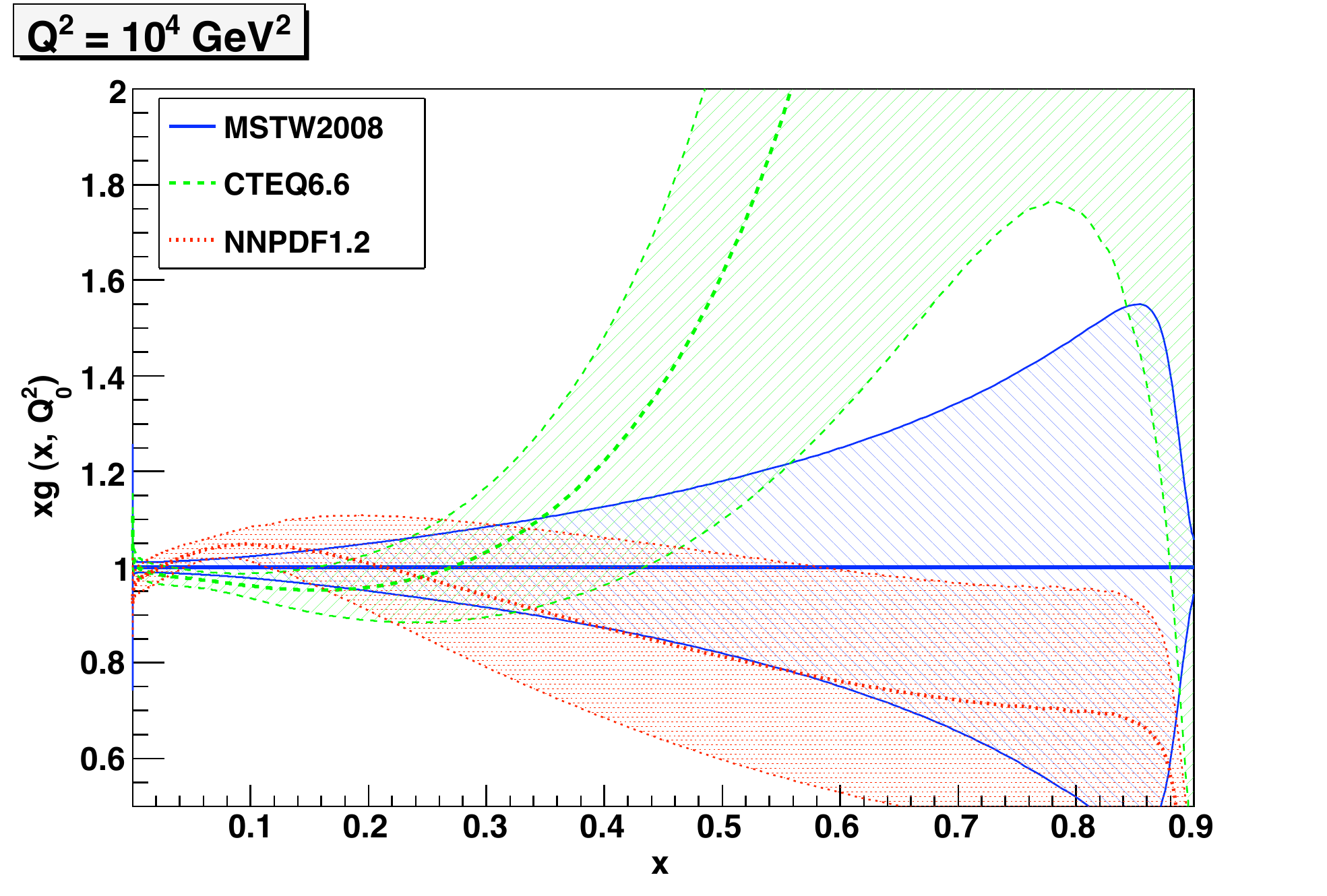}
\includegraphics[width=.49\linewidth]{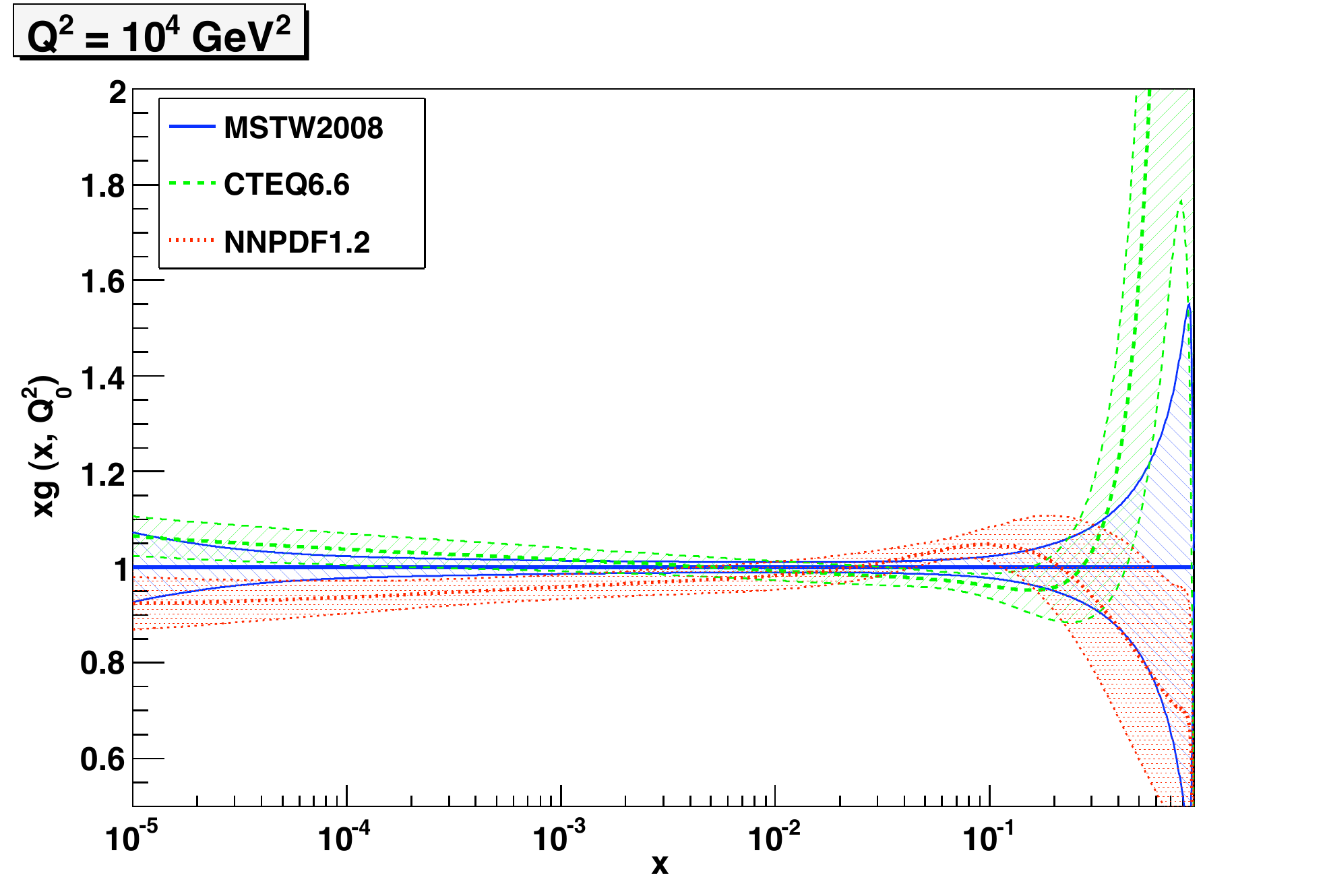}
\caption{\label{fig:pdf-cfr} \small
Comparison of the gluon PDFs from
MSTW08, CTEQ6.6  
and NNPDF1.2  at the scales $Q^2=4$~GeV$^2$ (upper plots)
and
$Q^2=10^4$~GeV$^2$ (lower plots), 
each determined with the value of $\as$ listed in
Eq.~(\ref{asrefcoll}),  all 
normalized to the central MSTW08 set.}
\end{center}
\end{figure}
Given 
sets of replicas determined with different values of $\alpha_s$,
it is possible to perform statistics in which $\alpha_s$ is varied,
by assuming a distribution of values for $\alpha_s$. 
For instance, the average over Monte Carlo replicas 
of a general quantity which depends on both
$\alpha_s$ and the PDFs, $\mathcal{F}\lp  {\rm PDF},\alpha_s\rp$
can be computed as
\be
\label{eq:avrep}
\la \mathcal{F}\ra_{\rep} =\frac{1}{N_{\rep}}\sum_{j=1}^{N_{\alpha}}
\sum_{k_j=1}^{N_{\rm rep}^{\alpha_s^{(j)}}} 
\mathcal{F}\lp  {\rm PDF}^{(k_j,j)},\alpha_s^{(j)}\rp \ ,
\ee
where ${\rm PDF}^{(k_j,j)}$ stands for  the $k_j$--th replica of the
PDF fit with $\alpha_s=\alpha_s^{(j)}$, and  the numbers $N_{\rm rep}^{\alpha_s^{(j)}}$ of
replicas for each value of $\alpha_s$ in the total sample are
determined by the probability distribution of values of $\alpha_s$,
with the constraint 
\be
\label{eq:repnumnorm}
N_{\rm rep} = \sum_{j=1}^{N_{\alpha_s}}N^{\alpha_s^{(j)}}_{\rm rep}.
\ee
Specifically, assuming that 
global fit values of $\alpha_s$ (such as Eqs.~(\ref{asbethke}-\ref{aspdg}))
are gaussianly distributed, the number of replicas is
\be
\label{eq:gaussianrepdist}
N^{\alpha_s^{(j)}}_{\rm rep}\propto \exp\lp 
-\frac{\lp \alpha_s^{(j)}- \alpha_s^{(0)}\rp^2}{
2 \lp \delta_{\as}^{(68)}\rp^2}\rp \ .
\ee
with the normalization condition Eq.~(\ref{eq:repnumnorm}).

\subsection{Comparison of global PDF sets with variable $\alpha_s$}
\label{sec:pdf}

\begin{figure}
\begin{center}
\includegraphics[width=.49\linewidth]{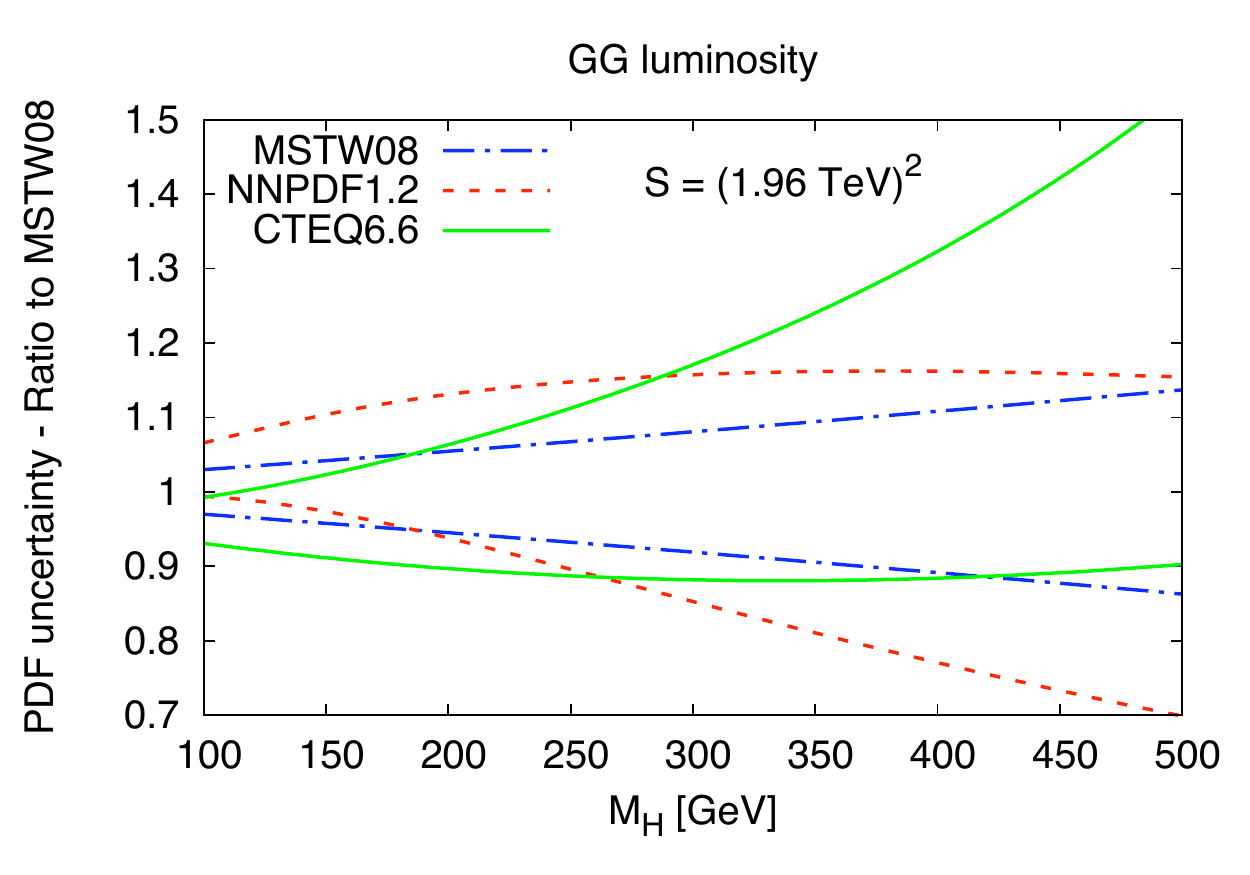}
\includegraphics[width=.49\linewidth]{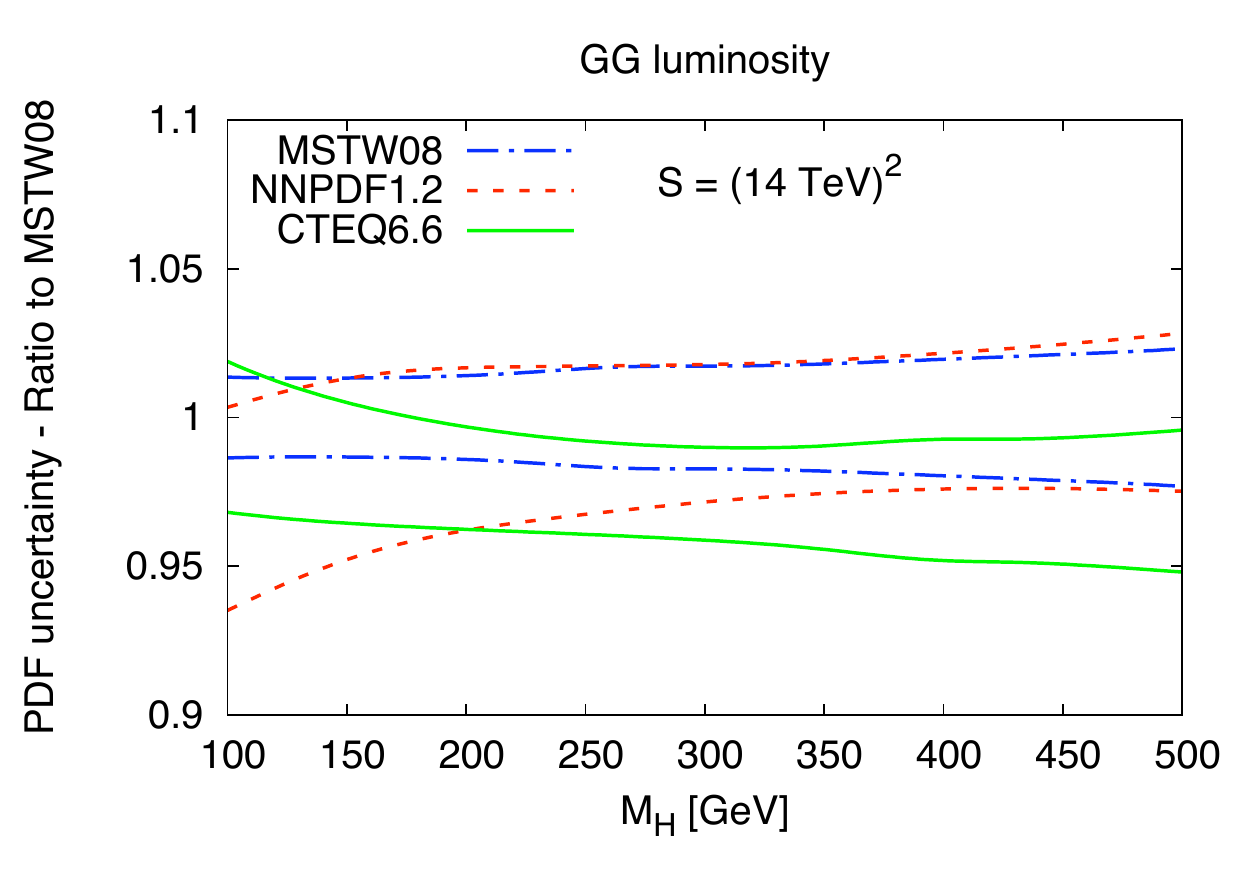}
\includegraphics[width=.49\linewidth]{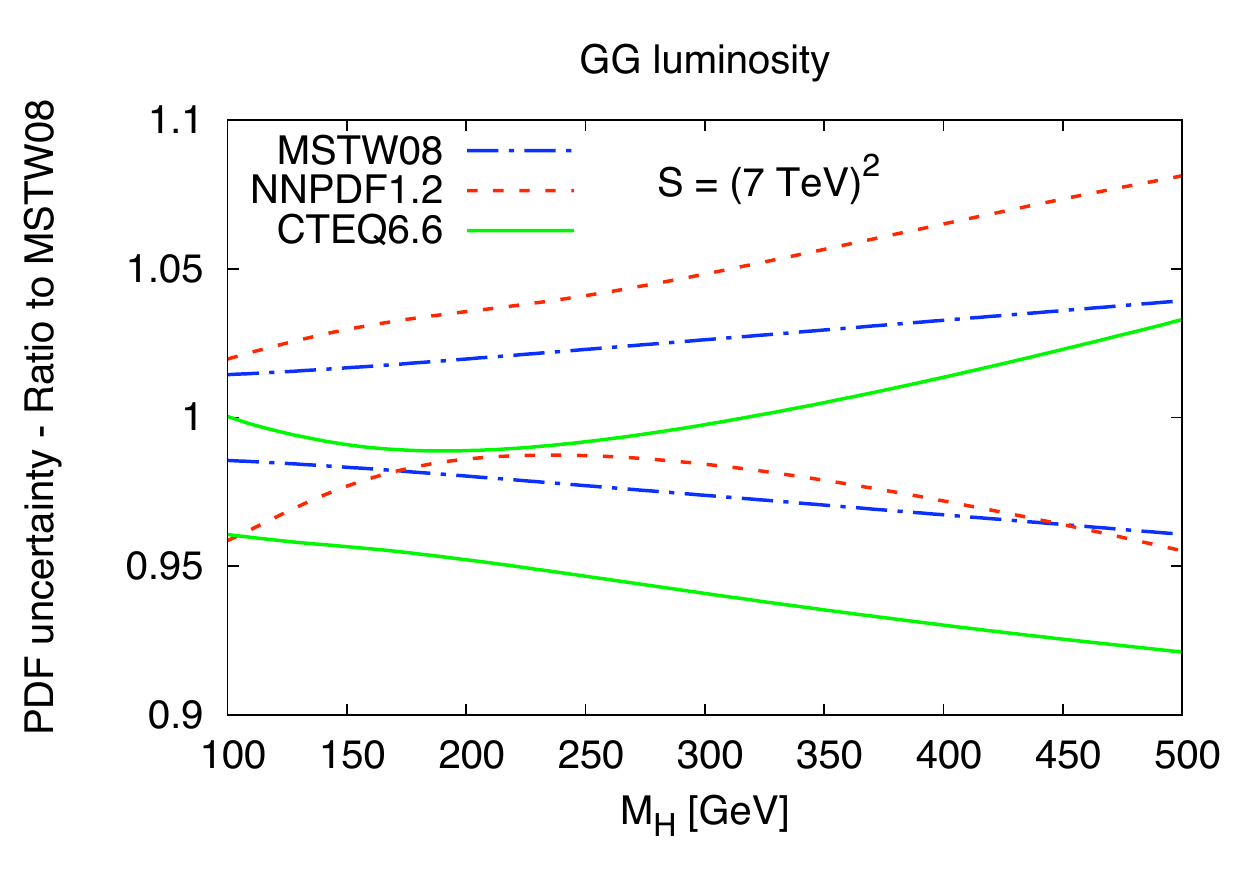}
\caption{\label{fig:lumi-cfr} \small
Comparison of the gluon luminosities from
MSTW08, CTEQ6.6  
and NNPDF1.2  at the Tevatron (upper left plot), 
LHC~7~TeV (upper right plot) and 
LHC~14~TeV (lower plot)
and
$Q^2=10^4$~GeV$^2$, each determined with the value of $\as$ listed in
Eq.~(\ref{asrefcoll}),  all 
normalized to the central MSTW08 set}
\end{center}
\end{figure}

The dependence of the gluon distribution on the value of $\alpha_s$ is
summarized in Fig.~\ref{fig:pdf2-100}, where results obtained using the CTEQ6.6,
MSTW08 and NNPDF1.2  $\alpha_s$ series are compared, both at 
the scale 
$Q^2=$~4 GeV$^2$, close to the scale at which
PDFs are parametrized, and at the high scale  $Q^2=10^4$ GeV$^2$, 
typical of electroweak final states. 
The three central sets  of Fig.~\ref{fig:pdf2-100} are then compared
directly in Fig.~\ref{fig:pdf-cfr}. The gluon luminosities computed
from these three central sets are then finally compared in
Fig.~\ref{fig:lumi-cfr} at Tevatron and LHC energies. The luminosities
are plotted as a function of $m_H$ for given energy using the
leading--order kinematic relation $x=m^2_H/s$. It should be born in
mind that, because of soft--gluon dominance~\cite{Kramer:1996iq}, the
NLO contribution is strongly peaked at the endpoint, and thus probes
mostly the luminosity at the same value of $x$ as the LO.

Notable features of these comparisons are the following:
\begin{itemize}
\item The same
correlation pattern between the gluon and $\alpha_s$ discussed
for NNPDF in Sect.~\ref{sec:nnpdfseries} is also apparent  for  other sets.
\item At $Q^2=4$~GeV$^2$ uncertainties at very small $x$ for
MSTW and NNPDF are large enough to swamp the $\alpha_s$
dependence. This does not happen for CTEQ, likely due to the more
restrictive gluon parametrization used. However, at $Q^2=10^4$~GeV$^2$
the region at which uncertainties blow up is pushed to much smaller 
values of $x$.
\item Uncertainties at large $x$ also swamp the $\alpha_s$ dependence,
  due to the scarcity or (for NNPDF1.2) lack of data in this region.
\item Even in the medium $x$ region where the gluon is best known PDF
  uncertainty bands are rather larger than the gluon variation due to
  the variation of $\alpha_s$ within its uncertainty 
  Eq.~(\ref{asbethke})  or even Eq.~(\ref{aspdg}).
\item The three gluons in Fig.~\ref{fig:pdf-cfr} 
overlap  to one--$\sigma$ in most of the kinematic region. However, at
low scale they disagree significantly at large $x$, and
at high scale they disagree, though by less than about two $\sigma$,
both at very large and very small $x$. 
\item Once gluons are convoluted into a parton luminosity, most of the
  disagreements seen in Fig.~\ref{fig:pdf-cfr} are washed out: indeed,
  the parton luminosities Fig.~\ref{fig:lumi-cfr} computed from
  CTEQ6.6, MSTW08 and NNPDF1.2 all agree within uncertainties, in the
  sense that their one--$\sigma$ error bands always overlap (though
  sometimes just about).
\end{itemize}

\subsection{Correlation between PDFs and $\alpha_s$}
\label{sec:nnpdfcorr}

Correlations between different PDFs, or between PDFs and physical
observables, have been computed by CTEQ using a Hessian
approach~\cite{Nadolsky:2008zw}, and by NNPDF using a Monte Carlo
approach~\cite{Ball:2009by}. 
Within a Monte Carlo approach it is in fact easy to
estimate the correlation between any pair of quantities by computing
their covariance  over the Monte Carlo sample. Statistics
involving the value of $\alpha_s$ can then be performed provided only
replicas with different values of $\alpha_s$ are available, as
discussed in Sect.~\ref{sec:nnpdfseries} above, Eq.~(\ref{eq:avrep}).

Indeed, in a Monte Carlo approach 
the correlation between the strong coupling and the
gluon (or any other PDF) is given by
\be
\label{eq:gcorr}
\rho \lc  \alpha_s\lp M_Z^2\rp,g\lp x,Q^2\rp\rc=
\frac{\la \alpha_s\lp M_Z^2\rp g\lp x,Q^2\rp \ra_{\rep}-
\la \alpha_s\lp M_Z^2\rp\ra_{\rep}\la g\lp x,Q^2\rp \ra_{\rep}
}{\sigma_{\alpha_s\lp M_Z^2\rp}\sigma_{g\lp x,Q^2\rp}},
\ee
where the distribution of values of $\alpha_s$ is automatically
reproduced if one picks $N_{\rm rep}^{\alpha_s^{(j)}}$ of
replicas for each value of $\alpha_s$  according to
Eqs.~(\ref{eq:repnumnorm}-\ref{eq:gaussianrepdist}) above.

\begin{figure}[t]
\begin{center}
\includegraphics[width=0.49\linewidth]{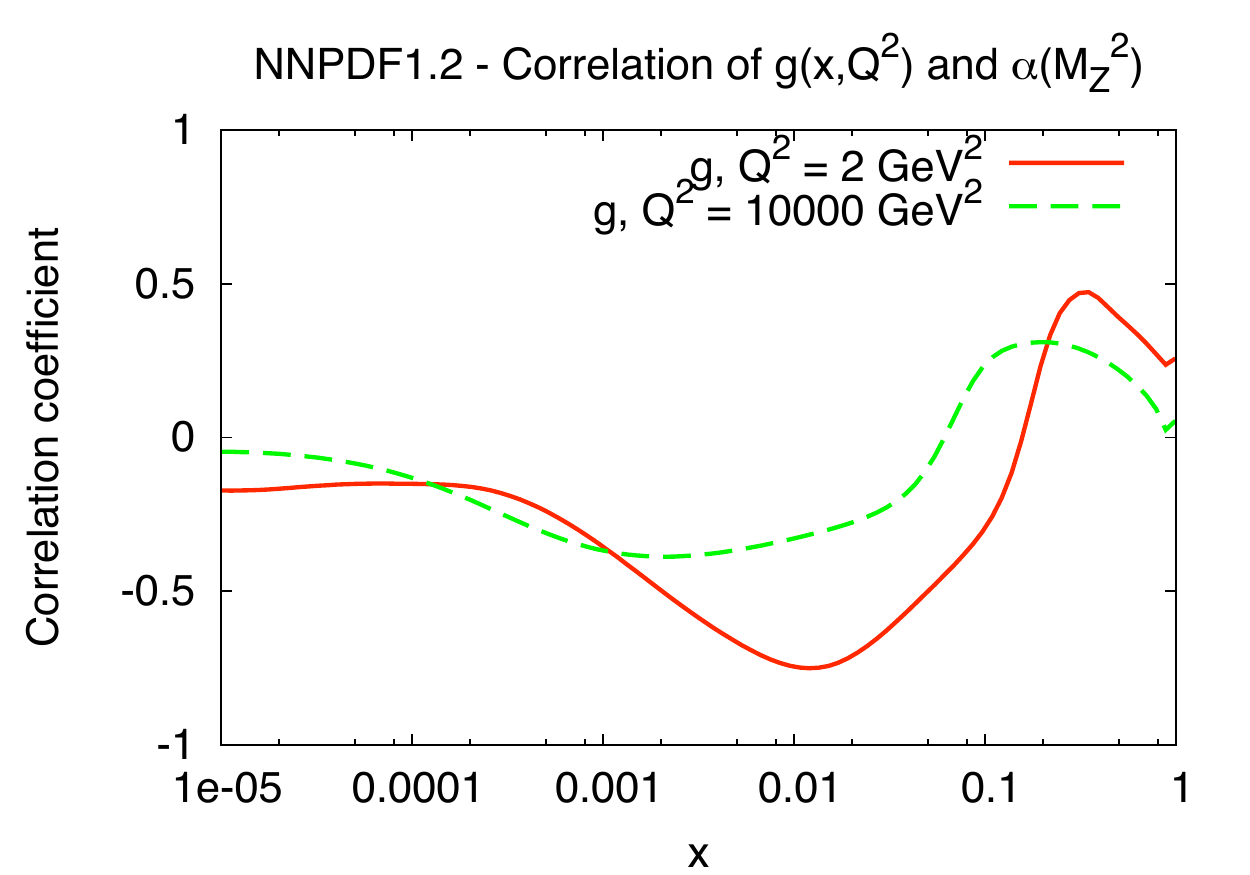}
\includegraphics[width=0.49\linewidth]{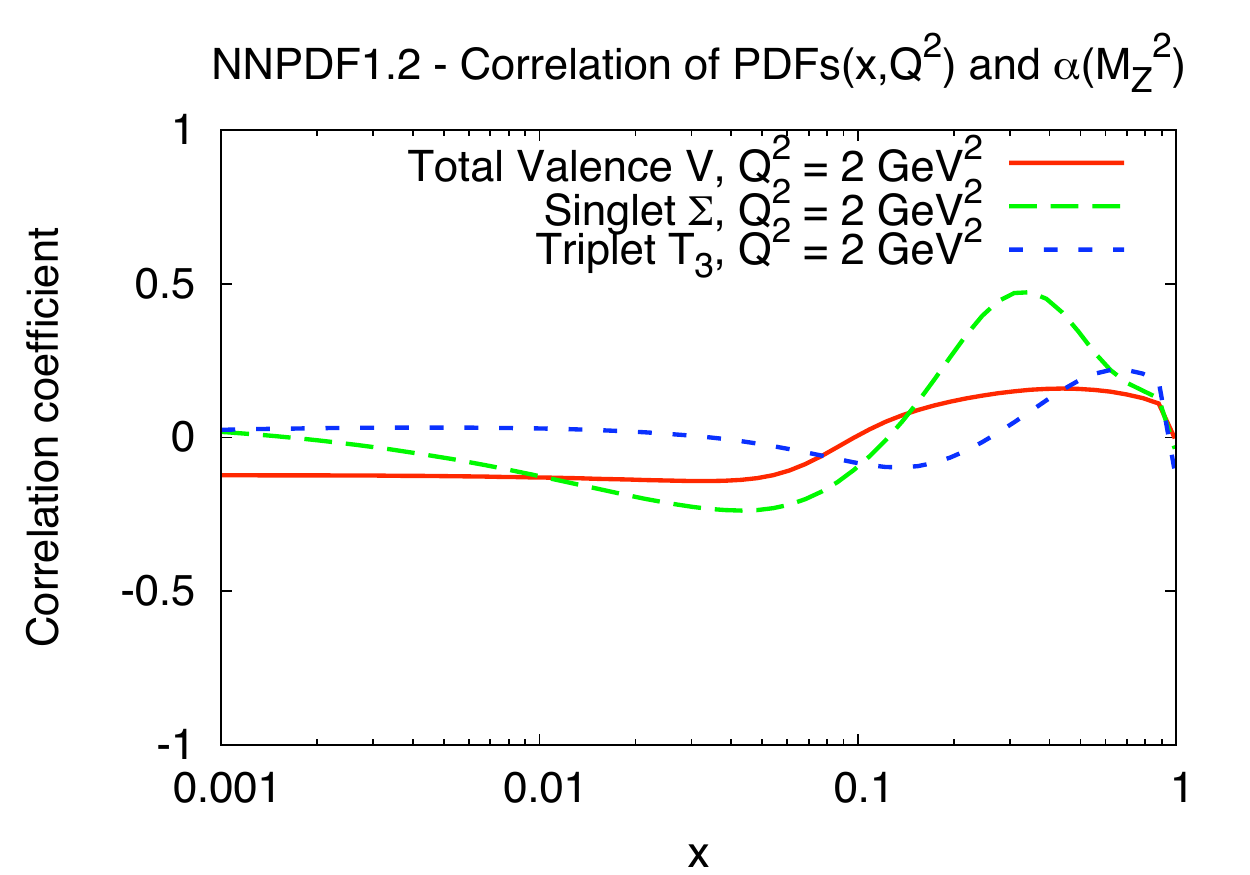}
 \caption{\small The correlation
Eq.~(\ref{eq:gcorr}) between PDFs and $\alpha_s(M_Z)$
as a function of $x$. Left: gluon PDF
at $Q^2=2$ GeV$^2$ and $Q^2=10^4$
GeV$^2$; right:
singlet, triplet and total valence PDFs for $Q^2=2$ GeV$^2$. 
}
\label{fig:gluon-alphas-corr}
\end{center}
\end{figure}
Our results for the correlation coefficient
 between the gluon and $\alpha_s(M_Z)$
as a function of $x$, computed using Eq.~(\ref{eq:gcorr}), with the
NNPDF1.2 PDFs of Sect.~\ref{sec:nnpdfseries} and 
Eq.~(\ref{eq:alphasref}) for $\alpha_s$,
both at a low scale
$Q^2=2$ GeV$^2$ and at a typical LHC scale $Q^2=10^4$ GeV$^2$,
are shown in 
Fig.~\ref{fig:gluon-alphas-corr}. It is interesting to note 
how evolution
decorrelates the gluon from the strong coupling. 
We also show in Fig.~\ref{fig:gluon-alphas-corr} the correlation
coefficient for other PDFs: as expected for the triplet and
valence PDFs it is essentially zero, that is, in NNPDF1.2 these
PDFs show essentially no sensitivity to $\alpha_s$. The
correlation coefficient Fig.~\ref{fig:gluon-alphas-corr}
quantifies the qualitative observations of
Fig.~\ref{fig:xg_comp}.

\clearpage
\section{The cross section: PDF uncertainties}
\label{sec:pdf-unc}

\begin{figure}
\begin{center}
\includegraphics[width=.32\linewidth]{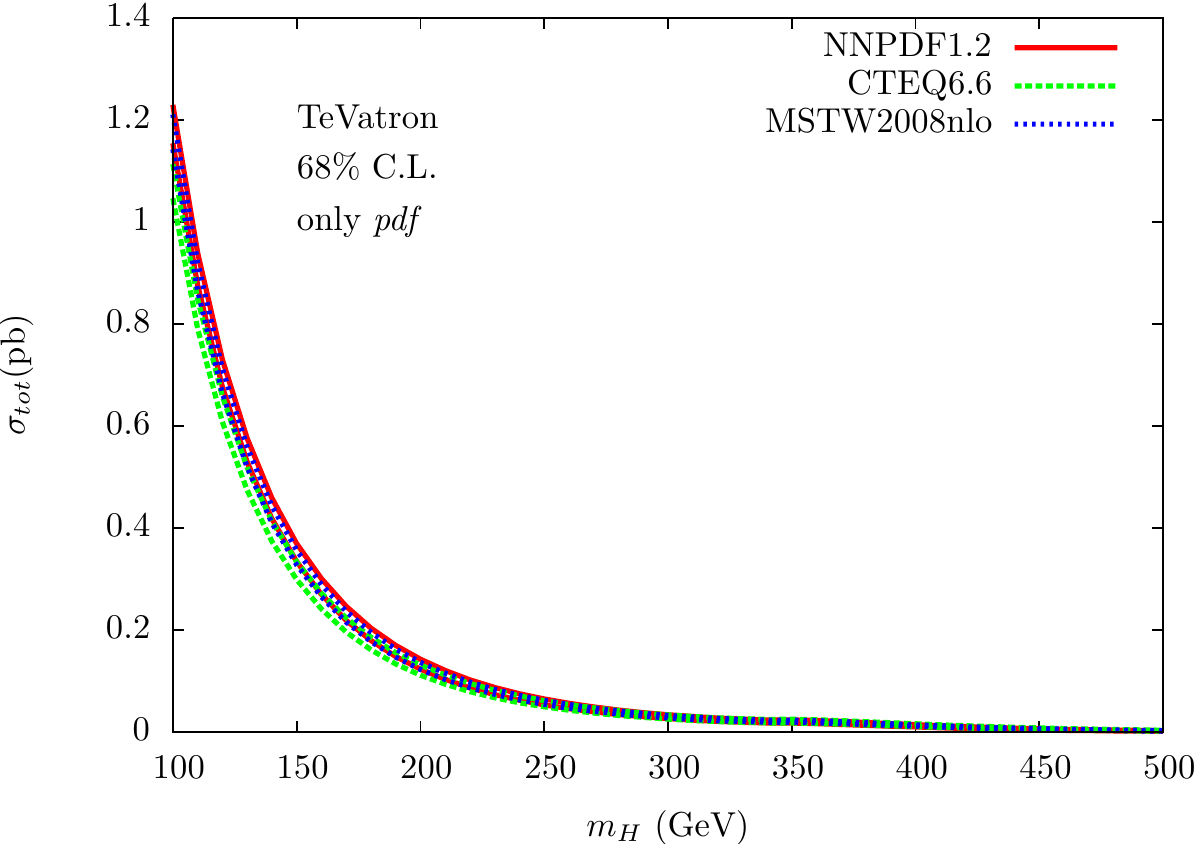}
\includegraphics[width=.32\linewidth]{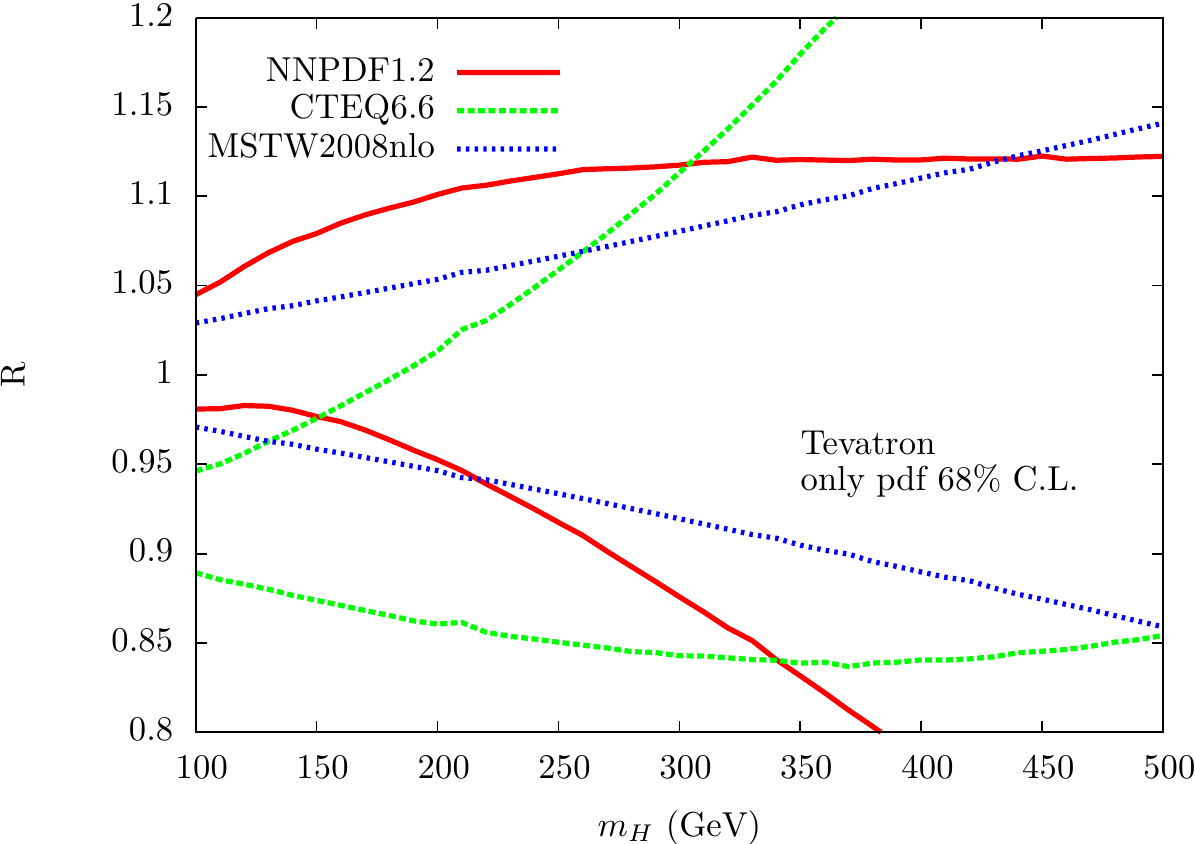}
\includegraphics[width=.32\linewidth]{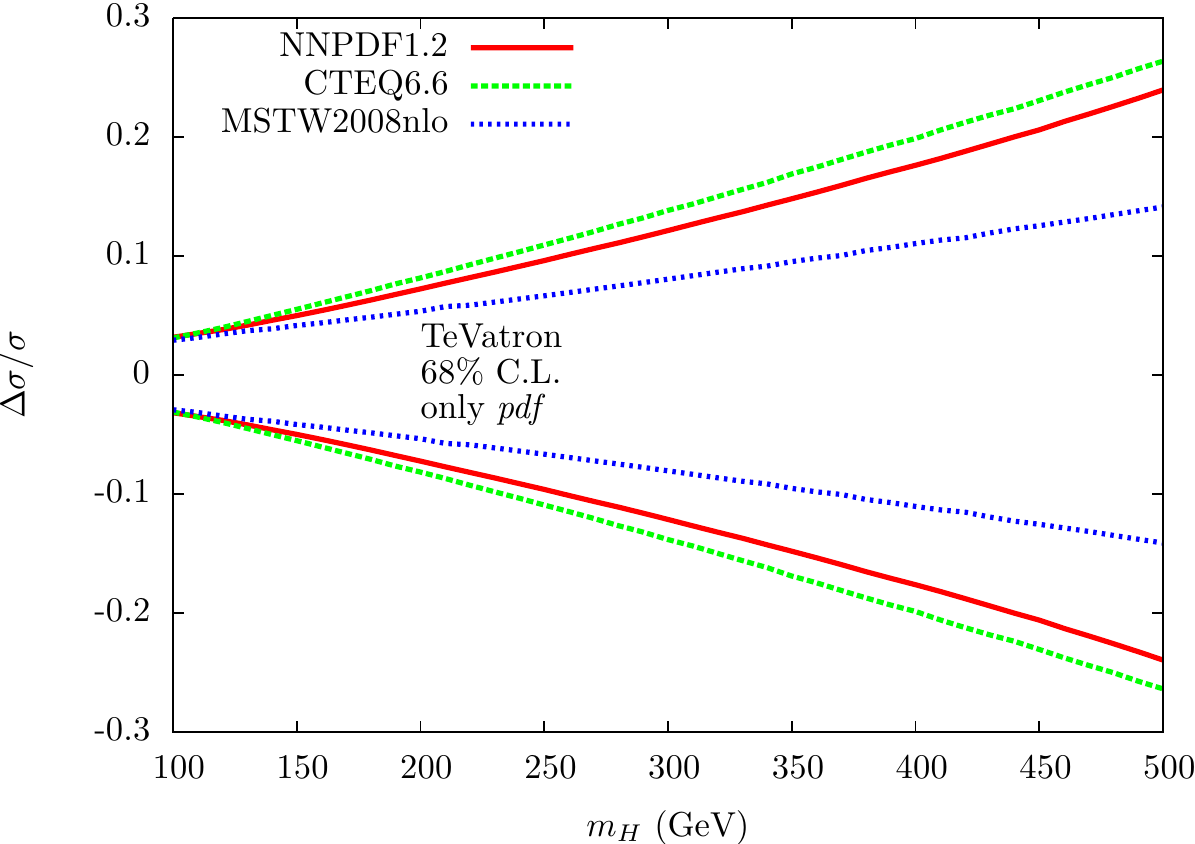}\\
\includegraphics[width=.32\linewidth]{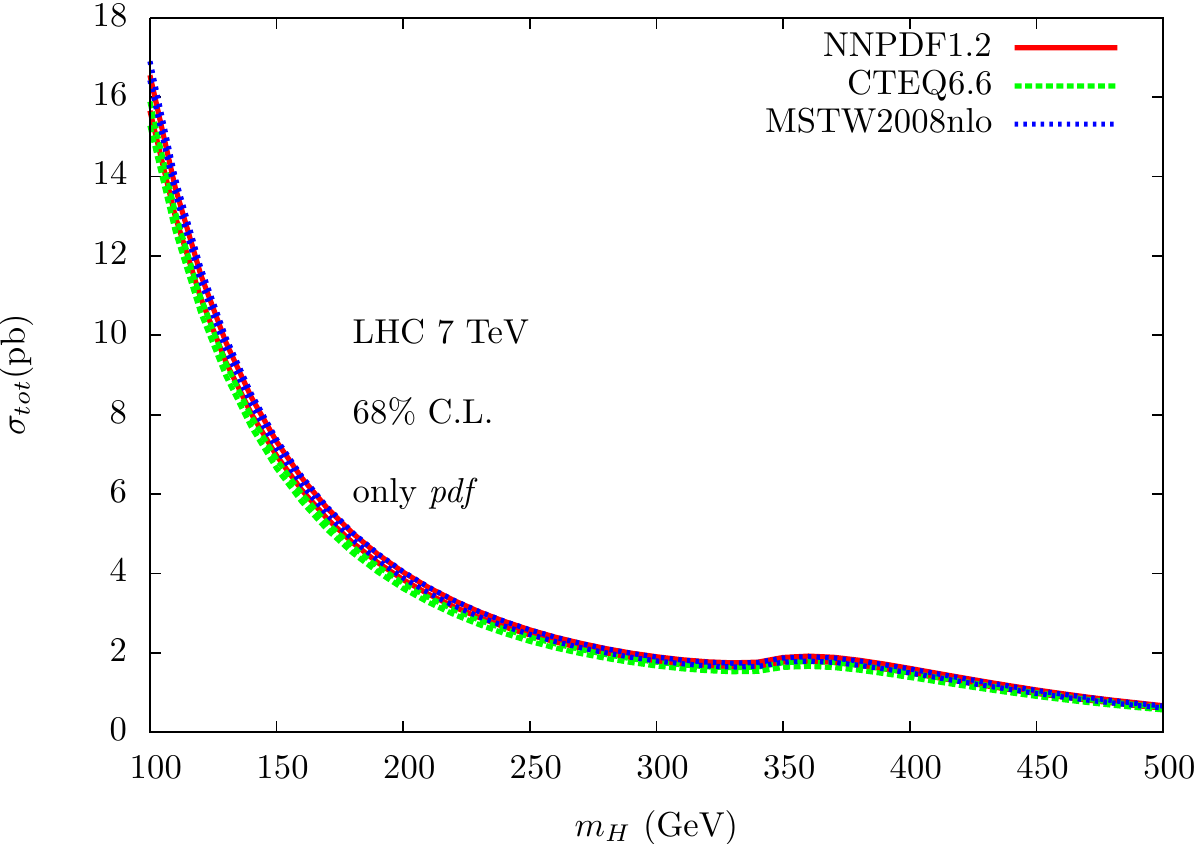}
\includegraphics[width=.32\linewidth]{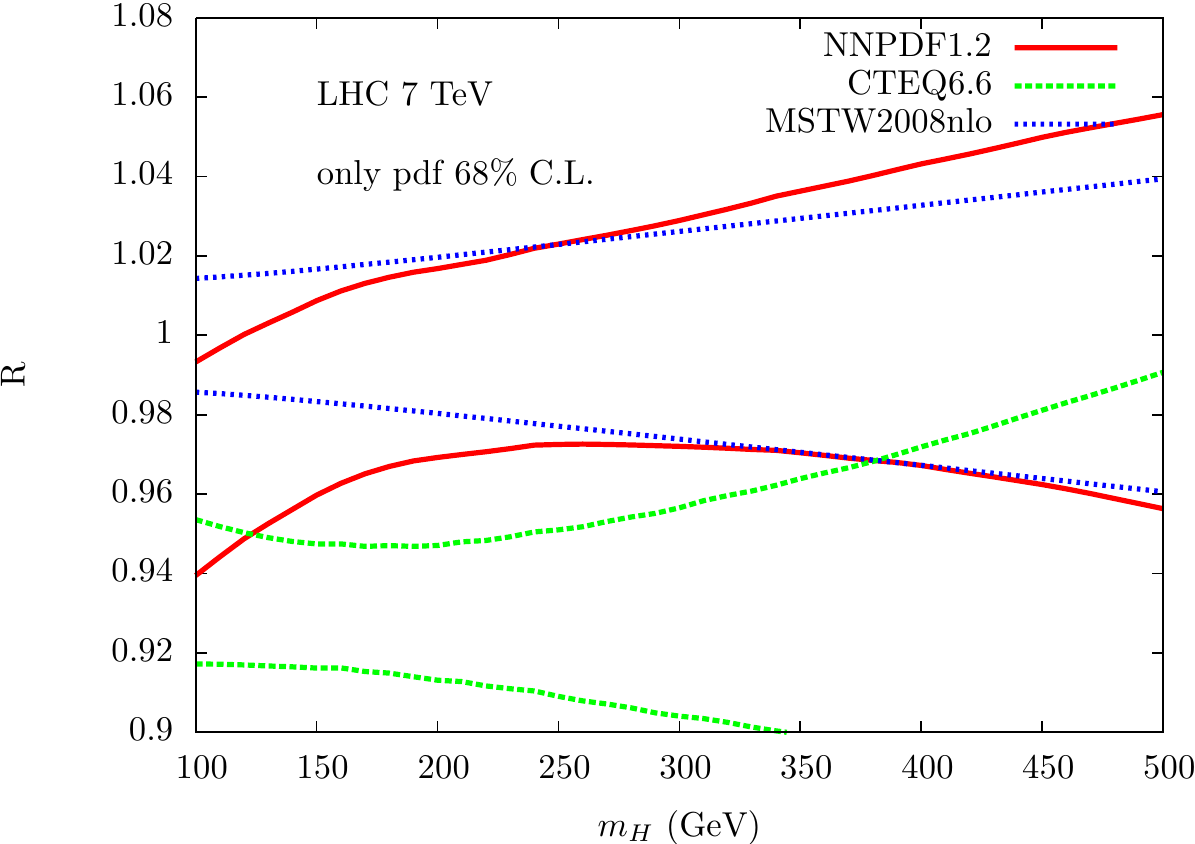}
\includegraphics[width=.32\linewidth]{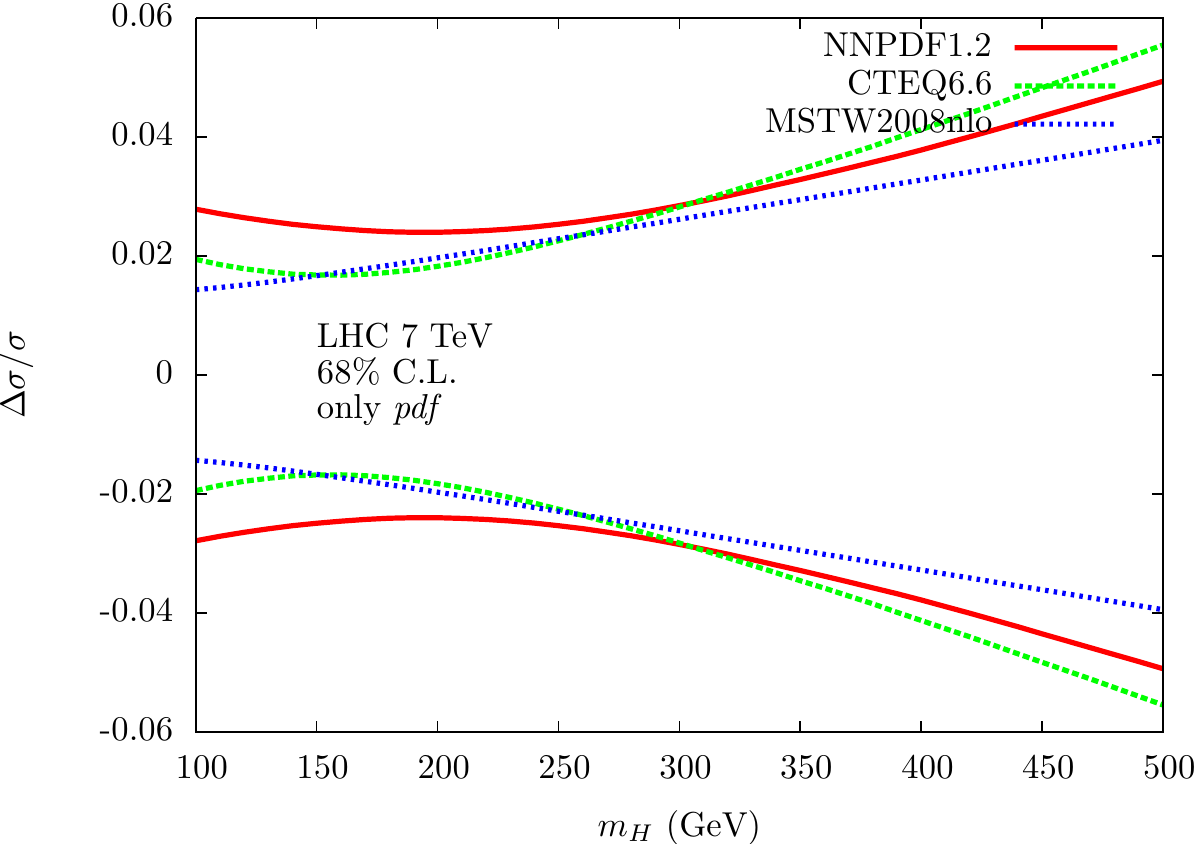}\\
\includegraphics[width=.32\linewidth]{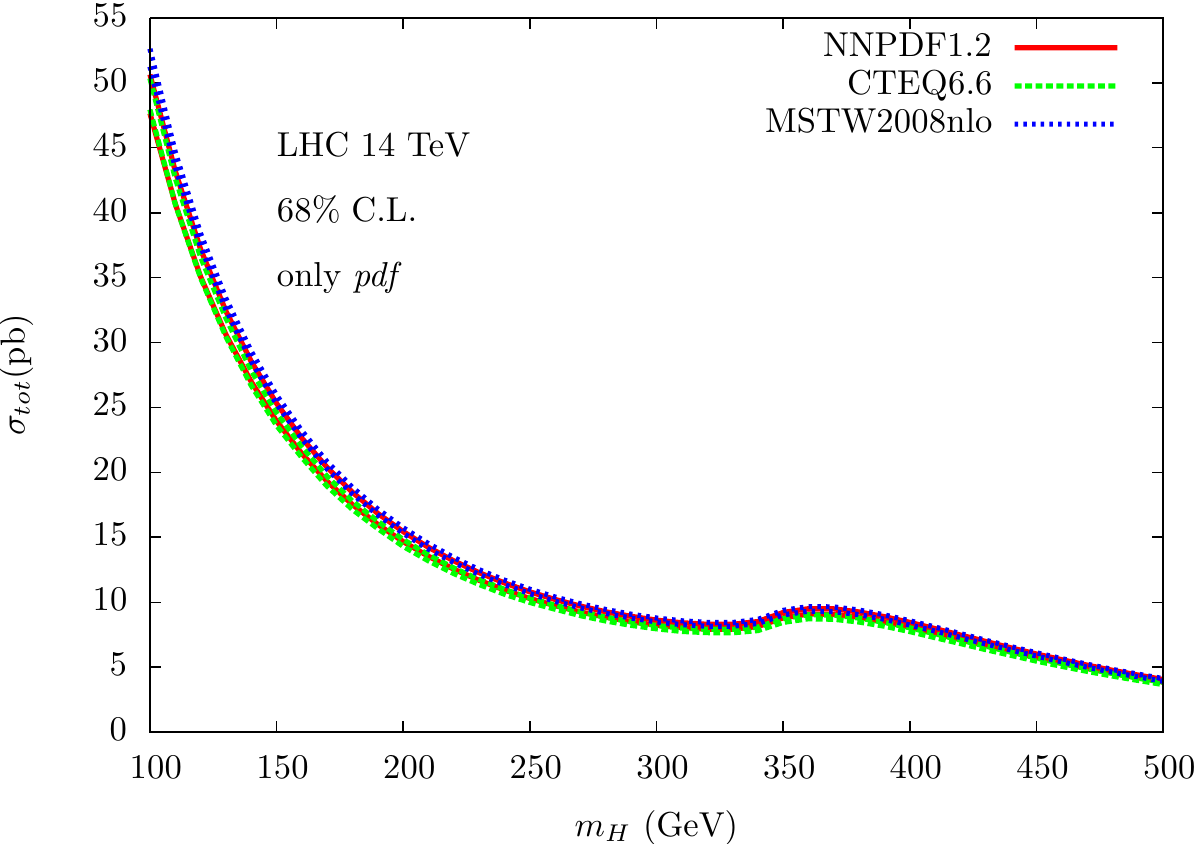}
\includegraphics[width=.32\linewidth]{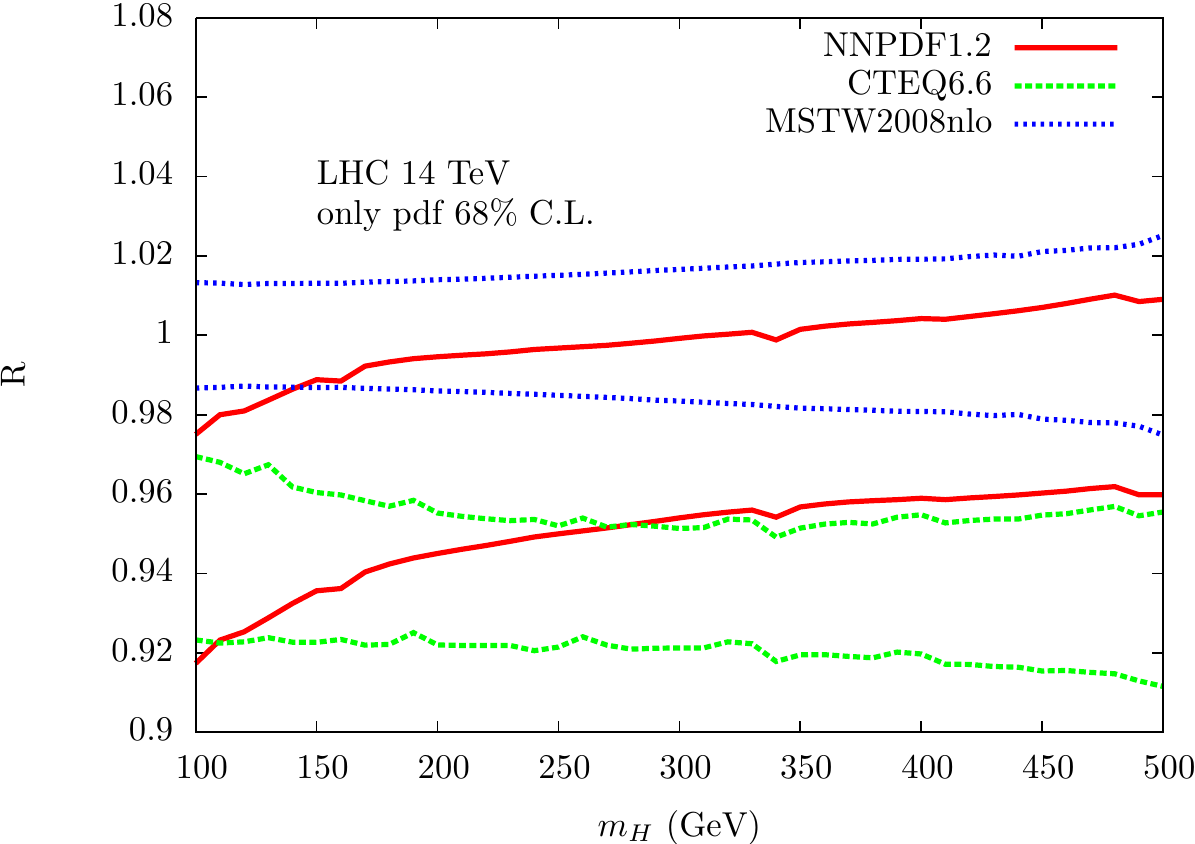}
\includegraphics[width=.32\linewidth]{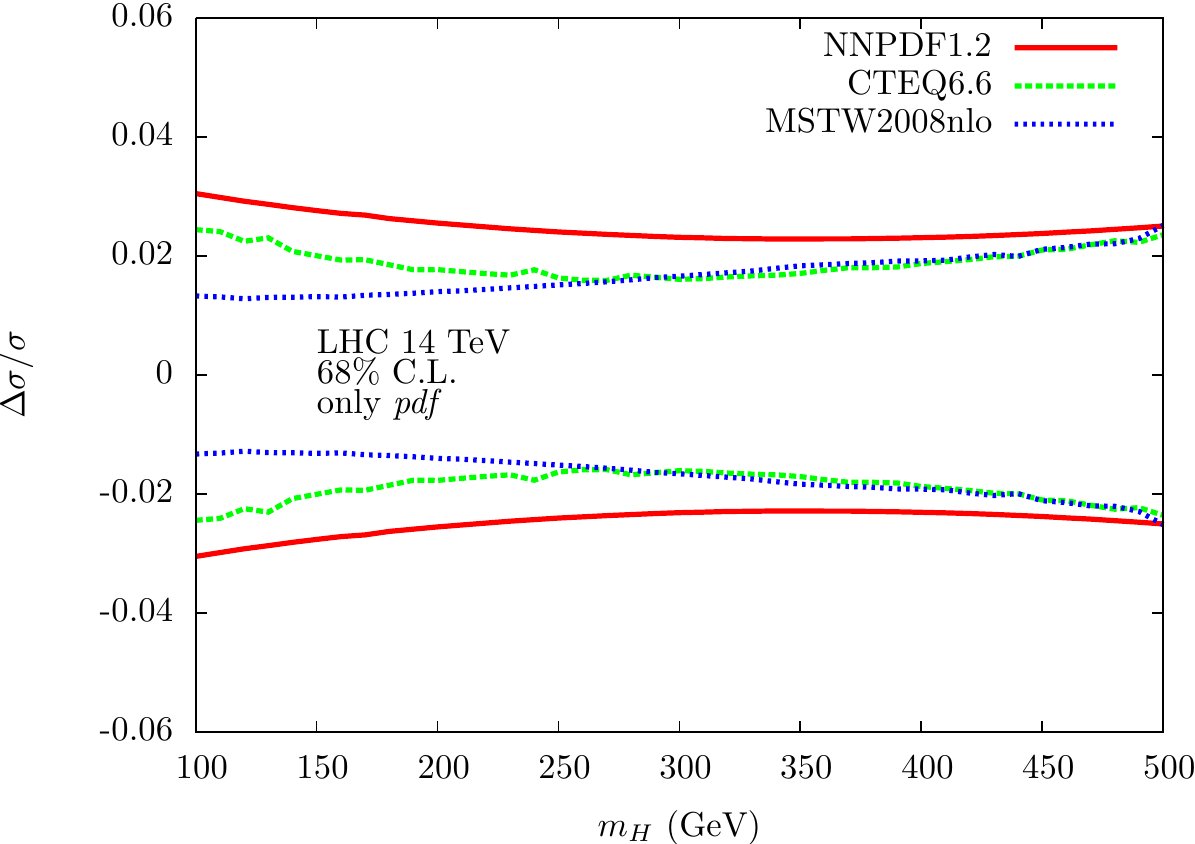}
\caption{\label{fig:onlypdf} Cross sections for Higgs production from
  gluon--fusion the Tevatron (top), LHC~7~TeV (center)
and LHC~14~TeV (bottom). All uncertainty bands are one--$\sigma$ PDF
  uncertainties, with $\alpha_s$ fixed according to
  Eq.~(\ref{asrefcoll}). The left column shows absolute results, the
  central column results normalized to the MSTW08 result, and the
  right column results normalized to each group's central result.
}
\end{center}
\end{figure}


We now turn to the computation of the Higgs 
production cross section and associate
uncertainty band due to PDF variation, with $\alpha_s$ kept fixed at
each group's preferred value, as given by Eq.~(\ref{asrefcoll}), and
uncertainties consistently determined as one--$\sigma$ intervals using
each groups recommended method, hence in particular Hessian methods
for CTEQ and MSTW and Monte Carlo methods for NNPDF. We will not
address here the issue of comparison of the various methods, and we will
simply take each group's results at face value, in particular by
taking as a  68\% confidence
level the interval determined as such by each group. When comparing
results it should be born in mind that, as discussed in
Sect.~\ref{sec:pdf} 
the cross sections probe the gluon luminosity at an essentially fixed
value of $x=m^2_H/s$.

\subsection{Comparison of cross sections and uncertainties}
\label{sec:compxsectunc}

In Fig.~\ref{fig:onlypdf} we compare the cross sections at Tevatron
and LHC  (7~TeV and 14~TeV) energies as a function of the Higgs mass;
each cross section is calculated using the best PDF set of each
group and the corresponding value of $\alpha_s$ of
Eq.~(\ref{asrefcoll}) in the determination of 
 the hard cross section.
Central values can differ by a sizable amount. Discrepancies are due
to three distinct reasons: the fact that the hard cross sections
(independent of the PDF) are different because of the different
values of $\alpha_s$; the fact that the PDFs are different because
they depend on $\alpha_s$ as shown in Figs.~\ref{fig:pdf2-100}; lastly
the fact that even when the same value of $\alpha_s$ is adopted PDF
determination from different groups do not coincide.

As discussed in Sect.~\ref{sec:nnpdf}, the first two effects tend to
compensate each other at small $x$ 
because of the anticorrelation between the gluon
and $\alpha_s$ (see Fig.~\ref{fig:gluon-alphas-corr}), while at large $x$
they go in the same direction. As we shall see in
Sect.~\ref{sec:alphaunc}, the transition between anticorrelation to
correlation happens for LHC~7~TeV for intermediate values of the Higgs
mass.

The relative impact the first effect (which affects the hard cross
section) and of the second two combined (which affect the parton
luminosity) can be assessed by comparing the cross sections of
Fig.~\ref{fig:onlypdf} with the luminosities of
Fig.~\ref{fig:lumi-cfr}: for instance, for $m_H=150$~GeV at the LHC
(7~TeV), the MSTW08 cross section is seen in Fig.~\ref{fig:onlypdf} to
be by about 7\% higher than the
CTEQ6.6. Of this, Fig.~\ref{fig:lumi-cfr} shows that about 3\% is due
to the different parton luminosity, hence about 4\% must be due to the
choice of $\alpha_s$ in the hard matrix element. In
Sect.~\ref{sec:alphaunc} we will determine this variation directly
(Fig.~\ref{fig:pdf_as_env}) and see that this is indeed the case.

Because the cross section starts at order $\alpha_s^2(m_H)$, with a
  NLO $K$--factor of order one, we expect  a percentage
change $\Delta\alpha_s$ in $\alpha_s$ to change the cross section by about 
$2.5\Delta\alpha_s$, which indeed suggests a 4\% change
of the hard cross section when 
$\alpha_s$ is changed from the MSTW08 to the CTEQ6.6 value. In fact,
comparison of Fig.~\ref{fig:onlypdf} to Fig.~\ref{fig:lumi-cfr}
shows that this simple estimate works generally quite well: the
difference in hard matrix elements is $2.5\Delta\alpha_s$ so 4\% when
moving from the MSTW08 to the CTEQ6.6 value, with the rest of the
differences seen in the cross sections in Fig.~\ref{fig:onlypdf} due
to the gluon luminosities displayed in Fig.~\ref{fig:lumi-cfr}.
Because the latter are compatible to one--$\sigma$, this is already
sufficient to show that nominal uncertainties on PDF sets are
sufficient to accommodate the different central values of NNPDF1.2,
CTEQ6.6 and MSTW08 once a common value of $\alpha_s$ is adopted.

Uncertainties turn out to be quite similar for all groups and of order
of a few percent, with uncertainties largest 
at the Tevatron for large Higgs mass. The  growth of 
uncertainties as the Higgs mass is raised or the energy is lowered is
due to the fact that larger $x$ values are then probed, where
knowledge of the gluon is less accurate,
 as shown in  Fig.~\ref{fig:pdf-cfr}.

\subsection{The uncertainty of the PDF uncertainty bands}
\label{sec:uncunc}

In order to answer the question of the compatibility
of different determinations of PDFs or of physical observables
extracted from them it is important that the uncertainties are
provided on the quantities which are being compared.
Whereas this is standard for central values, it is less frequently
done for uncertainties themselves. A systematic way of doing so in a
Monte Carlo approach has been introduced in
Refs.~\cite{Ball:2009by,Ball:2010de} (see in particular
Appendix~A of~\cite{Ball:2010de}): the difference between two
determinations of a central value or an uncertainty is compared to the
sum in quadrature of the uncertainty on each of the two quantities. 
Only when this ratio is significantly larger
than one is the difference significant.

\begin{figure}
\begin{center}
\includegraphics[width=100mm,angle=0]{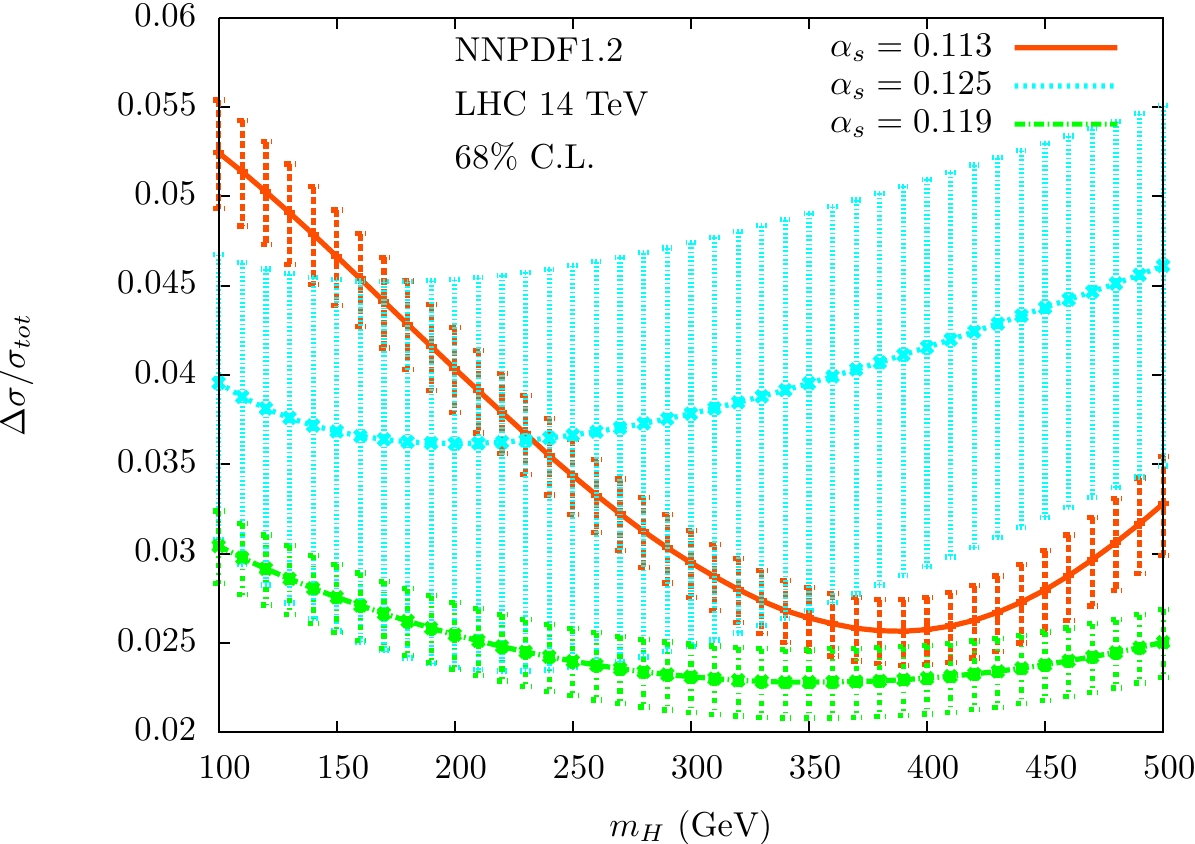}
\caption{\small \label{fig:pdf_unc_nnpdf}
 The uncertainty on the 
Higgs cross section determined using NNPDF1.2 with different values of
$\as$.
The width of the band shown denotes the statistical uncertainty on the
mean uncertainty of the sample, 
$\sigma[\sigma^2]$, Eq.~(\ref{eq:jacksigma}). Note that the
uncertainty on the uncertainty of the PDFs is $\sqrt{N_{\rep}}=10$
  times larger.}
\end{center}
\end{figure}
It is important to observe that when addressing the compatibility of
two determinations two inequivalent questions can be asked: whether
the two determination come from statistically indistinguishable
underlying distributions, or whether they come from statistically 
distinguishable distributions, but are nevertheless compatible. To
elucidate the difference, consider two different determinations of the
same quantities based on two independent sets of data. If the data
sets are compatible, the two determinations will be consistent within
uncertainties in the sense that the central values are compatible
within uncertainties ({\it i.e.} the central values will be distributed
compatibly to the given uncertainty if many determinations are
compared). However, the underlying distribution of results will in
general be different: for example, if one of the two datasets is
more accurate than the other, the two distributions will certainly be
statistically inequivalent (they will have different width), yet they
may well be compatible. We expect complete statistical equivalence of
two different PDF determinations only if they are based on the same
data and methodology: for example, NNPDF verifies that its PDF
determination is statistically independent of the architecture of the
neural networks which are used in the analysis. However,  when comparing
determinations which use different datasets, PDF parametrization,
minimization algorithm {\it etc.} we only expect them to be
compatibile, 
but not
statistically indistinguishable.

The determination of the of the uncertainty on the uncertainty appears
to be nontrivial in a Hessian approach, and it has not been addressed
so far in the context of Hessian PDF determination to the best of our
knowledge. We will thus only discuss it within a Monte Carlo
framework. It seems plausible then to assume that uncertainties on
uncertainties are of similar size in existing parton fits, so that a
difference in uncertainties between two given fits is only significant
if it is rather larger than about $\sqrt2 \sigma[\sigma^2]$, where
$\sigma[\sigma^2]$ is the typical uncertainty which we will now
determine in the Monte Carlo framework using NNPDF1.2 PDFs.

In a Monte Carlo approach, the uncertainty which should be used when
assessing statistical equivalence of two quantities determined from
two sets of $N_{\rm rep}$ replicas is the uncertainty of the mean of
the replica sample, while the uncertainty to be used when assessing
compatibility is the uncertainty of the sample itself, which is larger
than the former by a factor $N_{\rm
  rep}$~\cite{Ball:2009by,Ball:2010de}. 
The former vanishes in the
limit $N_{\rm rep}$, while the latter does not: indeed, if the two
replica sets come from the same underlying distributions all
quantities computed from them should coincide in the infinite--sample
size limit, while if they merely come from compatible distributions
even for very large sample quantities computed from them should remain
different. 

Given a set of $N_{\rm
  rep}$ replicas, the variance of the variance can be determined
as~\cite{Amsler:2008zzb}
  \begin{equation}
    \label{eq:ases}
    \sigma^2[\sigma^2]= 
    \frac{1}{N_{\mathrm{rep}}}\left[m_4[q]
      -\frac{N_{\mathrm{rep}}-3}{N_{\mathrm{rep}}-1}
      \left(\bar\sigma^2\right)^2\right], 
  \end{equation}
where $\sigma^2$ and $m_4$ are respectively the variance and fourth
moment of the replica sample. Equivalently, we may determine
$\sigma^2[\sigma^2]$ by the jackknife method (see
{\it e.g.} Ref.~\cite{efron}), {\it i.e.}
removing one of the replicas from the sample and determining the
$N_{\rm rep}$ variances
\be\label{eq:jacksigma}
\sigma^2_{i}(x)=
\frac{1}{N_{\rm rep}-2}
\sum_{j=1,j\neq i}^{N_{\rm rep}}(x_j-\mu_i(x))^2;\quad i=1,\dots,N_{\rm rep}. 
\ee
The variance of the variance is then given by
\be\label{eq:jacksigsigma}
\sigma^2[\sigma^2] =
\sum_{i=1}^{N_{\rm rep}}\left[(\sigma^2_i)^2-\overline{\sigma^4_i}\right],
\ee
where $\sigma^2_i$ is given by Eq.~(\ref{eq:jacksigma}) and 
\be\label{eq:jackmeansigma}
\overline{\sigma^4_i} =\frac{1}{N_{\rm rep}} \sigma^4_i.
\ee

The quantity $\sigma[\sigma^2]$ 
estimated using  Eq.~(\ref{eq:ases}) or equivalently
Eq.~(\ref{eq:jacksigsigma}) provides the uncertainty of the sample, and
indeed it vanishes in the limit $N_{\rm rep}\to\infty$: it should thus be
used to assess statistical equivalence. The quantity which should be
used to assess consistency is $\sqrt{N_{\rm rep}}\sigma[\sigma^2] $.

In Fig.~\ref{fig:pdf_unc_nnpdf}
we plot $\sigma[\sigma^2]$  determined using
Eq.~(\ref{eq:jacksigsigma})  using the NNPDF1.2 sets with different
values of $\alpha_s$. We can immediately conclude from this figure
that the  uncertainties on the cross sections does not change 
in a statistically significant way when $\alpha_s$ is varied within
its uncertainty Eq.~(\ref{asbethke}) or even Eq.~(\ref{aspdg}):
$\alpha_s$ has to be be varied by about three times the uncertainty
Eq.~(\ref{aspdg}) for the uncertainty on the cross section to change
in a statistically significant way. Incidentally, this plot also shows
that uncertainties are not necessarily smaller (and in fact are mostly
larger) when $\alpha_s$ is varied away from its preferred value, as
already seen in Sect.~\ref{sec:nnpdfseries} and
Fig.~\ref{fig:xg_comp}.

When assessing compatibility (as opposed to statistical equivalence),
the size of the uncertainty bands shown in
Fig.~\ref{fig:pdf_unc_nnpdf} must be rescaled by a factor
$\sqrt{N_{\rm rep}}=10$. This rescaled uncertainties
on the uncertainty are then more then half of the uncertainty itself. This
means that all uncertainties shown in Fig.~\ref{fig:onlypdf} for the
three PDF fits under investigation
are in
fact compatible with each other at the one--$\sigma$ level, since they
differ at most by a factor two (NNPDF vs. MSTW at LHC~14~TeV and the
lowest values of $m_H$), and in fact usually rather less than that.

\clearpage

\section{The cross section: $\alpha_s$ 
uncertainties}
\label{sec:as}

We have seen in the previous Section (see Fig.~\ref{fig:onlypdf})
that once differences in hard matrix elements due to the different
choice of $\alpha_s$ are accounted for, cross sections 
computed using different parton sets agree to one--$\sigma$ because
parton luminosities do.
However, parton luminosities compared in Fig.~\ref{fig:lumi-cfr}
were determined using the different values of
$\alpha_s$ Eq.~(\ref{asrefcoll}). 

For a fully consistent
comparison, we must determine central parton luminosities (and thus
cross sections) with a common value of $\alpha_s$, thereby isolating
the differences which are genuinely due to PDFs. The uncertainty
related to the choice of $\alpha_s$ must be then determined by
variation around the central value.  This then raises the question of the
correlation between this $\alpha_s$ variation and  the values of the
PDFs and their uncertainties.
 We now address all these issues in turn.

\subsection{The cross section with a common  $\alpha_s$ value}
\label{sec:impcvalpha}

\begin{figure}
\begin{center}
\includegraphics[width=.49\linewidth]{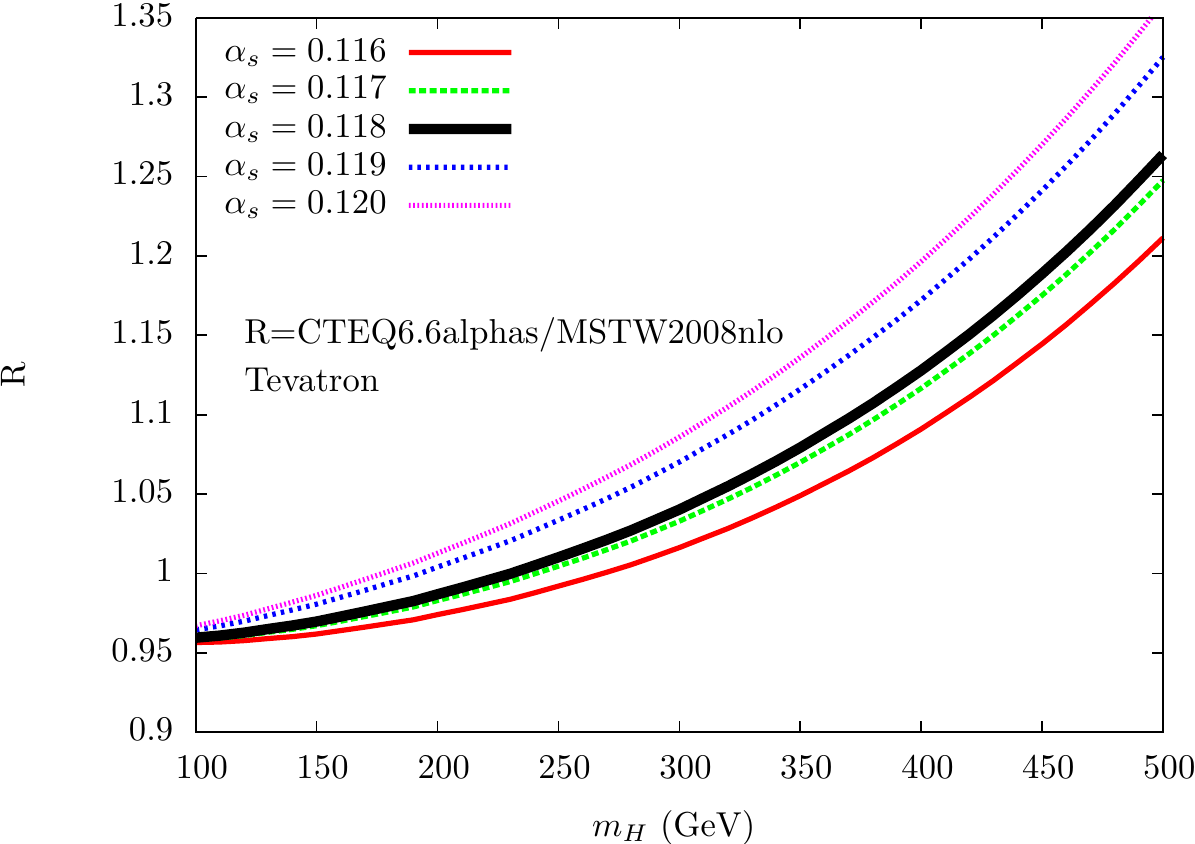}
\includegraphics[width=.49\linewidth]{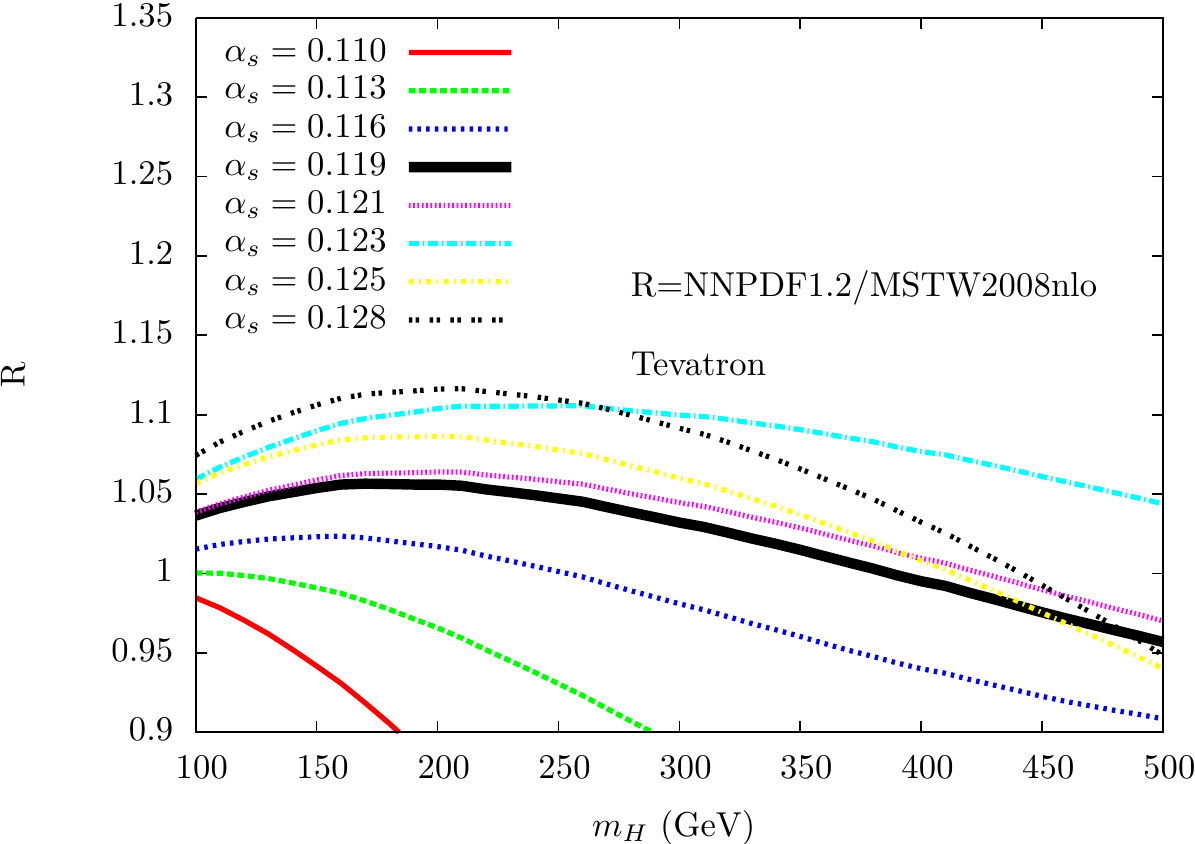}\\
\includegraphics[width=.49\linewidth]{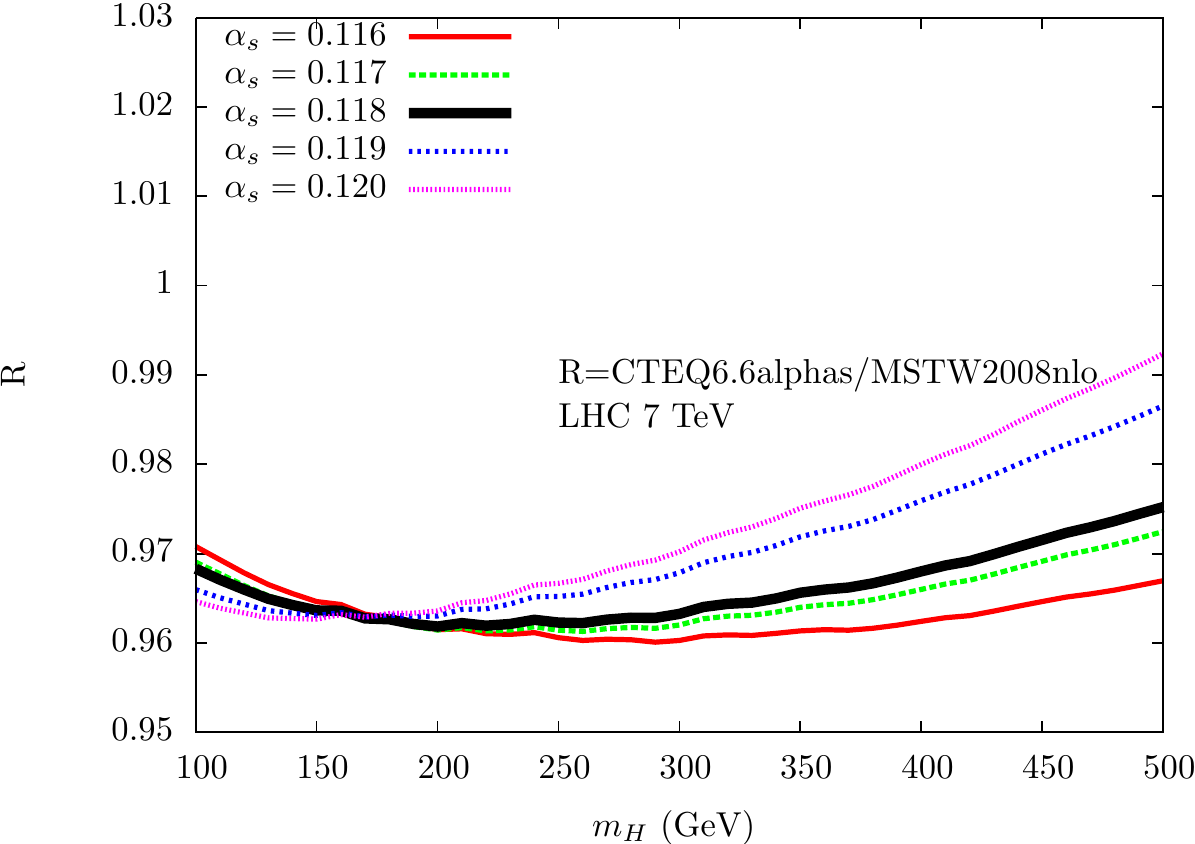}
\includegraphics[width=.49\linewidth]{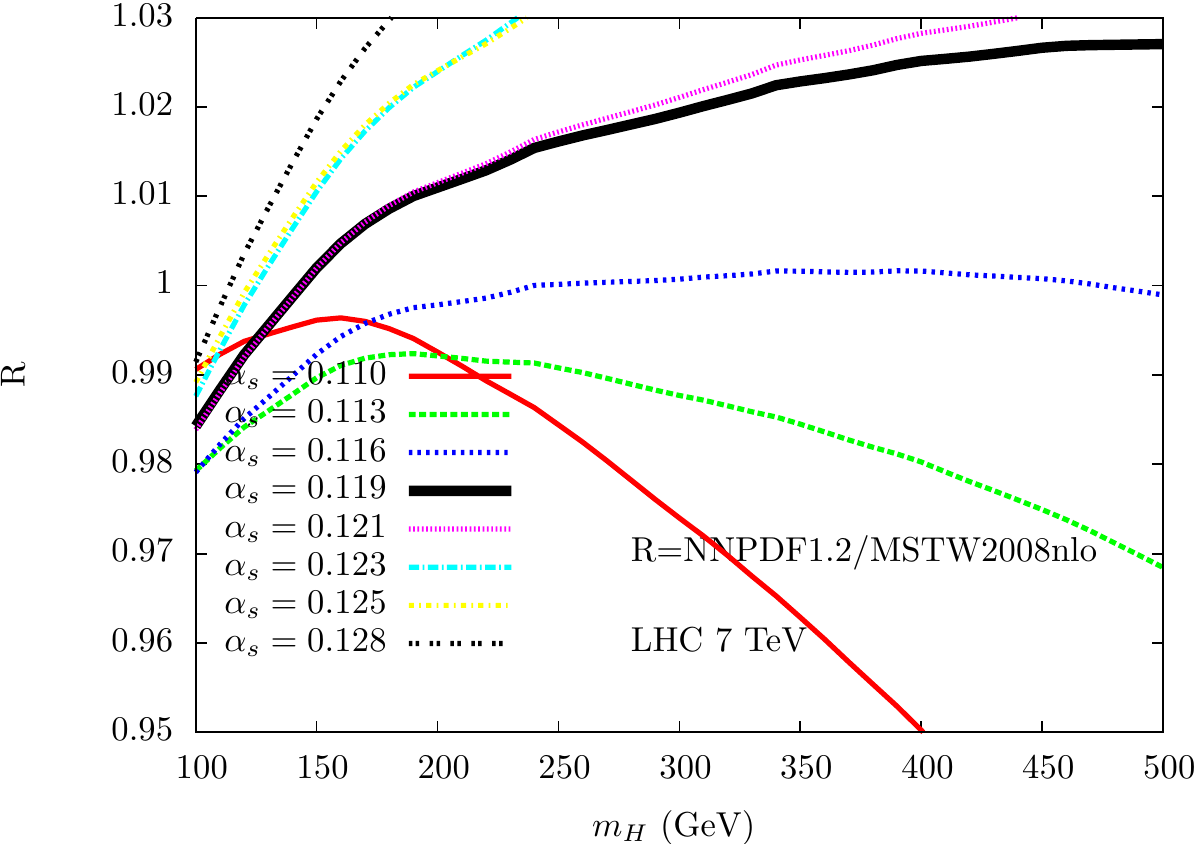}\\
\includegraphics[width=.49\linewidth]{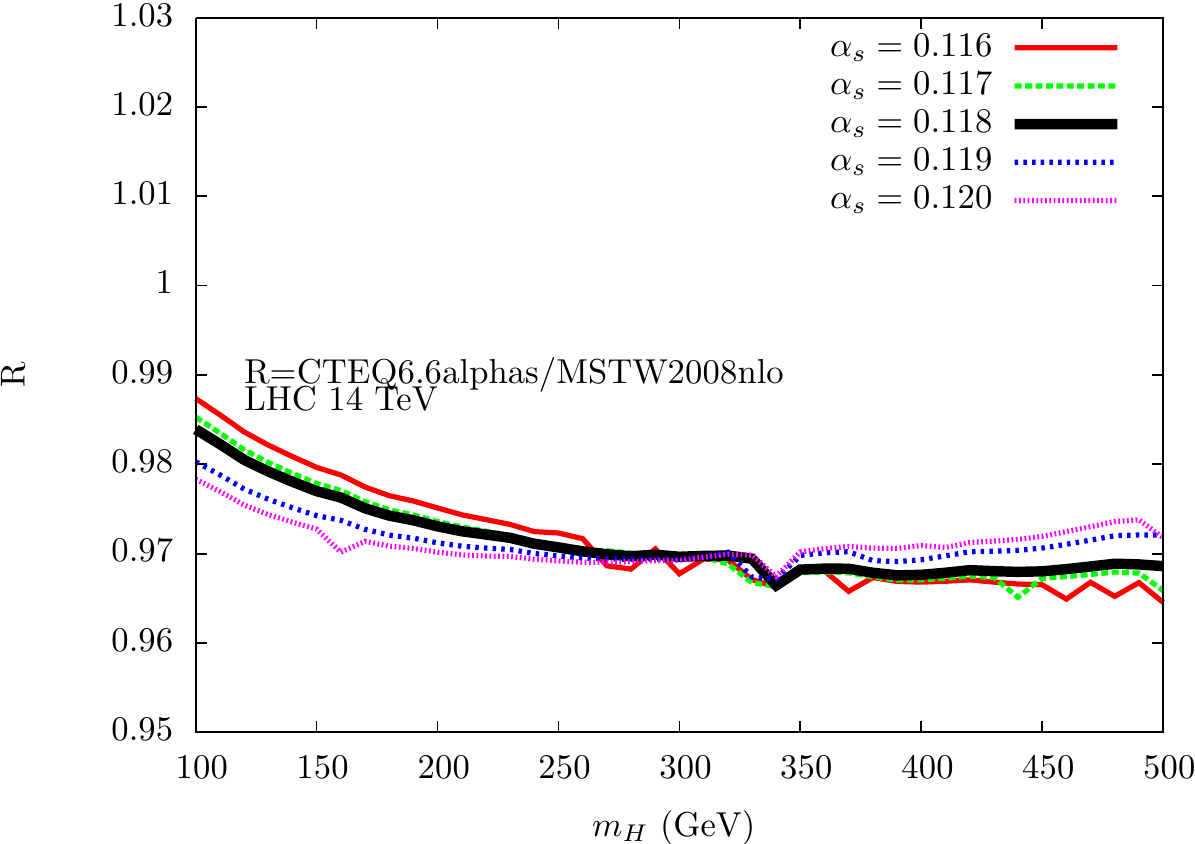}
\includegraphics[width=.49\linewidth]{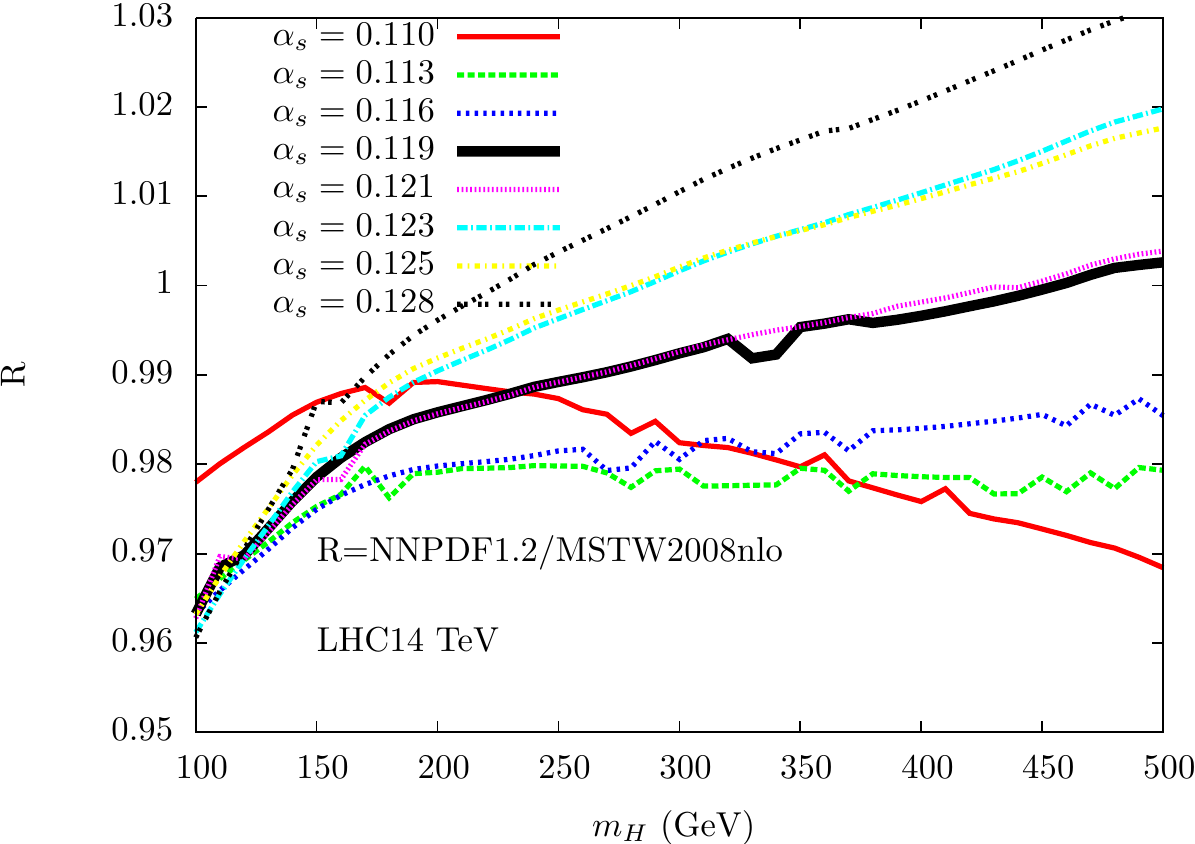}
\caption{\small \label{fig:pdf_same_as}
 Ratio of the Higgs production cross sections determined using
 PDFs from different groups, but
obtained with the same value of $\alpha_s$: CTEQ6.6/MSTW08
 (left) and NNPDF1.2/MSTW08 (right). Results are shown for different 
 values of $\alpha_s$, and for the Tevatron (top), LHC~7~TeV (center)
 and LHC~14~TeV (bottom).} 
\end{center}
\end{figure}
The cross sections obtained from the three PDF sets under
investigation    at the same value of $\alpha_s$ are compared
in Fig.~\ref{fig:pdf_same_as}, where the ratios of cross
sections computed using in the numerator and denominator  
two different PDF sets
(NNPDF1.2 vs. MSTW08 and CTEQ6.6 vs. MSTW08) are shown for 
different choices of $\alpha_s$.

Comparison of these ratios to the PDF uncertainties
shown in Fig.~\ref{fig:onlypdf} show that they are typically of a
similar size: namely, two--three percent at the LHC, while at the Tevatron 
they are about twice as large for light Higgs mass, rapidly growing more or
less linearly, up to  around 10\% around the top threshold. The largest
discrepancy in comparison to the PDF uncertainties is found at the
lowest Higgs masses at the Tevatron between the CTEQ and MSTW sets,
whose uncertainty bands barely overlap there. This can be traced to
the behaviour of the gluon luminosities of Fig.~\ref{fig:lumi-cfr}.

Hence, the spread of central values obtained using
the PDF sets under investigation is consistent with their
uncertainties, which are thus unlikely to be incorrectly estimated. Of
course, it should be born in mind that these uncertainties are in turn
estimated with a finite accuracy, unlikely to be much better than  50\%,
as seen in Sect.~\ref{sec:uncunc}.

\subsection{Uncertainty due to the choice of $\alpha_s$}
\label{sec:alphaunc}
\begin{figure}[h]
\begin{center}
\includegraphics[width=.49\linewidth]{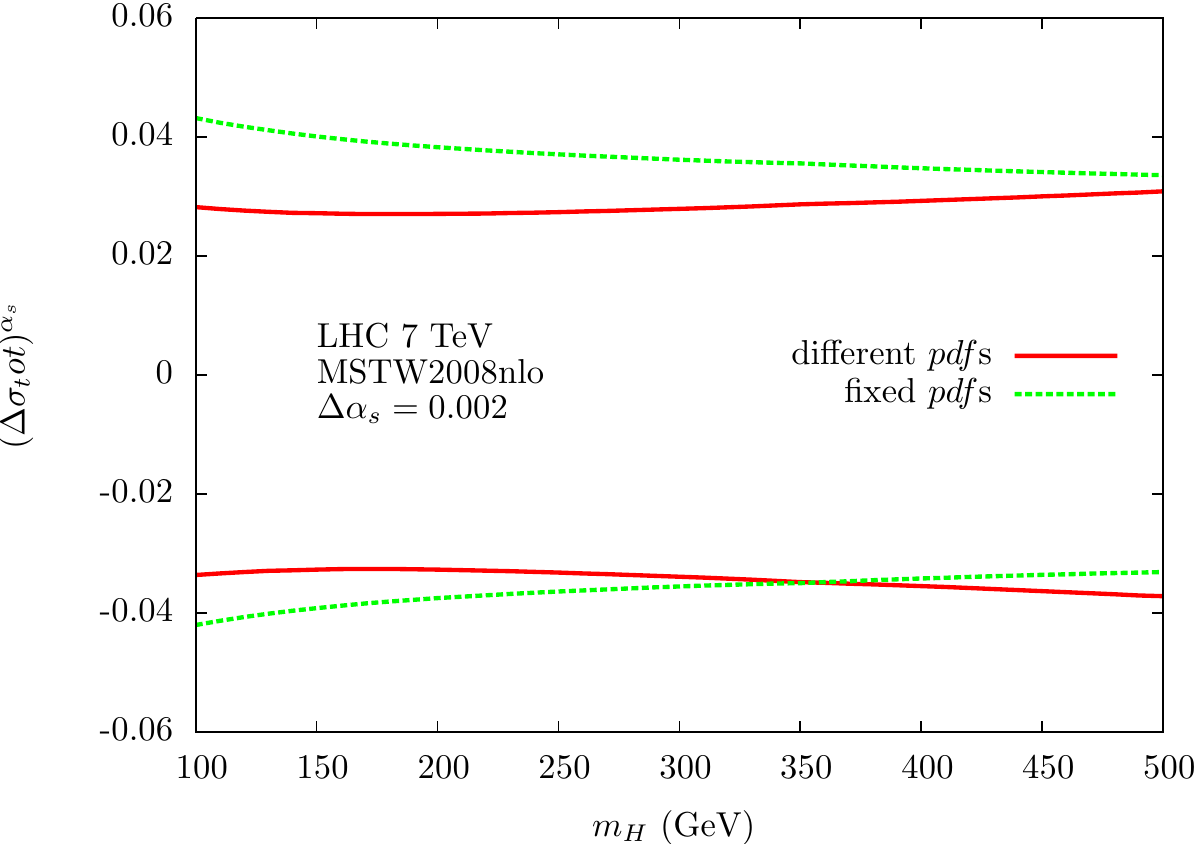}
\includegraphics[width=.49\linewidth]{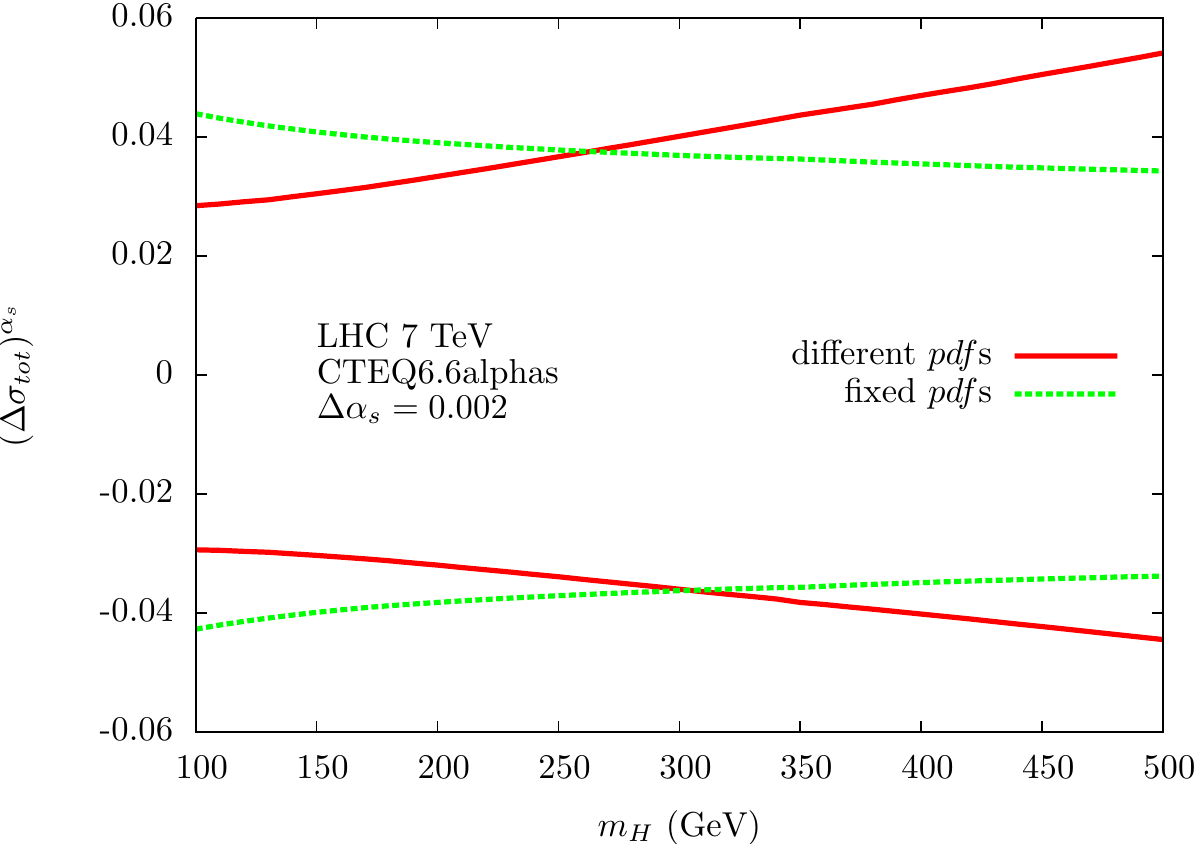}
\includegraphics[width=.49\linewidth]{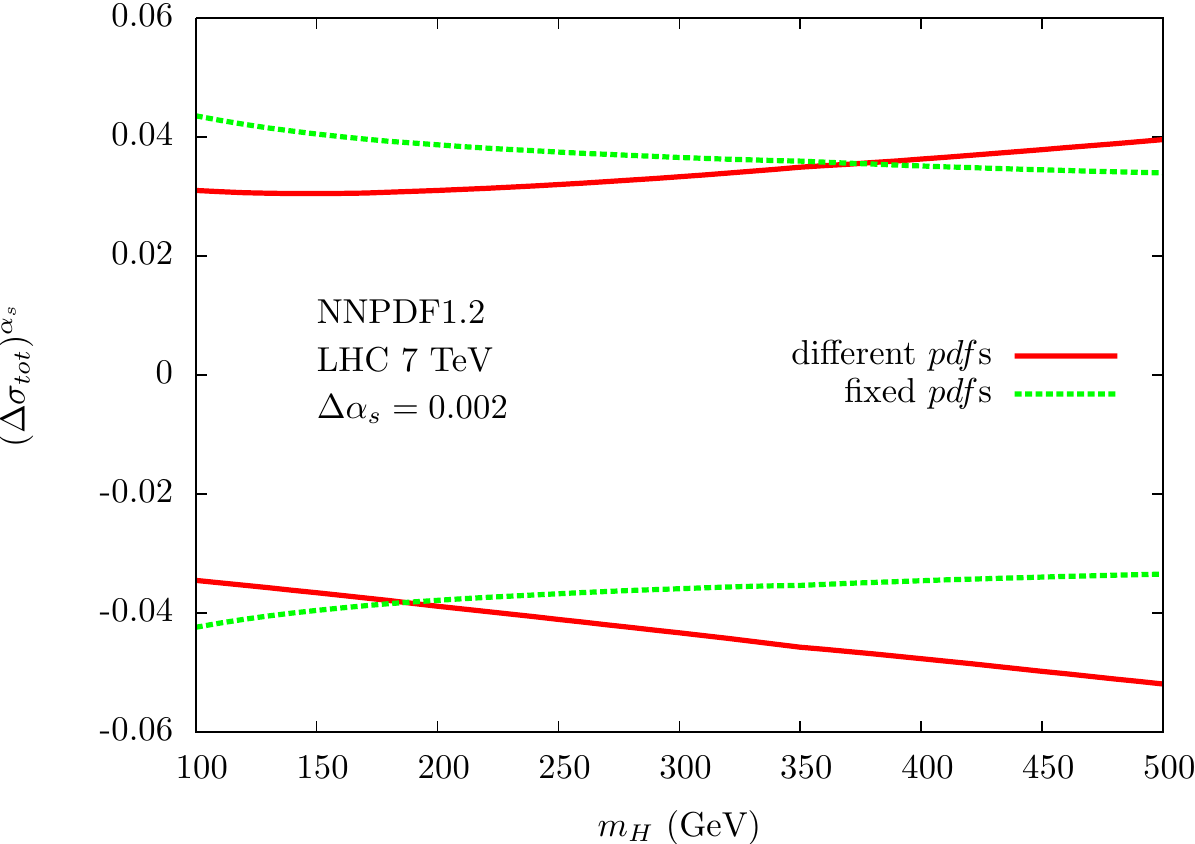} 
\caption{\label{fig:pdf_as_env}
Relative variation $\Delta\sigma/\sigma$
Eq.~(\ref{eq:as1}) of the cross 
section due a variation $\Delta\alpha_s=0.002$ about the preferred 
value Eq.~(\ref{eq:alphasref}) adopted by each PDF set. Results are
shown both with PDFs kept fixed (dashed bands), 
or with the best--fit PDF set
corresponding to each value of $\alpha_s$ (solid bands).
}
\end{center}
\end{figure}

The simplest way to estimate the uncertainty due to $\alpha_s$ is to
take it as uncorrelated to the PDF uncertainty, and determine the
variation of the cross section as $\alpha_s$ is varied. This, in turn,
can be done either by simply keeping the PDFs fixed, or else by also taking
 into account their correlation to the value of $\alpha_s$ discussed
 in Sect.~\ref{sec:nnpdf}, {\it i.e.} by using for each value of  $\alpha_s$
 the corresponding best--fit PDF set. In either case, the 
 uncertainty on the cross section due to the variation of $\alpha_s$ is
\be
(\Delta\sigma)^\pm_{\as}=\sigma(\as\pm\Delta \as)-\sigma(\as),
\label{eq:as1}
\ee 
with, in the two cases, the cross section computed  either with a
fixed set of PDF, or with the sets of PDFs corresponding to the three
values  $\as\pm\Delta\as$.

We have determined $(\Delta\sigma)^\pm_{\as}$ Eq.~(\ref{eq:as1})
using the central value of $\alpha_s$ of
Eq.~(\ref{eq:alphasref}) and $\Delta\as=0.002$, which would
correspond to a 90\% C.L. variation of $\alpha_s$ according to
Eq.~(\ref{sas90}), and it is (almost exactly) equal to
the difference between the preferred values of
$\alpha_s$ for CTEQ6.6 and MSTW08, Eq.~(\ref{asrefcoll}). For
NNPDF1.2 a PDF set with $\alpha_s=0.117$ is not available, hence the
lower cross section in the case in which the PDFs are varied has been
determined by a suitable rescaling of that which corresponds to $\as=0.116$. 

Results for LHC~7~TeV are shown in Fig.~\ref{fig:pdf_as_env}: 
upper and lower cross sections correspond to
the upper and lower values of $\alpha_s$, but the variation turns out
to be symmetric about the central value to good approximation. If the
PDF is kept fixed, we find
$(\Delta\sigma)^\pm_{\as}/\sigma\approx1.04$ when
$\Delta\as=0.002$,  {\it i.e.} a variation of about 4\%, in excellent
agreement with the simple estimate of 
Sect.~\ref{sec:compxsectunc}. If the PDFs are also varied, the width of the
uncertainty band on  the cross section for light Higgs becomes
smaller, 
because of
the anticorrelation of the small $x$ gluon to $\alpha_s$ discussed in
Sect.~\ref{sec:nnpdf} (see Fig.~\ref{fig:gluon-alphas-corr}), but it
becomes wider as the the Higgs mass is raised and larger $x$ values
are probed, for which the correlation of  gluon to $\alpha_s$ is
positive.

\begin{figure}[t]
\begin{center}
\includegraphics[width=.49\linewidth]{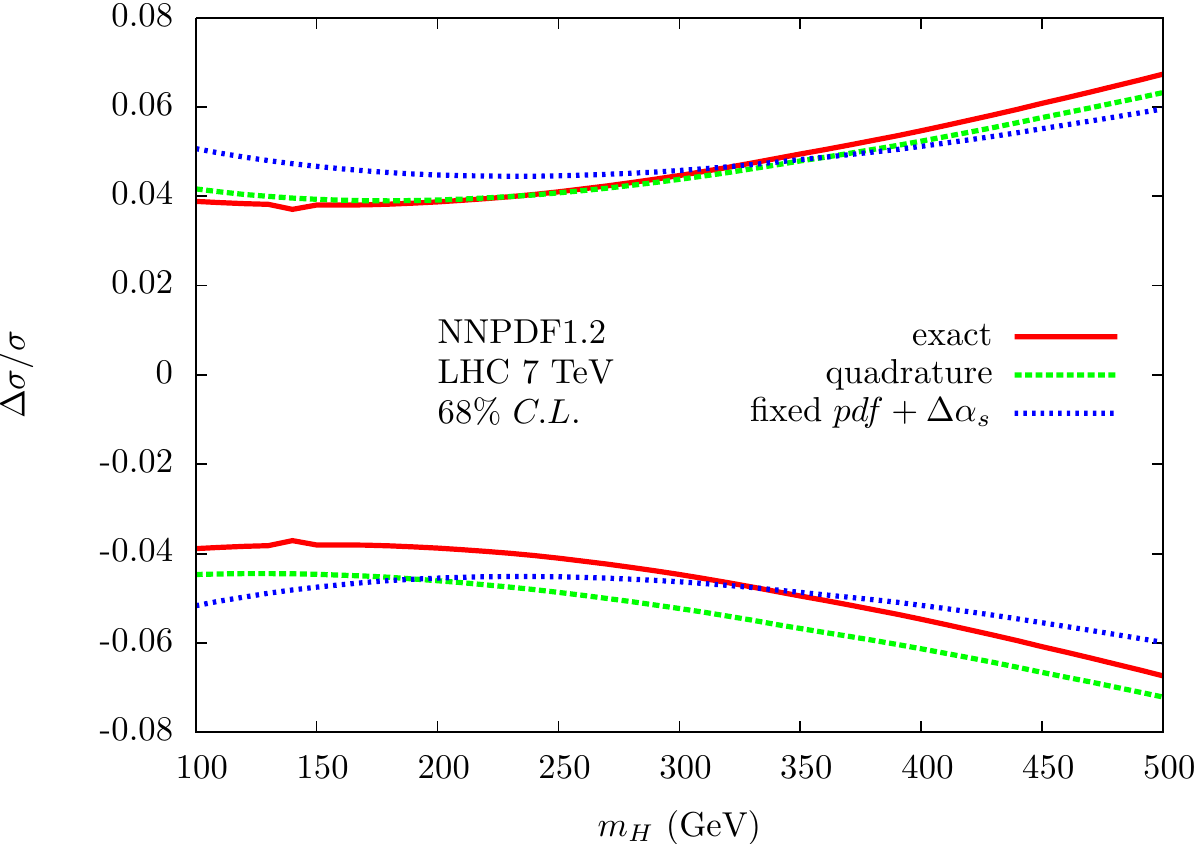}
\includegraphics[width=.49\linewidth]{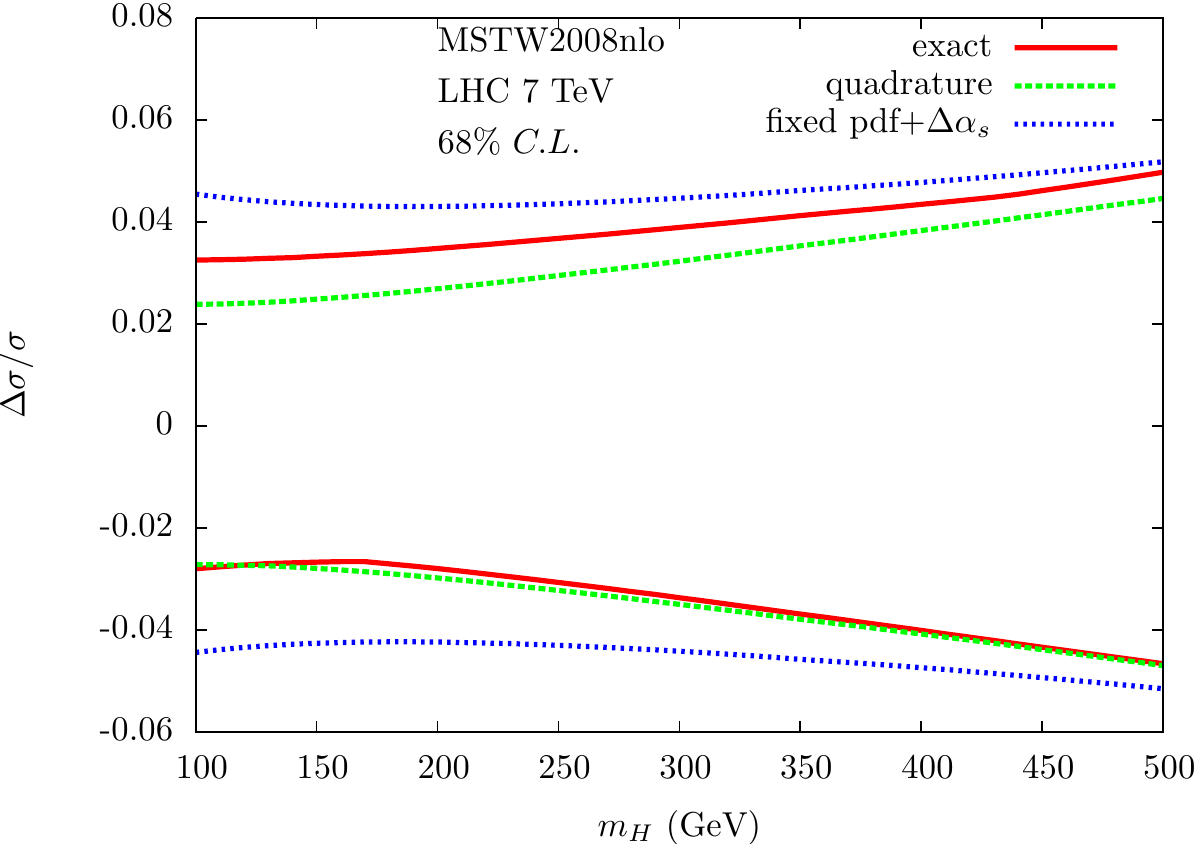}
\caption{\label{fig:exact_quadr}  \small
The combined PDF+$\alpha_s$ relative one--$\sigma$ 
uncertainty on the Higgs cross section
  with NNPDF1.2 (left) and MSTW08 (right). The three bands
  correspond to results obtained keeping into account the full
  correlation between $\alpha_s$ and PDF uncertainties (exact), by
  adding in quadrature PDF uncertainties and $\alpha_s$ uncertainties
  in turn determined by keeping into account the correlation between  $\alpha_s$
  and PDF central values (quadrature), and finally by adding in
  quadrature PDF uncertainties and $\alpha_s$ uncertainties determined
  with PDFs fixed at their central value (fixed
  PDF+$\Delta\alpha_s$). The central values of $\alpha_s$ are given by
  Eq.~(\ref{asrefcoll}), and its one--$\sigma$ range is in
  Eq.~(\ref{sas68}) in all cases except the MSTW08 exact for which it
  is given in Eq.~(\ref{eq:mstwdalpha}).}
\end{center}
\end{figure}

\subsection{Impact of $\alpha_s$ on uncertainties}
\label{sec:alphauncunc}

So far we have considered the uncertainties due to PDFs and due to the
value of $\alpha_s$ separately. However, as seen in
Sect.~\ref{sec:nnpdf}, when $\alpha_s$ changes, not only the central
values but also the uncertainties on PDF change, and this leads to a
correlation between PDF and $\alpha_s$ uncertainties. The effects of
this correlation on the determination of combined PDF+$\alpha_s$
uncertainties is likely to be moderate: it is a higher order effect,
and the dependence of uncertainties on $\alpha_s$ is weak, especially
if compared to the their own uncertainty, see
Fig.~\ref{fig:pdf_unc_nnpdf}.  

We will now quantify this correlation
by computing the total PDF+$\alpha_s$ uncertainty when the correlation
is kept into account, and comparing results to those obtained when PDF
and $\alpha_s$ uncertainties are added in quadrature.
As discussed in Sect.~\ref{sec:nnpdf}, the correlated uncertainty can
be determined
both in the Hessian approach using MSTW08 and in a
Monte Carlo approach using NNPDF. In the MSTW methodology this is done
by relying on a simultaneous determination of PDFs and $\alpha_s$. As
a consequence, the value and range of variation of
$\alpha_s$ must be those obtained in this
determination. This will not hamper our analysis in that the MSTW08
value and
range for $\alpha_s$ of Ref.~\cite{Martin:2009bu}
are close to those under consideration. 
With the NNPDF methodology we are free to choose any value and
range for $\alpha_s$, inasmuch as the corresponding Monte Carlo PDF
replicas are available.

We have thus computed joint PDFs+$\alpha_s$ uncertainties on the Higgs
cross section. For MSTW08 we have followed the procedure of
Ref.~\cite{Martin:2009bu}: the total upper and lower (generally
asymmetric) uncertainties  on an observable $F$
are determined as
\bea
(\Delta F)_+^{\rm PDF+\as}
&=&\max_{\as}
\left(
\{F^{\as}(S_0)+(\Delta F^{\as}_{\rm PDF})_+ \}
\right)-
F^{\as^0}(S_0)
\label{daspdf}
\\
(\Delta F)_-^{\rm PDF+\as}
&=&
F^{\as^0}(S_0)-
\min_{\as}
\left(
\{F^{\as}(S_0)-(\Delta F^{\as}_{\rm PDF})_- \},
\right)
\nonumber
\eea
where $F^{\as}(S_0)$ is the observable computed using the central PDF
set $S_0$ and the value $\alpha_s$ of the strong coupling,
$(\Delta F^{\as}_{\rm PDF})_\pm$ is the PDF uncertainty on the
observable for given fixed value of $\alpha_s$, as determined  from
the  Hessian PDF
eigenvectors~\cite{Martin:2009iq,Martin:2009bu}, and the maximum and
minimum are determined from a set of five results, each computed with
one distinct value of $\alpha_s$ (central,$\pm$ half confidence level,
$\pm$ confidence 
level). 
The value of $\alpha_s^0$ is as given in
Eq.~(\ref{asrefcoll}), while its one--$\sigma$ upper and lower
variations are 
\begin{equation}
\label{eq:mstwdalpha}
\Delta^{(68)} \as=^{+0.0012}_{-0.0015}.
\end{equation}

For NNPDF, the uncertainty is simply given by the standard deviation
of the joint distribution of PDF replicas and $\alpha_s$ values
\begin{equation}
\Delta F^{\rm PDF+\as}=\sigma_F
\equiv
\left[
\frac{1}{N_{\rm rep}-1}
\sum_{j=1}^{N_{\alpha}}
\sum_{k_j=1}^{N_{\rm rep}^{\alpha_s^{(j)}}}
(
F[\{ q^{(k_j,j)}   \}]
-
F[\{ q^0 \}]
)^2
\right]^{1/2}
\label{mcnnpdf}
\end{equation}
where the number of replicas $N_{\rm rep}^{\alpha_s^{(j)}}$ for each
value $\alpha_s^{(j)}$ of the strong coupling is determined in the
gaussian case by Eq.~(\ref{eq:gaussianrepdist}). In this case, we have  taken
as central value and uncertainty on $\alpha_s$ those given in
Eqs.~(\ref{asrefcoll}-\ref{sas68}) respectively. 

The results for the uncertainty 
obtained in this way are shown in Fig.~\ref{fig:exact_quadr},
 each normalized to the corresponding central
cross section.  They are compared to results obtained adding in
quadrature the PDF uncertainties displayed in Fig.~\ref{fig:onlypdf}
and the $\alpha_s$ uncertainties Eq.~(\ref{eq:as1}) displayed in
Fig.~\ref{fig:pdf_as_env}, in turn obtained either with fixed PDFs, or
by taking the PDF set that corresponds to each value of
$\alpha_s$. Note that the range of $\alpha_s$ variation 
 for the MSTW08 curve with full correlation kept into
account, given in Eq.~(\ref{eq:mstwdalpha}), differs slightly from
that, Eq.~(\ref{sas68}), used in all other cases.
The effect of the correlation between $\alpha_s$ and PDF uncertainties is
indeed quite small: as one might expect, it is in fact smaller than the effect
of the correlation between $\alpha_s$ and PDF central values shown in
Fig.~\ref{fig:pdf_as_env}, and much smaller than the uncertainty on
the uncertainty discussed in Sect.~\ref{sec:uncunc}.

\clearpage

\section{The cross section: final results}
\label{sec:finres}

\begin{figure}
\begin{center}
\includegraphics[width=0.49\linewidth]{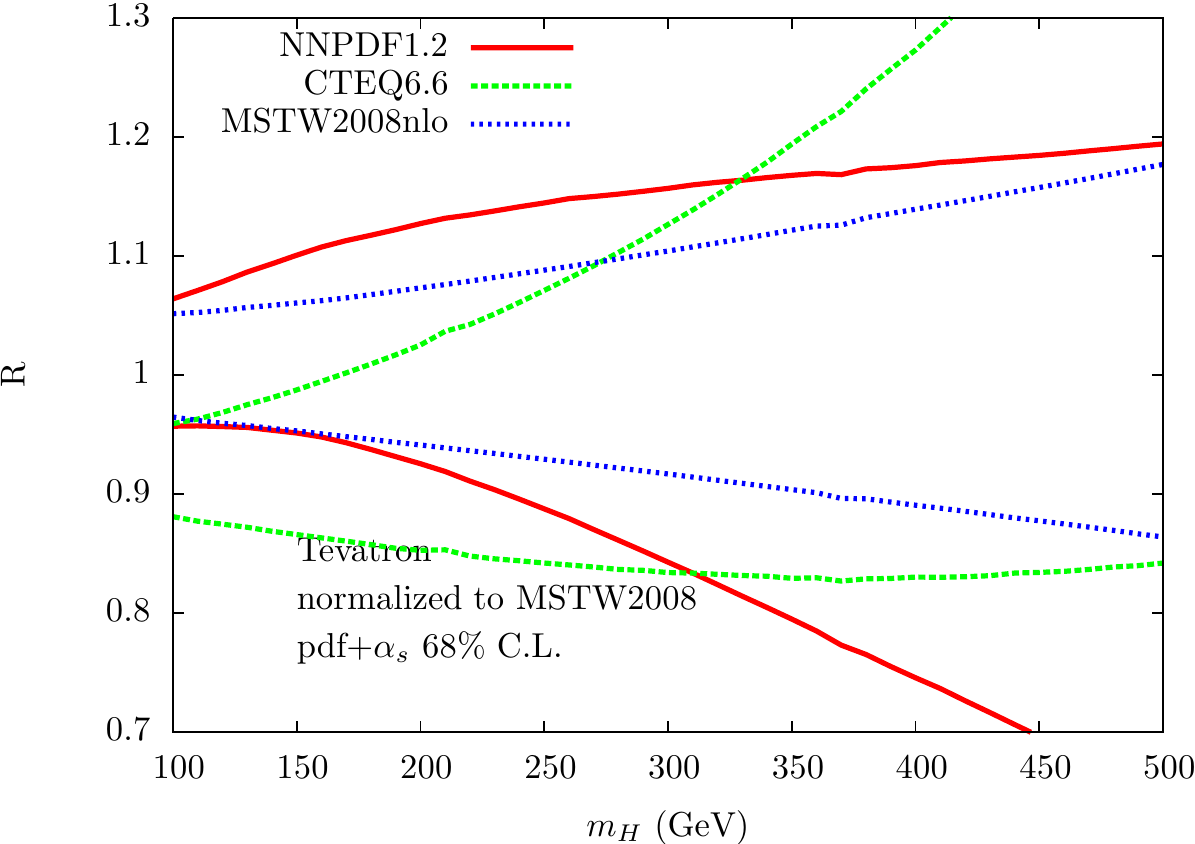}
\includegraphics[width=0.49\linewidth]{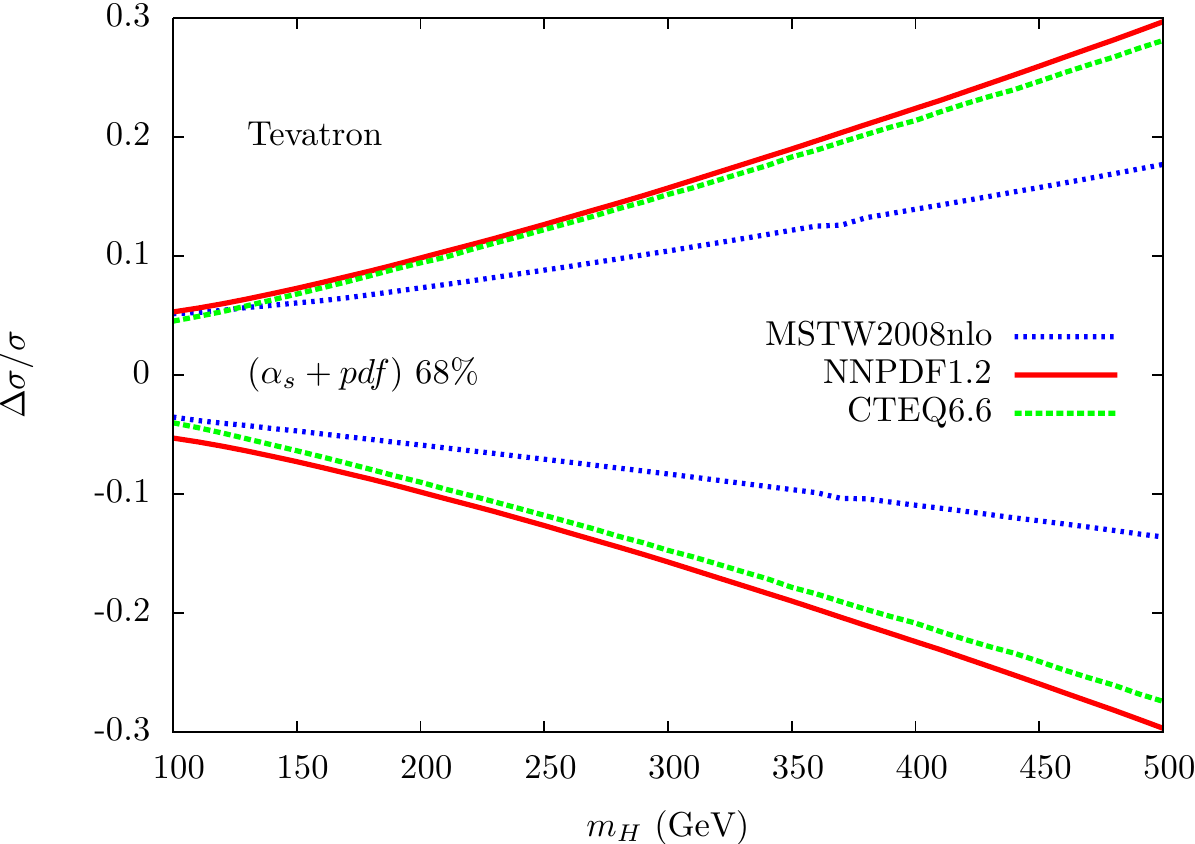}\\
\includegraphics[width=0.49\linewidth]{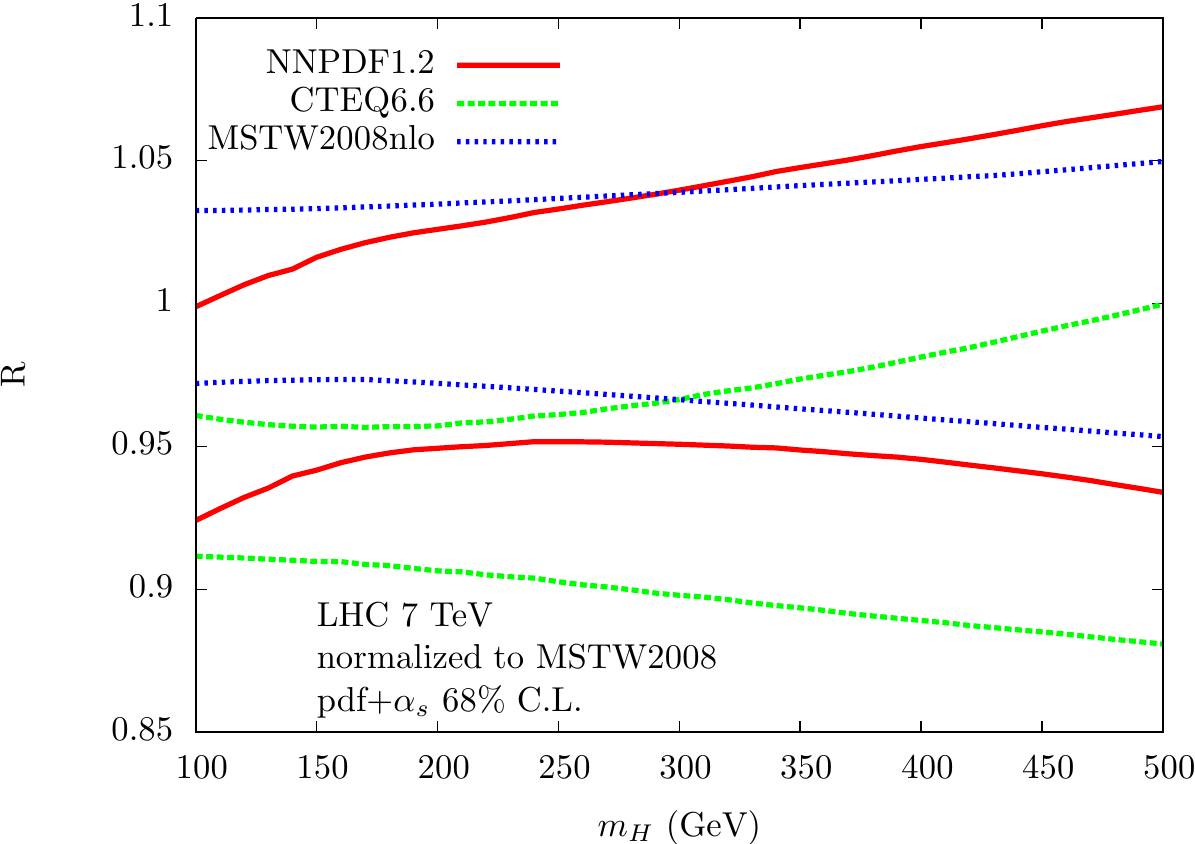}
\includegraphics[width=0.49\linewidth]{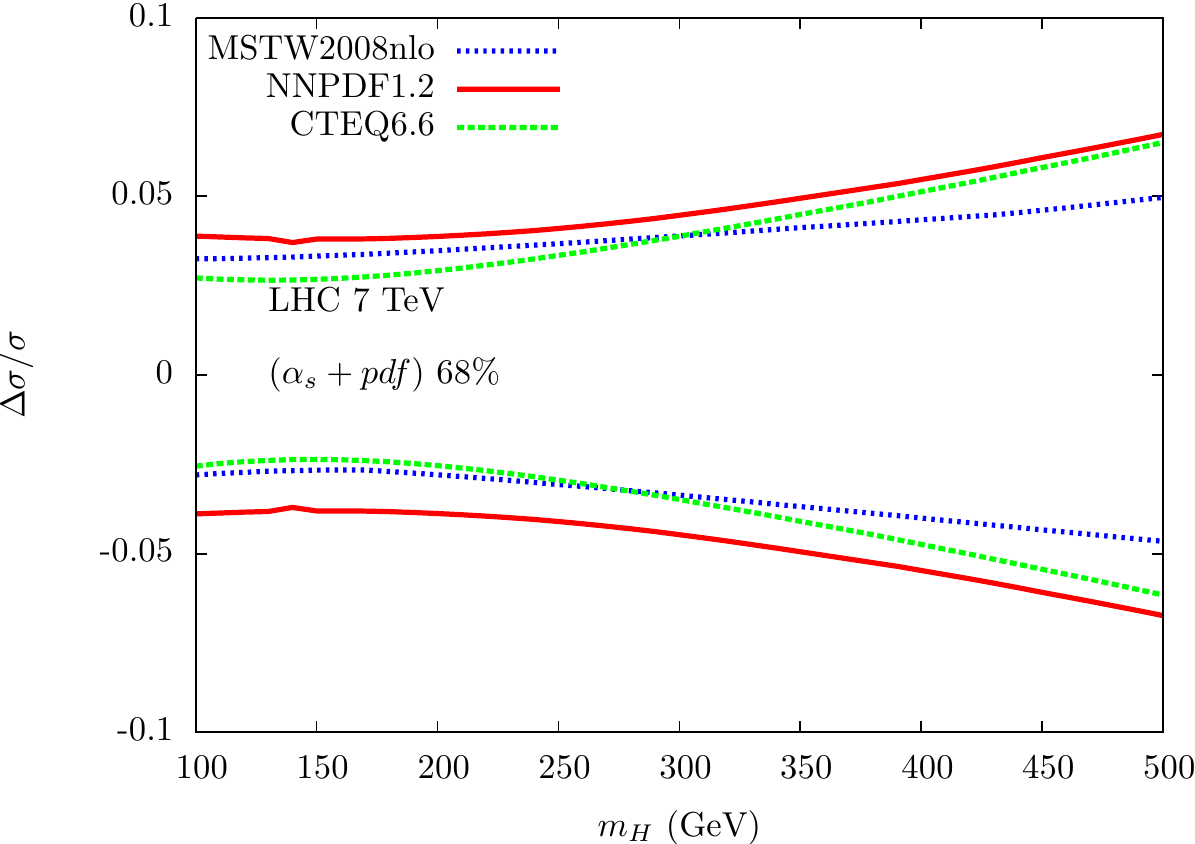}\\
\includegraphics[width=0.49\linewidth]{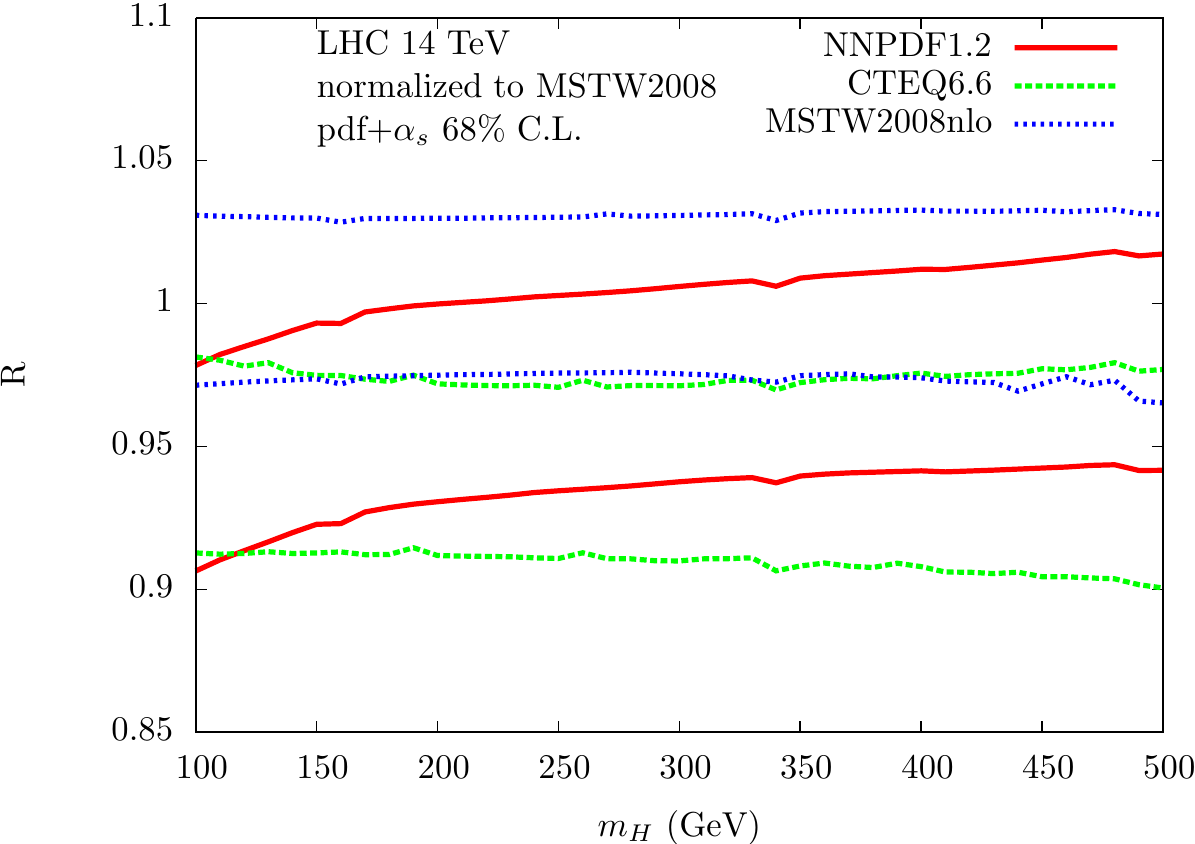}
\includegraphics[width=0.49\linewidth]{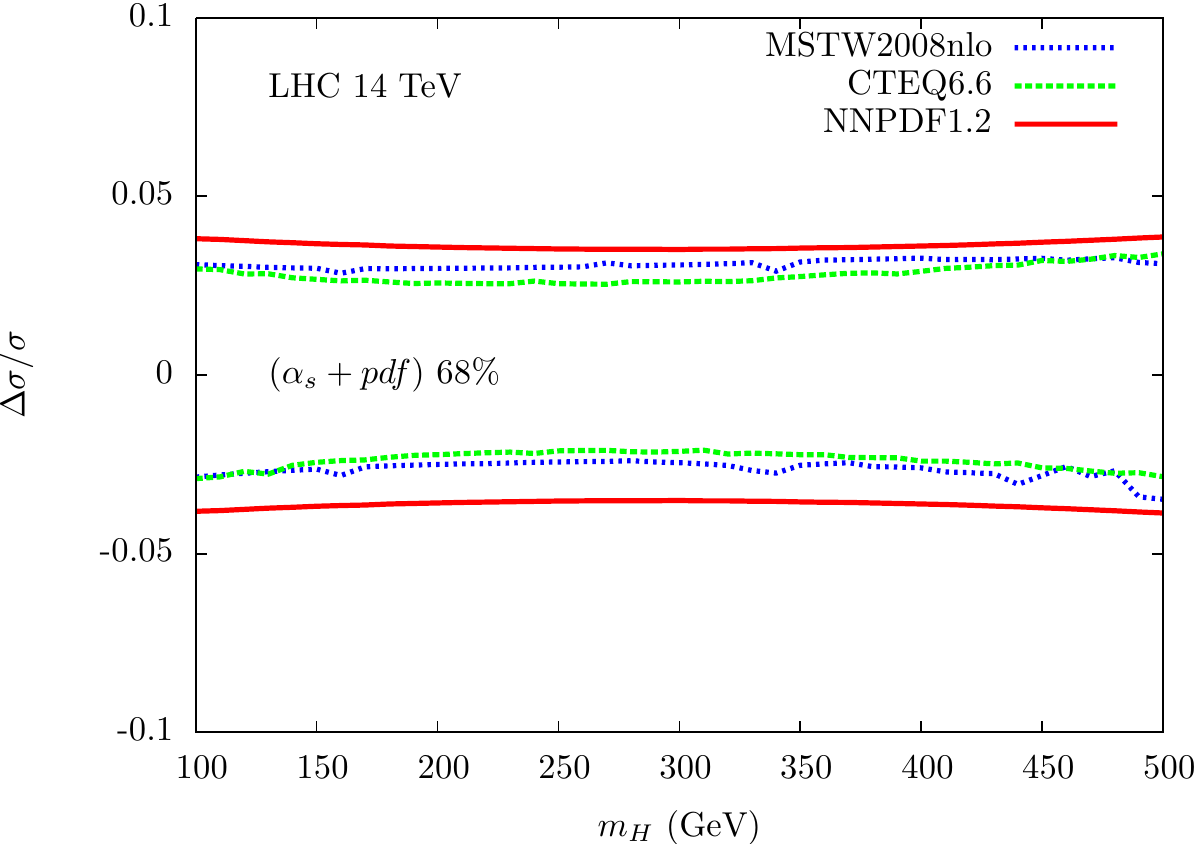}
\caption{\small 
Cross sections for Higgs production from
  gluon--fusion the Tevatron (top), LHC~7~TeV (center)
and  LHC~14~TeV (bottom). All uncertainty bands are  one--$\sigma$ combined
  PDF+$\alpha_s$ 
  uncertainties, as  in Fig.~\ref{fig:exact_quadr} (exact) for MSTW and NNPDF,
  and as in Fig.~\ref{fig:pdf_as_env} for CTEQ, 
with the central value of $\alpha_s$ of
  Eq.~(\ref{asrefcoll}). The left
  column shows results normalized to the MSTW08 result, and the
  right column results normalized to each group's central result
  (relative uncertainty).
\label{fig:finres1}
}
\end{center}
\end{figure}
Our final results for the Higgs cross section are collected in 
 Fig.~\ref{fig:finres1}: the cross sections are the same of
 Fig.~\ref{fig:onlypdf}, but now with the total
 PDF+$\alpha_s$ uncertainty. This is computed taking fully into
 account the correlation between $\alpha_s$, PDF central values and
 uncertainties for NNPDF and MSTW (as discussed
 in Sect.~\ref{sec:alphauncunc} and shown in
 Fig.~\ref{fig:pdf_as_env}).
For CTEQ, which does not provide
 error sets for each value of $\alpha_s$, the PDF uncertainties 
(Fig.~\ref{fig:onlypdf})  
and $\alpha_s$
 uncertainties (Sect.~\ref{sec:alphaunc} and
 Fig.~\ref{fig:pdf_as_env}) 
are added in quadrature; however, as seen in
Sect.~\ref{sec:alphauncunc}, this makes little difference.

The main features of these final results are the following:
\begin{itemize}
\item The total uncertainty on the cross section found by various
  groups are in reasonable agreement, especially if one recalls that
  they are typically affected by an uncertainty of order of about half
  their size, as seen in Sect.~\ref{sec:uncunc}. This follows from
  the fact that the total uncertainty is close to the
  sum in quadrature of the PDF and $\alpha_s$ uncertainties, with
 the $\alpha_s$ uncertainty essentially the same for
  all PDF sets, and PDF uncertainties on the input parton luminosity
  quite close to each other,
\item The predictions obtained using the three given sets all agree within
  the total PDF+$\alpha_s$ one--$\sigma$ uncertainty. 
This is a consequence of the fact that the
  central values of the cross section computed using the same value of
  $\alpha_s$ for all sets agree within PDF uncertainties, and that the
  spread of central values of $\alpha_s$ used by the three groups
  Eq.~(\ref{asrefcoll}) is essentially the same as the $\alpha_s$
  uncertainty under consideration.
\end{itemize}

The results shown in Fig.~\ref{fig:finres1} can be combined into a
determination of the cross section and its combined PDF+$\alpha_s$
uncertainty. We have seen that
there is reasonably good agreement
between the parton sets under investigation: apparent
disagreement is only found if one compares results obtained with values of
$\alpha_s$ which differ more than the uncertainty on
$\alpha_s$. However, in some cases (for example at the Tevatron for
light Higgs) the agreement is marginal: the one--$\sigma$ uncertainty
bands just about overlap. Ideally, this situation should be 
resolved by the PDF fitting groups by investigating the origin of the
underlying imperfect agreement of parton luminosity. However, until this
is done, a common determination of the cross section with
a more conservative estimate of the
uncertainty may be obtained by suitably inflating the PDF uncertainty. 

We have considered two different procedures which lead to such a common
determination, based on the idea of combining  a common $\alpha_s$
uncertainty together with a PDF uncertainty suitably enlarged in
order to keep into account the spread of PDF central values obtained
using different PDF sets. 
\begin{description}
\item[Procedure A:]
This
procedure 
is based on the observation 
that the change in PDFs and uncertainties
when $\alpha_s$ is varied by $\Delta\alpha_s\sim0.001$ is small, as
shown in Fig.~\ref{fig:pdf2-100}, and implicitly demonstrated by the
weak effect of the keeping it into account when evaluating $\alpha_s$
and PDF uncertainties, Figs.~\ref{fig:pdf_as_env}-\ref{fig:exact_quadr}.
Therefore, we can obtain a prediction with a common value of
$\alpha_s$ for all groups by simply using the common intermediate
value of $\alpha_s=0.119$ in the computation of the hard cross section
Eq.~(\ref{Geq}), and then using each group's PDFs and full
PDF+$\alpha_s$ uncertainties, despite the fact that strictly speaking
they correspond to the slightly different values of $\alpha_s$ listed
in Eq.~(\ref{asrefcoll}). Because all
predictions are given at the same value of $\alpha_s$, their spread
reflects differences in underlying PDFs. Hence, we can take
 as 
a conservative estimate of the one--$\sigma$ total PDF+$\alpha_s$
uncertainty on this process the envelope of these predictions,
{\it i.e.} the band between the highest and the lowest prediction. 
This procedure would be easiest to implement if all PDF groups were to
provide PDFs with a common $\alpha_s$ value and uncertainty. It is
still viable provided the central $\alpha_s$ values are not too
different, and the $\alpha_s$ uncertainty can be taken as the same or
almost the same for all groups.

\begin{figure}[t]
\begin{center}
\includegraphics[width=0.49\linewidth]{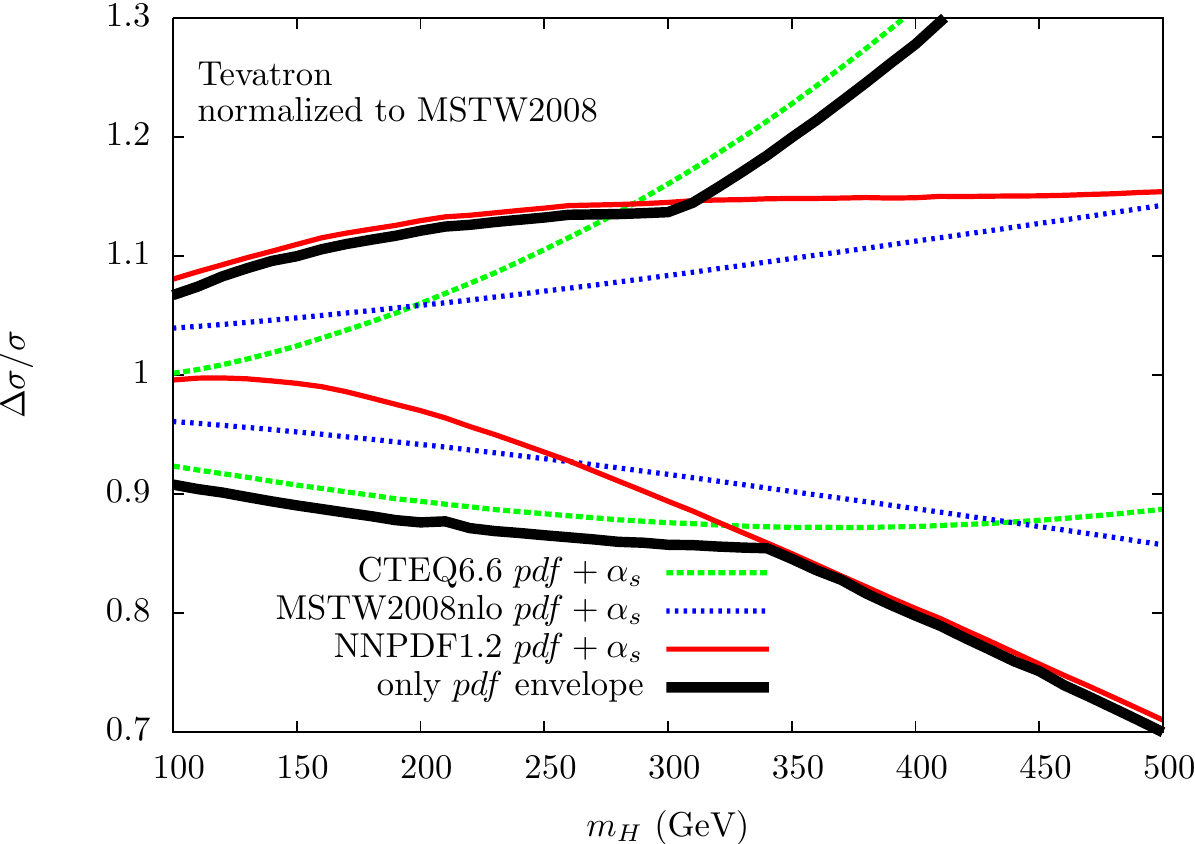}
\includegraphics[width=0.49\linewidth]{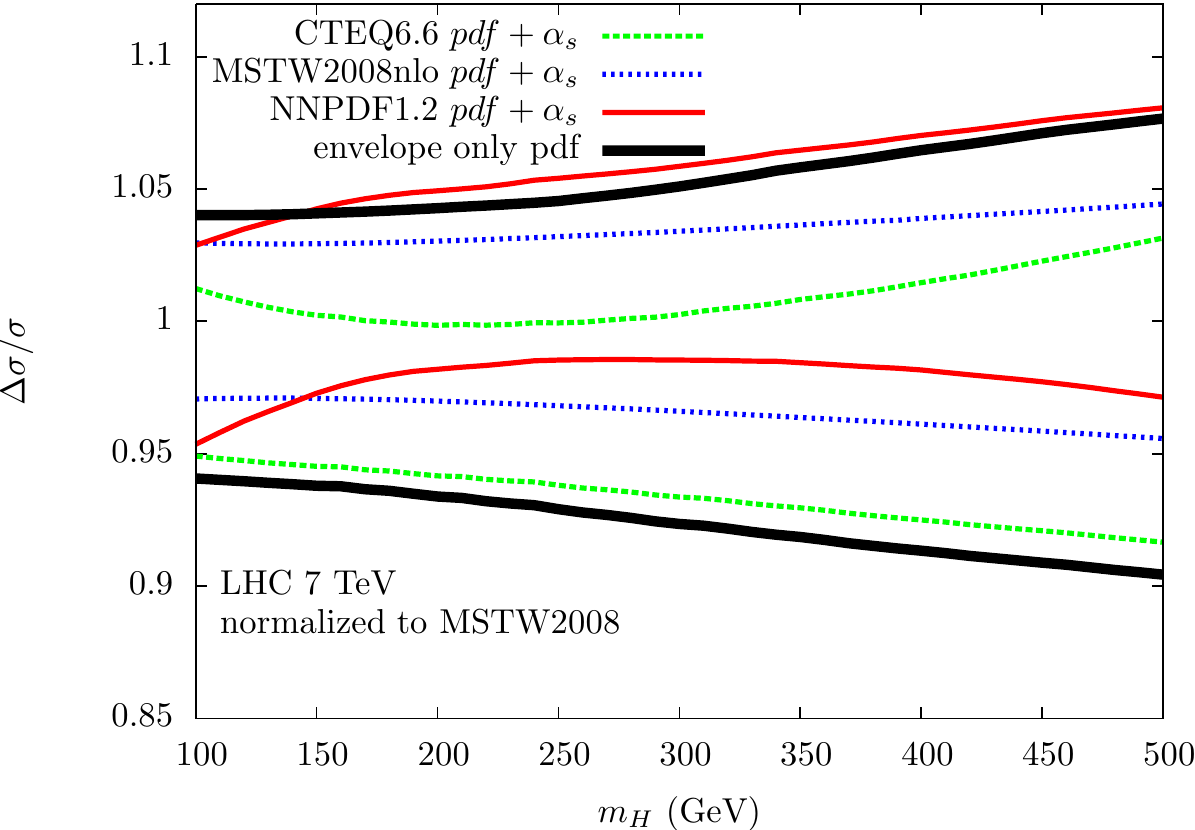}
\includegraphics[width=0.49\linewidth]{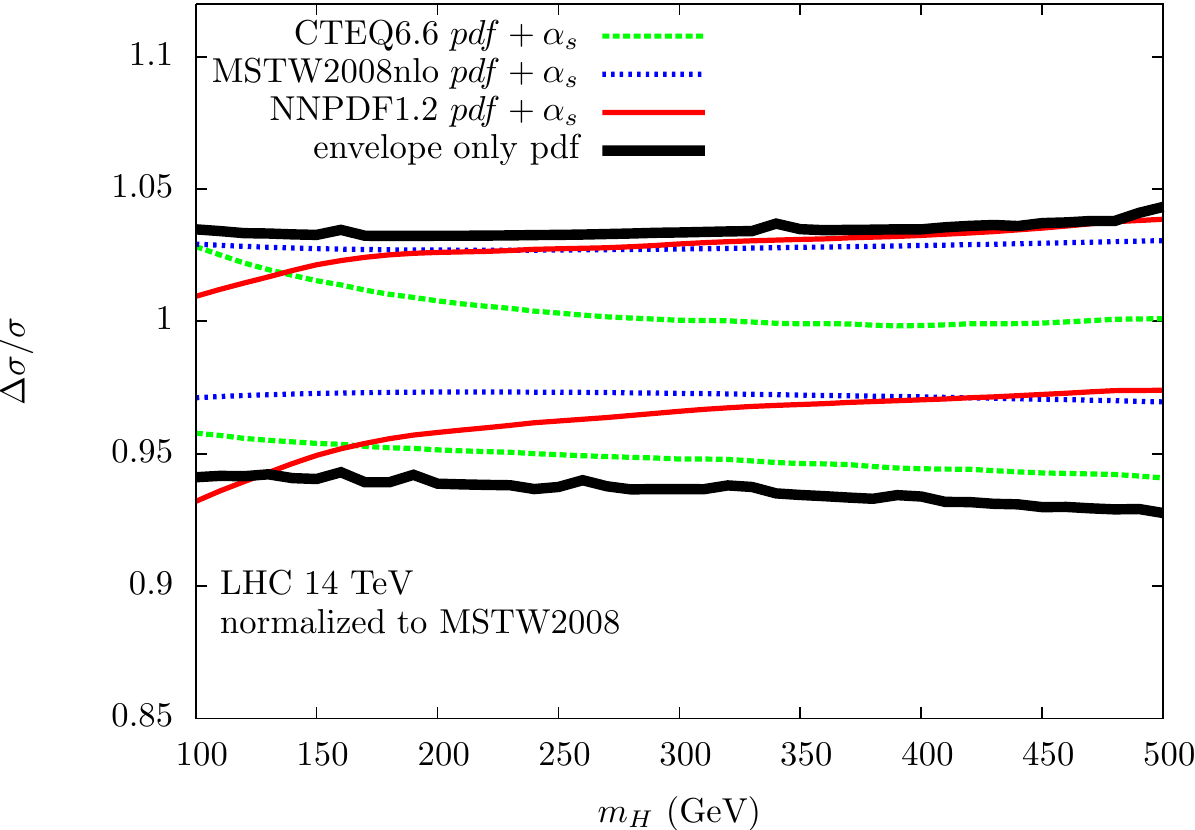}
\caption{\small The
  CTEQ, MSTW and NNPDF curves obtained using the common value
  $\alpha_s=0.119$ in the cross section Eq.~(\ref{Geq}), but with the
  PDF sets and uncertainties corresponding to each group's value of
  $\alpha_s$ Eq.~(\ref{asrefcoll}). All curves are normalized to the
  central MSTW curve  obtained in this way.
The common prediction can be taken
  as the envelope of these curves ({\bf Procedure A}). 
The envelope curve shown ({\bf Procedure B}) is instead
  the envelope of the CTEQ, MSTW and NNPDF 
predictions showed in Fig.~\ref{fig:onlypdf},
  obtained including PDF uncertainties only, but with each  group's value of
  $\alpha_s$ used both in the PDFs and cross section, also normalized
  as the other curves.
\label{fig:finres2}
}
\end{center}
\end{figure}
\item[Procedure B:]
This procedure is based on the observation that in fact the
spread of central values of $\alpha_s$  Eq.~(\ref{asrefcoll}) is
essentially the same as the width as the one--$\sigma$ error band
Eq.~(\ref{sas68}). Because the total uncertainty is well approximated
by the combination of $\alpha_s$ and PDF uncertainties, this suggests
that we can simply substitute the $\alpha_s$ uncertainty by the spread
of values obtained with the three sets. Hence, a conservative estimate
of the one--$\sigma$ total PDF+$\alpha_s$ 
is obtained by taking the envelope of the PDF--only uncertainty bands
obtained using each of the three sets, each at its preferred value 
of $\alpha_s$ Eq.~(\ref{asrefcoll}), {\it i.e.} the envelope of the bands
shown in Fig.~\ref{fig:onlypdf}.
\end{description}

These two conservative estimates are shown in Fig.~\ref{fig:finres2},
where we display the uncertainty bands  obtained from the three MSTW, CTEQ and
NNPDF sets whose envelope corresponds to the first method, as well
as the envelope of the bands of Fig.~\ref{fig:onlypdf}, corresponding
to the second method.
The results turn out to be in near--perfect
agreement, and we can  take them as a conservative estimate of
the PDF+$\alpha_s$ uncertainty.  Note that if any of the three sets
were discarded, 
the prediction would change in a not insignificant way. 

Typical
uncertainties are, for light Higgs,
of order of 10\% at the Tevatron and 5\% at the LHC. Very large
uncertainties are only found for very heavy Higgs at the Tevatron,
which is sensitive to the poorly known large $x$ gluon. 
As a central prediction one may take the midpoint between the upper
and lower bands: in practice, this turns out to be extremely close to the MSTW08
prediction found adopting the previous method, {\it i.e.} using the MSTW08
PDFs but with $\alpha_s=0.119$ in the matrix element.

These results for the combined PDF+$\alpha_s$ uncertainties should
be relevant for Higgs searches both at the Tevatron and at the
LHC. For instance, the latest combined Tevatron analysis on
Higgs production via gluon--fusion~\cite{Aaltonen:2010yv},
which excludes a SM Higgs in the mass range
162-166 GeV at 95\% C.L., quotes a 11\% systematic uncertainty
from PDF uncertainties and higher order variations. It would be interesting
to reassess the above exclusion limits if the combined PDF+$\alpha_s$
uncertainties are estimated as discussed in this work.

\clearpage

\section{Conclusions}
\label{sec:conclusions}

We have presented a systematic study of the impact of PDF and
$\alpha_s$ uncertainties in the total NLO cross section for the
production of standard model Higgs in gluon--fusion. Whereas a
full estimate of the uncertainty on this process would also require a
discussion of other sources of uncertainty, such as electroweak
corrections and the uncertainties related to higher order QCD
corrections (NNLO, soft gluon resummation etc.) our investigation has
focussed on PDF uncertainties, which are likely to be dominant for
many or most LHC standard candles, and the $\alpha_s$ uncertainty
which is tangled with them. The process considered here is one for
which these uncertainties are especially large, and thus it provides a
useful test case.

Our main findings can be summarized as follows:
\begin{itemize}
\item
Parton distributions are correlated to the value of $\alpha_s$ in a
way which is visible, but of moderate significance. In particular, if
$\alpha_s$ is varied within a reasonable range, not much larger than
the current global uncertainty, uncertainties due to PDFs and the
variation of $\alpha_s$ can be considered to good approximation
independent and the total uncertainty can be found adding them in
quadrature.
\item  The gluon luminosities  determined from MSTW08, CTEQ6.6 and
  NNPDF1.2 agree to one $\sigma $ in the sense that their uncertainty
  bands always overlap (though just so for light Higgs at the
  Tevatron). As a consequence, the Higgs cross sections determined using
  these PDF sets agree provided only the same value of $\alpha_s$ is
  used in the computation of the hard cross section. The spread of
  central values between sets is of the order of the PDF uncertainty.
\item The PDF uncertainties determined using these sets are in
  reasonable agreement and always differ by a factor less than two,
  while  being affected by an uncertainty which is likely to be
  about half their size. The $\alpha_s$ uncertainties are essentially
  independent of the PDF set, and provide more than half of the
  combined uncertainty. The combined uncertainties determined using
  the  sets under investigation are in thus in good agreement with
  each other.
\item A conservative estimate of the total uncertainty can be obtained
  from the envelope of the PDF+$\alpha_s$ uncertainties obtained from
  each set, all evaluated with a common central $\alpha_s$ value.
  Equivalently, it can be obtained 
from the envelope of the PDF--only uncertainties of sets
  evaluated each at different value of $\alpha_s$, within a range of
  values which covers the accepted $\alpha_s$ uncertainty.
\item A typical conservative PDF+$\alpha_s$ uncertainty is, for light
  Higgs, 
of order
  10\% at the Tevatron and 5\% at the LHC. This is at most a factor
  two larger than the PDF+$\alpha_s$ uncertainty obtained using each
  individual parton set. Exclusion of any of the three sets
  considered here would lead to a total uncertainty which is rather
  closer to that of individual parton sets.
\end{itemize}

Further improvements in accuracy could be obtained by accurate
benchmarking and cross--checking of PDF determinations in order to
isolate and understand the origin of existing disagreements. 
However, the overall
agreement of existing sets appear to be satisfactory even for this
worst--case scenario.

\bigskip
\noindent
{\bf Acknowledgements}: 
We thank all the members of the NNPDF collaboration, which has
developed the PDF fitting methodology and code which has been used to
produce the Monte Carlo varying--$\alpha_s$ PDF  sets used in this
study.
S.F. thanks the  members of the PDF4LHC workshop for various
discussions on PDF uncertainties and their impact on LHC observables, 
in particular Joey Huston and
Robert Thorne.
This work was partly supported 
by the European network HEPTOOLS under contract MRTN-CT-2006-035505.

\bigskip
\bigskip

\noindent
{\bf Note added:} As this paper was being finalized, a
study~\cite{Baglio:2010um} of uncertainties on Higgs production has
appeared. This paper presents detailed investigations of uncertainties
which are not being discussed by us, specifically electroweak and
scale (higher order QCD) uncertainties, while it addresses only marginally the
issue on which we have concentrated, namely the interplay of
$\alpha_s$ and PDF uncertainties.


\end{document}